\documentclass[a4paper,preprint]{imsart}
\usepackage[OT1]{fontenc}
\usepackage[utf8]{inputenc}
\usepackage{amsmath,amsthm,amssymb}
\usepackage{graphicx}
\usepackage[pdftex,colorlinks]{hyperref}

\usepackage{natbib}
\bibpunct{(}{)}{,}{a}{}{;}

\startlocaldefs
\usepackage[ruled,vlined]{algorithm2e}
\theoremstyle{definition}
\newtheorem{proposition}{Proposition}
\theoremstyle{remark}

\newcommand{\argmax}{\mathop{\mathrm{arg\ max}}}

\newcommand{\set}[1]{\left\{#1\right\}}

\newcommand\xylabel[3]{
  \begin{tabular}{@{}r@{\hspace{0.25em}}l@{}}
    \rotatebox{90.0}{\makebox[0.16cm]{#3}} &
    \raisebox{-1.1cm}{\includegraphics[width=0.16\textwidth]{#1}} \\
    & \multicolumn{1}{c}{#2}
  \end{tabular}
}

\endlocaldefs

\begin{document}

\begin{frontmatter}

  \title{Inferring Multiple Graphical Structures}
  
  \runtitle{Inferring Multiple Graph Structures}

  \begin{aug}
    \author{\fnms{Julien} \snm{Chiquet}\corref{}},
    \author{\fnms{Yves} \snm{Grandvalet}},
    \author{\fnms{Christophe} \snm{Ambroise}}.
    \ead[label=e1]{[julien.chiquet,christophe.ambroise]@genopole.cnrs.fr}
    \ead[label=e3]{yves.grandvalet@utc.fr}
    \ead[label=u]{http://stat.genopole.cnrs.fr}

    \address{Laboratoire Statistique et G\'enome\\
      523, Place des Terrasses, 91000 \'Evry\\
      \printead{e1,e3}\\
      \printead{u}} 
    
    \affiliation{Universit\'{e} d'\'{E}vry  Val d'Essonne --  CNRS UMR
      8071 Statistique et G\'{e}nome \and 
      Universit\'{e} de Technologie de Compi\`{e}gne -- CNRS UMR 6599 Heudiasyc}
    \runauthor{Chiquet, Grandvalet and Ambroise} 
  \end{aug}
  
  \begin{abstract}
    Gaussian  Graphical  Models  provide  a convenient  framework  for
    representing dependencies between  variables.  Recently, this tool
    has  received a  high  interest for  the  discovery of  biological
    networks.   The literature  focuses  on the  case  where a  single
    network is  inferred from  a set of  measurements, but,  as wetlab
    data is  typically scarce, several assays,  where the experimental
    conditions  affect interactions,  are  usually merged  to infer  a
    single  network.  In  this paper,  we propose  two  approaches for
    estimating  multiple related  graphs, by  rendering  the closeness
    assumption into an empirical prior or group penalties.  We provide
    quantitative  results demonstrating the  benefits of  the proposed
    approaches. The methods presented in this paper are embeded in the
    \texttt{R} package \texttt{simone} from version 1.0-0 and later.
  \end{abstract}
  
  \tableofcontents
 
\end{frontmatter}

\sloppy

\section{Motivations}
\label{sec:introduction}

Systems biology provides a large amount of data sets that aim to
understand  the complex relationships  existing between  the molecular
entities that drive any  biological process. Depending on the molecule
of  interest, various  networks  can be  inferred, e.g.,  gene-to-gene
regulation networks or  protein-protein interaction networks.  The basic
idea  is to  consider that  if two  molecules interact,  a statistical
dependency between their expression should be observed.

A  convenient model  of multivariate  dependence patterns  is Gaussian
Graphical  Modeling  (GGM).  In  this  framework,  a  multidimensional
Gaussian  variable  is characterized  by  the so-called  concentration
matrix, where  conditional independence between pairs  of variables is
characterized by a  zero entry.  This matrix may  be represented by an
undirected graph, where each vertex represents a variable, and an edge
connects two  vertices if the  corresponding pair of  random variables
are dependent, conditional on the remaining variables.

Merging different experimental conditions from wetlab data is a common
practice         in         GGM-based        inference         methods
\citep{2002_Toh,2005_SAGMB_Schafer}.  This process enlarges the number
of observations  available for inferring  interactions.  However, GGMs
assume  that the  observed data  form an  independent  and identically
distributed (i.i.d.) sample.  In the aforementioned paradigm, assuming
that the merged  data is drawn from a single
Gaussian  component  is  obviously   wrong,  and  is  likely  to  have
detrimental side effects in the estimation process.

In this paper, we propose to remedy this problem by estimating
multiple GGMs, each of whom matching different modalities of the same
set of variables, which correspond here to the different experimental
conditions.  As the distributions of these modes have strong
commonalities, we propose to estimate these graphs jointly.
Considering several tasks at a time has been attracting much attention in the
machine learning literature, where the generic problem is usually referred to as
``multi-task learning'' \citep{Caruana97}.
\citet{2009_report_Efron} used the terms ``indirect evidence'' and ``learning
from the experience of others'' for similar ideas.  The
principle is to learn an inductive bias, whose role is to stabilize the
estimation process, hopefully enabling more accurate predictions in small sample
size regimes \citep{2000_JAIR_Baxter}.
The techniques comprise the empirical and hierarchical Bayes methodologies and
regularization schemes \citep[see][for example]{Argyriou08}, which may be
interpreted as approximations to the latter.
Here, we will mainly follow the penalty-based approach to leverage the inference of
several related graphs towards a common pattern.

A typical example of this
problem arises when inferring gene interactions from 
data measured on slightly different stem cells, such as wild and mutant. 
It is reasonable to assume that most, though not all, interactions will be
common to both types of cells. The  line of attack we propose  alleviates
the difficulties arising from the scarcity of data in each
experimental condition by coupling the estimation problems.  Our first
proposal biases the estimation of the concentration matrices towards a
common value.
Our second proposition focuses on the similarities in the sparsity pattern that
are more directly related to the graph itself.  We propose the Cooperative-LASSO, which
builds on the Group-LASSO \citep{2006_Yuan} to favor solutions with a common
sparsity pattern, but encodes a further preference towards solutions with
similar sign patterns, thus preserving the type of co-regulation (positive or
negative) across assays.

To our knowledge, the present work is the first to exploit the multi-task
learning framework for learning GGMs.  However, coupling the estimation of
several networks has recently been investigated for Markov random fields, in
the context of time-varying networks.  \citet{2009_AAS_Kolar} propose
two specific constraints, one for smooth variations over time, the other one for
abrupt changes.  Their penalties are closer to the Fused-LASSO
and total variations penalties than to the group penalties proposed here.


\section{Network Inference with GGM}
\label{sec:background}

In  the  GGM framework,  we  aim to  infer  the  graph of  conditional
dependencies among the $p$ variables  of a vector $X$ from independent
observations  $(X^1,   \dots,  X^n)$.   We   assume  that  $X$   is  a
$p$-dimensional           Gaussian           random           variable
$X\sim\mathcal{N}(\mathbf{0}_p,       {\boldsymbol\Sigma})$.       
Let $\mathbf{K} = {\boldsymbol\Sigma}^{-1}$ be the concentration matrix of the
model; the non-zero entries of $K_{ij}$ indicate a conditional dependency
between the variables $X_i$ and $X_j$, and thus define the graph $\mathcal{G}$
of conditional dependencies of~$X$.

The  GGM approach produces  the graph  $\mathcal{G}$ from  an inferred
$\mathbf{K}$.   The latter  cannot be  obtained by  maximum likelihood
estimation  that would  typically return  a full  matrix, and  hence a
useless   fully  connected   graph.   To   produce   sparse  networks,
\citet{2008_JMLR_Banerjee}   propose  to   penalize  the   entries  of
$\mathbf{K}$              by             an             $\ell_1$-norm.
\citet{2007_BS_Friedman} latter addressed  the   very
same  problem        with    an    elegant   algorithm    named    the
  \emph{graphical-LASSO}.   Their well-motivated approach  produces a
sparse, symmetric and  positive-definite estimate of the concentration
matrix.   However,  a cruder  though  more  direct  approach has  been
reported   to  be   more   accurate  in   terms   of  edge   detection
\citep{2008_SAGM_Villers,2008_preprint_Rocha}.      This     approach,
proposed by  \citet{2006_AS_Meinshausen} and known as
  \emph{neighborhood  selection},  determines  $\mathcal{G}$  via  an
iterative  estimation of  the  neighborhood of  its  nodes.  For  this
purpose,  it considers  $p$ independent  $\ell_1$-penalized regression
problems.   Let $\mathbf{X}$  be  the $n\times  p$  matrix of  stacked
observations,  whose   $k$th  row  contains   $(X^k)^\intercal$.   The
vertices adjacent to vertex $i$ are estimated by the non-zero elements
of $\boldsymbol\beta$ solving
\begin{equation}
  \label{eq:lasso_MB}
  \min_{\boldsymbol\beta\in\mathbb{R}^{p-1}}     \frac{1}{n}\left\|
    \mathbf{X}_i   -   \mathbf{X}_{\backslash   i}   \boldsymbol\beta
  \right\|_2^2 + \lambda \|\boldsymbol\beta\|_{1}
  \enspace,
\end{equation}
where  $\mathbf{X}_i$   is  the  $i$th  column   of  $\mathbf{X}$  and
$\mathbf{X}_{\backslash  i}$  is $\mathbf{X}$  deprived  of its  $i$th
column: the $i$th variable is ``explained'' by the remaining ones.  As
the  neighborhood of  the  $p$ variables  are  selected separately,  a
post-symmetrization must be  applied to manage inconsistencies between
edge  selections; \citeauthor{2006_AS_Meinshausen}  suggest AND  or OR
rules,  which  are both  asymptotically  consistent  (as  $n$ goes  to
infinity).

Solving  the  $p$   regression  problems  \eqref{eq:lasso_MB}  may  be
interpreted  as inferring  the  concentration matrix  in a  penalized,
\emph{pseudo}   maximum   likelihood   framework,  where   the   joint
distribution  of  $X$  is  approximated  by the  product  of  the  $p$
distributions  of   each  variable  conditional  on   the  other  ones
\citep{2008_preprint_Rocha,2009_article_Ambroise,2010_AS_Ravikumar},
that is
\begin{equation*}
  \mathcal{L}(\mathbf{K}|\mathbf{X}) =  \sum_{i=1}^p \left(\sum_{k=1}^n
    \log \mathbb{P}(X_i^k|X^k_{\backslash i};\mathbf{K}_i)\right)
  \enspace,
\end{equation*}
where $X_{\backslash i}^k$ is the $k$th realization of the vector $X$
deprived  of  the  $i$th  coordinate.
Considering the Gaussian assumption on the generation of the data $\mathbf{X}$, the
pseudo-log-likelihood admits a compact and simple expression (see derivation in
Appendix~\ref{sec:derivation:pseudolikelihood}):
\begin{multline}
  \label{eq:pseudologlikelihood}
  \mathcal{L}(\mathbf{K}|\mathbf{X})    =   \frac{n}{2}   \log\det
  (\mathbf{D}) -\frac{n}{2} 
  \mathrm{Tr}\left(\mathbf{D}^{-\frac{1}{2}}\mathbf{K}     \mathbf{S}
    \mathbf{K}\mathbf{D}^{-\frac{1}{2}}\right) \\ - \frac{np}{2}
  \log(2\pi)
\enspace,
\end{multline}
where $\mathbf{S} =n^{-1} \mathbf{X}^\intercal \mathbf{X}$ is the empirical
covariance matrix, and $\mathbf{D}$ is a $p \times p$ diagonal matrix with elements
$D_{ii} = K_{ii}$. In  the  sequel,  it  will  be
convenient to  use the  sufficiency of $\mathbf{S}$  for $\mathbf{K}$,
and,     by     a     slight     abuse     of     notations,     write
$\mathcal{L}(\mathbf{K}|\mathbf{X})=\mathcal{L}(\mathbf{K}|\mathbf{S})$.

Following \citet{2008_JMLR_Banerjee}, an $\ell_1$ penalty may be added
to     obtain      a     sparse     estimate      of     $\mathbf{K}$.
Nevertheless,   our   approach   to  maximizing   the
  pseudo-log-likelihood is  much simpler than the  optimization of the
  log-likelihood  proposed by \citet{2008_JMLR_Banerjee}.   Indeed, as
  stated  formally  in   the  following  proposition,  maximizing  the
  penalized  pseudo-log-likelihood on  the set  of  arbitrary matrices
  (not constrained to be  either symmetric or positive definite) boils
  down  to  solving $p$  independent  LASSO  problems  of size  $p-1$.
  Furthermore,  for the  purpose of  discovering the  graph structure,
  additional  computational  savings are  achieved  by remarking  that
  $\mathbf{D}$ needs not to be estimated.  We thus avoid the iterative
  scheme of  \citet{2008_preprint_Rocha} alternating optimization with
  respect  to  $\mathbf{D}$  and   to  the  off-diagonal  elements  of
  $\mathbf{D}^{-1} \mathbf{K}$.
\begin{proposition}
 \label{prop:blockwise_resolution} Consider the following
reordering of the rows and columns of $\mathbf{K}$ and $\mathbf{S}$:
\begin{equation}
  \label{eq:K_block}
   \begin{bmatrix}
    \mathbf{K}_{\backslash i  \backslash i} &  \mathbf{K}_{i  \backslash i} \\
    \mathbf{K}_{i  \backslash i}^\intercal &  K_{ii} \\
  \end{bmatrix}, \quad 
  \begin{bmatrix}
    \mathbf{S}_{\backslash i  \backslash i} & \mathbf{S}_{i  \backslash i} \\
    \mathbf{S}_{i  \backslash i}^\intercal & S_{ii} \\
  \end{bmatrix} \enspace,
\end{equation}
where $\mathbf{K}_{\backslash  i\backslash i}$ is  matrix $\mathbf{K}$
deprived  of  its   $i$th  column  and  its  $i$th   line,  and  where
$\mathbf{K}_{i  \backslash i}$  is  the $i$th  column of  $\mathbf{K}$
deprived of its $i$th element. The problem
\begin{equation}\label{eq:pen_pseudologlikelihood}
   \max_{\{K_{ij}:i\neq    j\}}    \;   \;
   \mathcal{L}(\mathbf{K}|\mathbf{S}) - \lambda \| \mathbf K\|_{1}
   \enspace,
\end{equation}
where $\|\mathbf{K}\|_{1}$ is the componentwise $\ell_1$-norm, 
can be solved column-wisely by considering
$p$ LASSO problems in form
\begin{equation}
  \label{eq:pseudolikelihood_penalized_blockwise}
  \min_{ \boldsymbol\beta \in \mathbb{R}^{p-1}}
  \frac{1}{2} \left\| \mathbf{S}_{\backslash i  \backslash i}^{1/2} \boldsymbol\beta +
    \mathbf{S}_{\backslash i  \backslash i}^{-1/2}\mathbf{S}_{i  \backslash i}\right\|_2^2 + 
  \frac{\lambda}{n} \left\| \boldsymbol\beta \right\|_1
  \enspace,
\end{equation}
where   the   optimal   $\boldsymbol\beta$   is   the   maximizer   of
\eqref{eq:pen_pseudologlikelihood}        with        respect       to
$K_{ii}^{-1}\mathbf{K}_{i    \backslash     i}$    as    defined    in
\eqref{eq:K_block}.  Hence, Problem \eqref{eq:pen_pseudologlikelihood}
may      be      decomposed       into      the      $p$      Problems
\eqref{eq:pseudolikelihood_penalized_blockwise}    of   size   ${p-1}$
generated by the $p$ possible permutations in \eqref{eq:K_block}.

The            full            solution            to
  Problem~\eqref{eq:pen_pseudologlikelihood}, in $\{K_{ij}:i\neq j\}$,
  also requires $K_{ii}$ \citep[see][]{2008_preprint_Rocha}.  However,
  since our  interest is to  unveil the graph structure,  the sparsity
  pattern of the penalized maximum likelihood estimate of $\mathbf{K}$
  is    sufficient,   and   the    latter   is    directly   recovered
  from~$\boldsymbol\beta$.
\begin{proof}
See appendix \ref{sec:blockdecomposition}.
\end{proof}
\end{proposition}%

From the definition of the covariance matrix $\mathbf{S}$, it is clear that
Problem \eqref{eq:lasso_MB} is a slight reparameterization of Problem
\eqref{eq:pseudolikelihood_penalized_blockwise}.  Hence, the graph produced by
the approach of \citet{2006_AS_Meinshausen} is identical to the one obtained by
maximizing the penalized pseudo likelihood \citep[Proposition
8]{2009_article_Ambroise}.


\section{Inferring Multiple GGMs}\label{sec:solving}

In transcriptomics, it  is a common practice to  conduct several assays
where  the experimental  conditions differ,  resulting in  $T$ samples
measuring the  expression of the  same molecules.  From  a statistical
viewpoint, we have $T$ samples belonging to different sub-populations,
hence with different distributions.
Assuming that each sample was drawn independently from a Gaussian distribution
$X^{(t)}\sim\mathcal{N}(\mathbf{0}_p,{\boldsymbol\Sigma}^{(t)})$, the $T$ samples
may be processed separately by following the approach described in Section
\ref{sec:background}.
The objective function is expressed compactly as a sum:
\begin{equation}
  \label{eq:T_indep}
  \max_{\{ K_{ij}^{(t)} : i\neq j \}_{t=1}^T} \;
  \sum_{t=1}^T \left( 
    \mathcal{L}(\mathbf{K}^{(t)}|\mathbf{S}^{(t)}) 
    - \lambda  \|\mathbf{K}^{(t)}\|_{1}
  \right) \enspace.
\end{equation}
Note that it is sensible to apply the same penalty parameter $\lambda$ for all
samples provided that the $T$ samples have similar sizes and originate from similar distributions,
in particular regarding scaling and sparseness.

Problem \eqref{eq:T_indep} ignores the relationships between regulation networks.
When
the tasks are known to have strong commonalities, the multi-task
learning framework is well adapted, especially for small sample sizes, where
sharing information may considerably improve estimation accuracy.
To couple the estimation problems, we have to break the separability in
$\mathbf{K}^{(1)},\dots,\mathbf{K}^{(T)}$ in Problem \eqref{eq:T_indep}.  This
may be achieved either modifying the data-fitting term or the penalizer.  These
two options result respectively in the graphical Intertwined-LASSO and the
graphical Cooperative-LASSO presented below.

\subsection{Intertwined Estimation}

In the \emph{Maximum A Posteriori} framework, the estimation of a concentration
matrix can be biased towards a specific value, say $\mathbf{S}_{0}^{-1}$.  From a
practical viewpoint, this is usually done by considering a
conjugate prior on $\mathbf{K}$ , that is, a Wishart distribution 
${\cal W}(\mathbf{S}_{0}^{-1},n)$.
The MAP estimate is then computed as if we had observed additional observations
of empirical covariance matrix $\mathbf{S}_{0}$.

Here, we would like to bias
each estimation problem towards the same concentration matrix, whose value is
unknown.  An empirical Bayes solution would be to set
$\mathbf{S}_{0} =\bar{\mathbf{S}}$, where $\bar{\mathbf{S}}$ is the weighted
average of the $T$ empirical covariance matrices.
As in the maximum likelihood framework, this approach would lead to a full
concentration matrix. Hence, we will consider here a penalized criterion, which 
does not exactly fit the penalized maximum likelihood nor the MAP frameworks, but
that will perform the desired coupling between the estimates of
$\mathbf{K}^{(1)},\dots,\mathbf{K}^{(T)}$ while pursuing the original sparseness
goal.

Formally, let $n_1,\dots,n_T$ be the sizes of the respective
samples,   whose  empirical   covariance  matrices   are   denoted  by
$\mathbf{S}^{(1)}, \dots,\mathbf{S}^{(T)}$. Also denote $n=\sum n_t$, we
consider the following problem:
\vspace{-1em}
\begin{equation}
  \label{eq:criterion1}
  \max_{\{ K_{ij}^{(t)} : i\neq j \}_{t=1}^T} \;
  \sum_{t=1}^T \left( 
  \mathcal{L}(\mathbf{K}^{(t)}|\tilde{\mathbf{S}}^{(t)})   -
  \lambda \|\mathbf{K}^{(t)}\|_{1}
  \right) \enspace,
\end{equation}
where   $\tilde{\mathbf{S}}^{(t)}    =   \alpha   \mathbf{S}^{(t)}   +
(1-\alpha)\bar{\mathbf{S}}$   and   $\bar{\mathbf{S}}  =   n^{-1}
\sum_{t=1}^T n_t\mathbf{S}^{(t)}$.
As this criterion amounts to consider that we observed a blend of the actual
data for task $t$ and data from the other tasks, we will refer to this approach 
as intertwined estimation.
The idea is reminiscent of the compromise between linear discriminant analysis
and its quadratic counterpart performed by the regularized discriminant analysis
of \citet{1989_JASA_Friedman}.
Although the tools are similar, the primary goals differ:
\citet{1989_JASA_Friedman} aims at getting a control on the number of effective
parameters, we want to bias empirical distributions towards a common model.

In order to avoid multiple hyper-parameter tuning, the results shown in the
experimental section were obtained with $\alpha$ arbitrarily set to
{\small$1/2$}. More refined strategies are left for future work.

\subsection{Graphical Cooperative-LASSO}
\label{sec:cooplasso}

The second approach consists in devising penalties that encourage similar
sparsity patterns across tasks, such as the Group-LASSO \citep{2006_Yuan}, which
has already inspired some multi-task learning strategies \citep{Argyriou08},
but was never considered for learning graph models.
We shortly describe how Group-LASSO may be used for inferring multiple graphs before
introducing a slightly more complex penalty that was inspired by the application
to biological interactions, but should be relevant in many other applications.

As in the single task
case,  sparsity  of the  concentration  matrices  is  obtained via  an
$\ell_1$  penalization  of their  entries.   An additional  constraint
imposes the  similarity between  the two concentration  matrices. Each
interaction is considered as a group.

The Group-LASSO is a mixed norm that encourages sparse solutions with respect to
groups, where groups form a pre-defined partition of variables.
In the GGM framework, by grouping the partial correlations between variables
across the $T$ tasks, such a penalty will favor graphs
$\mathcal{G}_1,\ldots,\mathcal{G}_T$ with common edges.
The learning problem is then
\begin{equation}
   \label{eq:criterion2}
  \max_{\{ K_{ij}^{(t)} : i\neq j \}_{t=1}^T} \;
  \sum_{t=1}^T
   \mathcal{L}(\mathbf{K}^{(t)}|\mathbf{S}^{(t)})  - \lambda
   \sum_{ i\neq  j} \bigg(\sum_{t=1}^T
     \left(K_{ij}^{(t)}\right)^2\bigg)^{1/2}.
\end{equation}
Though this formalization expresses some of our expectations regarding the
commonalities between tasks, it is not really satisfying here since we aim at
inferring the support of the solution (that is, the set of non-zero entries of
$\mathbf{K}^{(t)}$).  To enable the inference of different networks $(t,u)$, we must have
some $(i,j)$ such that $K_{ij}^{(t)}=0$ and $K_{ij}^{(u)}\neq0$.
This event occurs with probability zero with the Group-LASSO, whose
variables enter or leave the support group-wise \citep{2006_Yuan}. 
However, we may cure this problem by considering a regularization term that
better suits our needs. 
Namely, when the graphs represent the regulation networks of the same set of
molecules across experimental conditions, we expect a stronger similarity pattern
than the one expressed in \eqref{eq:criterion2}.
Specifically, the co-regulation encompasses up-regulation and
down-regulation and the type of regulation is not likely to be inverted across
assays: in terms of partial
correlations, sign swaps are very
unlikely. This additional constraint is formalized in the following learning problem:
\begin{multline}%
  \label{eq:criterion3}%
  \max_{\{ K_{ij}^{(t)} : i\neq j \}_{t=1}^T} \;
   \sum_{t=1}^T
   \mathcal{L}(\mathbf{K}^{(t)}|\mathbf{S}^{(t)}) 
   \\[-0.75ex] 
   - \lambda \sum_{i\neq j}
    \left( 
     \bigg( \sum_{t=1}^T \left( K_{ij}^{(t)} \right)_{\!+}^2 \bigg)^{1/2}
     \hspace{-1em} +
     \bigg( \sum_{t=1}^T \left(-K_{ij}^{(t)} \right)_{\!+}^2 \bigg)^{1/2}
    \right)
    \enspace,
\end{multline}
where $\left( u \right)_{\!\!+} = \max(0,u)$.

\begin{figure*}
  \centering
    \begin{small}
      \begin{tabular}{|*{4}{c@{\hspace*{0.25cm}}|}}
        \hline
        & $\beta_2^{(2)}=0$ & $\beta_2^{(2)}=0.1$ & $\beta_2^{(2)}=0.3$ \\ \hline
        & & & \\ 
        &
        \includegraphics[width=0.25\textwidth]{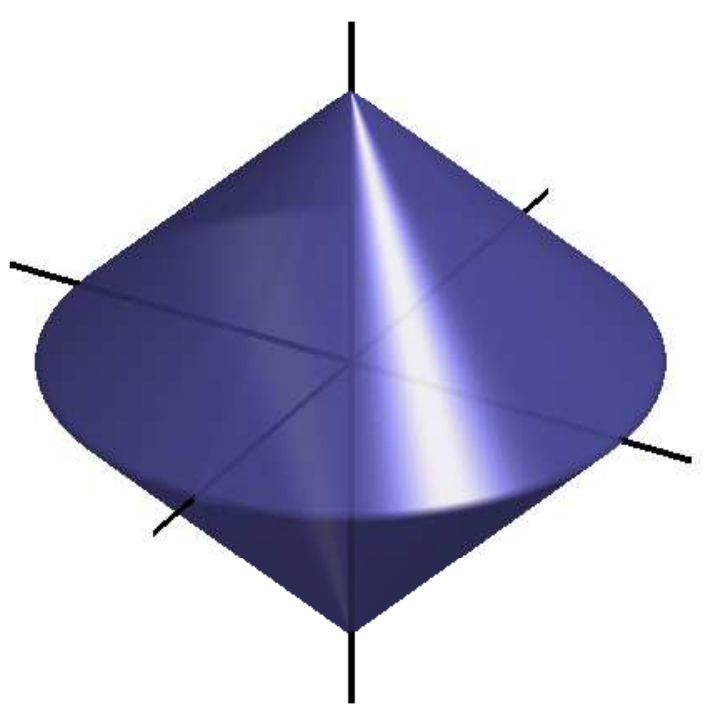} &
        \includegraphics[width=0.25\textwidth]{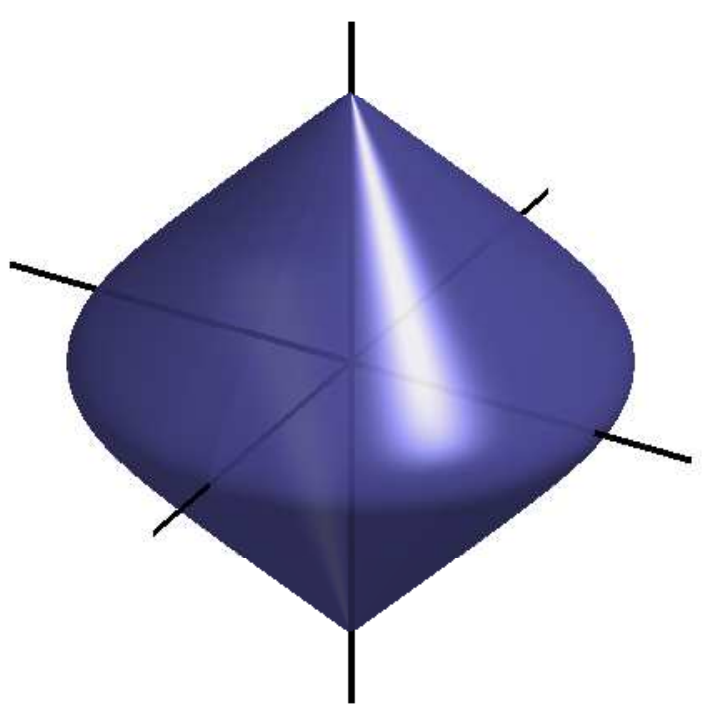} &
        \includegraphics[width=0.25\textwidth]{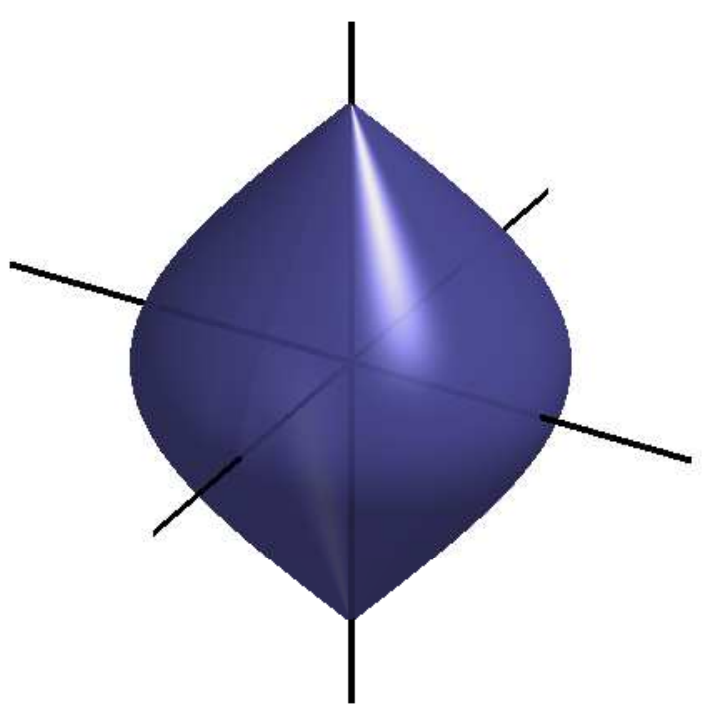} \\
        \hline & & & \\
        \rotatebox{90.0}{\makebox[0cm]{$\beta_1^{(2)}=0$}}&
        \xylabel{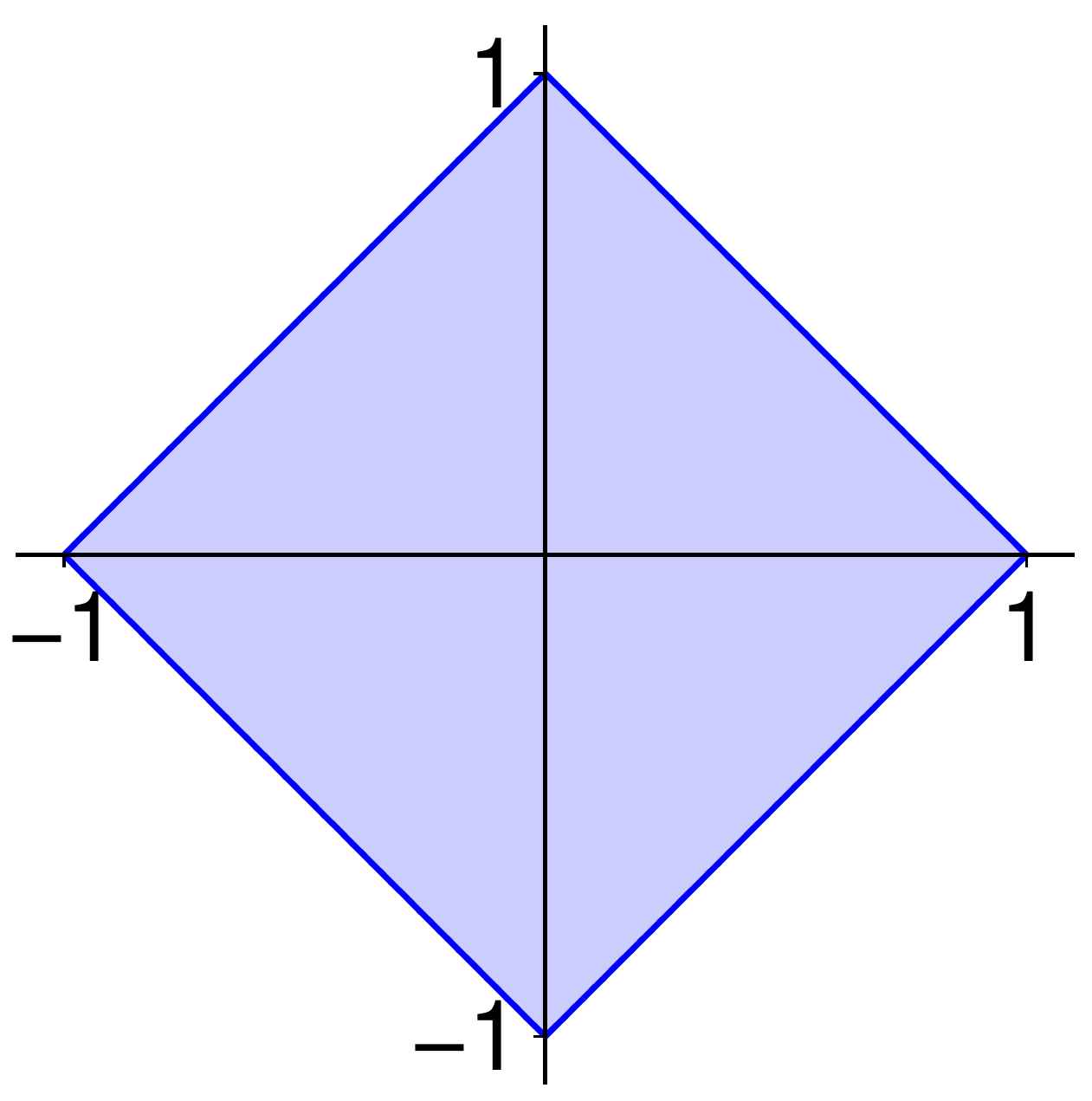}{$\beta_1^{(1)}$}{$\beta_2^{(1)}$} &
        \xylabel{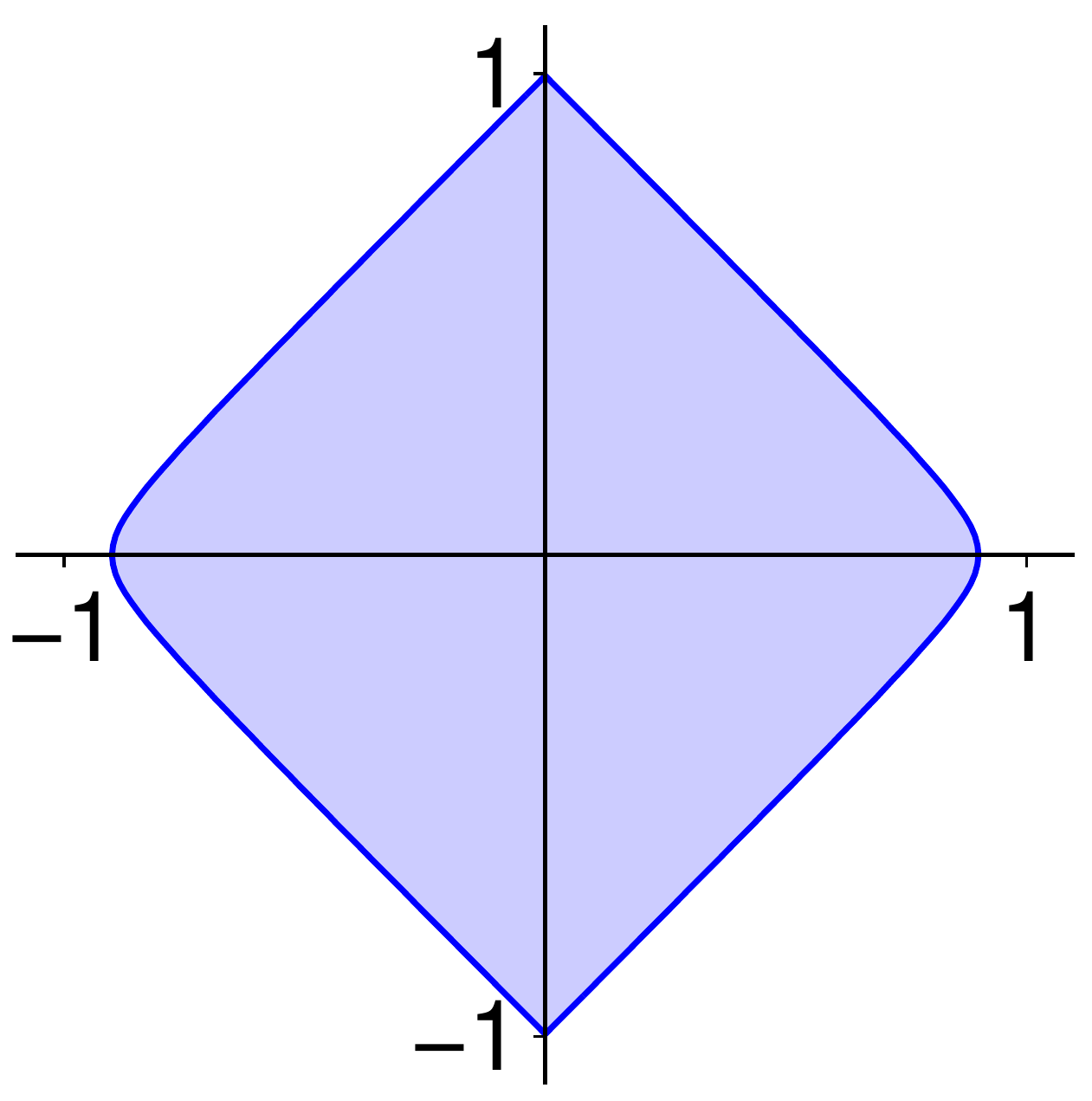}{$\beta_1^{(1)}$}{$\beta_2^{(1)}$} &
        \xylabel{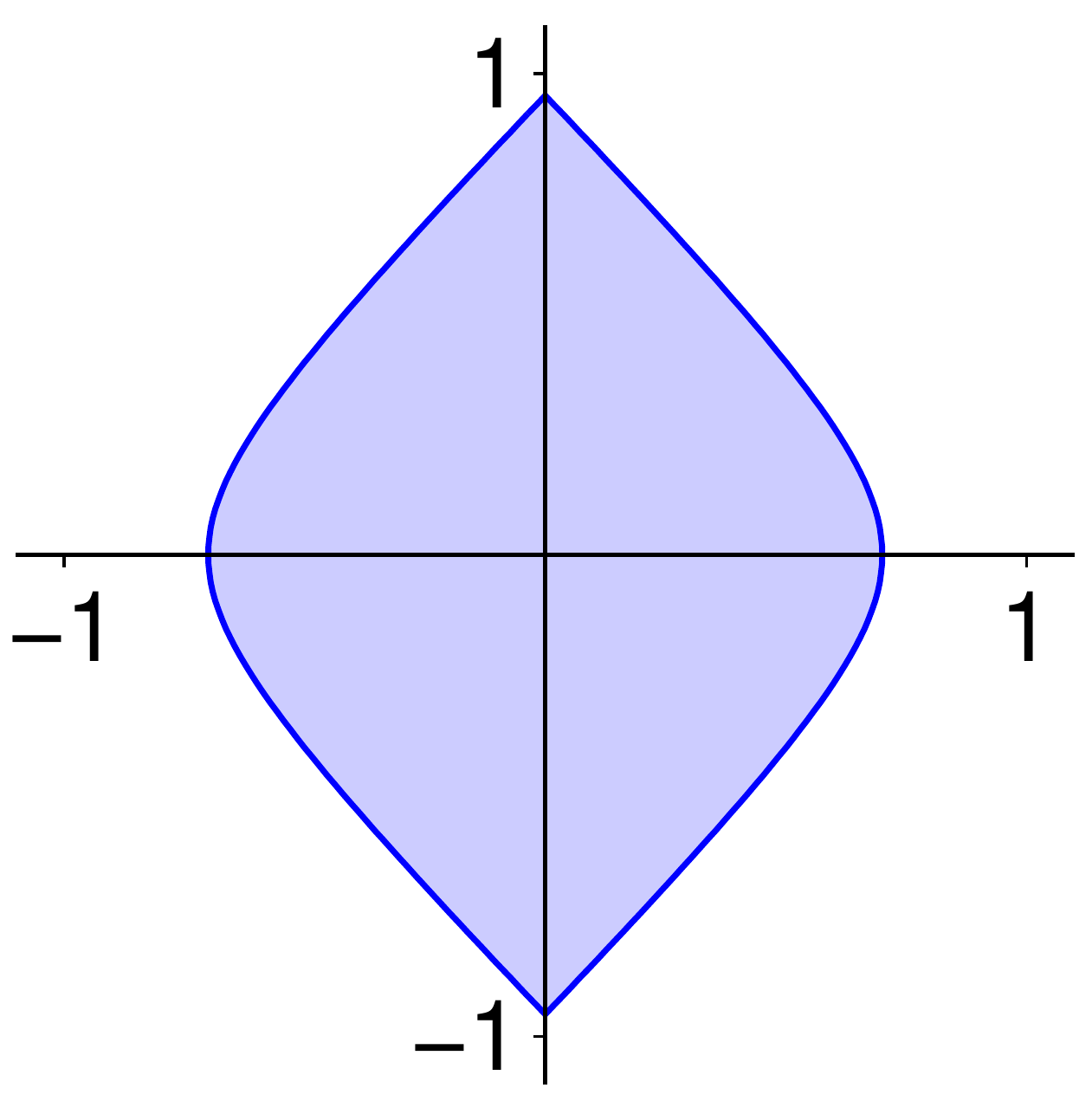}{$\beta_1^{(1)}$}{$\beta_2^{(1)}$}\rule[-5em]{0pt}{0pt} \\
        \hline & & & \\
        \rotatebox{90.0}{\makebox[0.3cm]{$\beta_1^{(2)}=0.1$}}&
        \xylabel{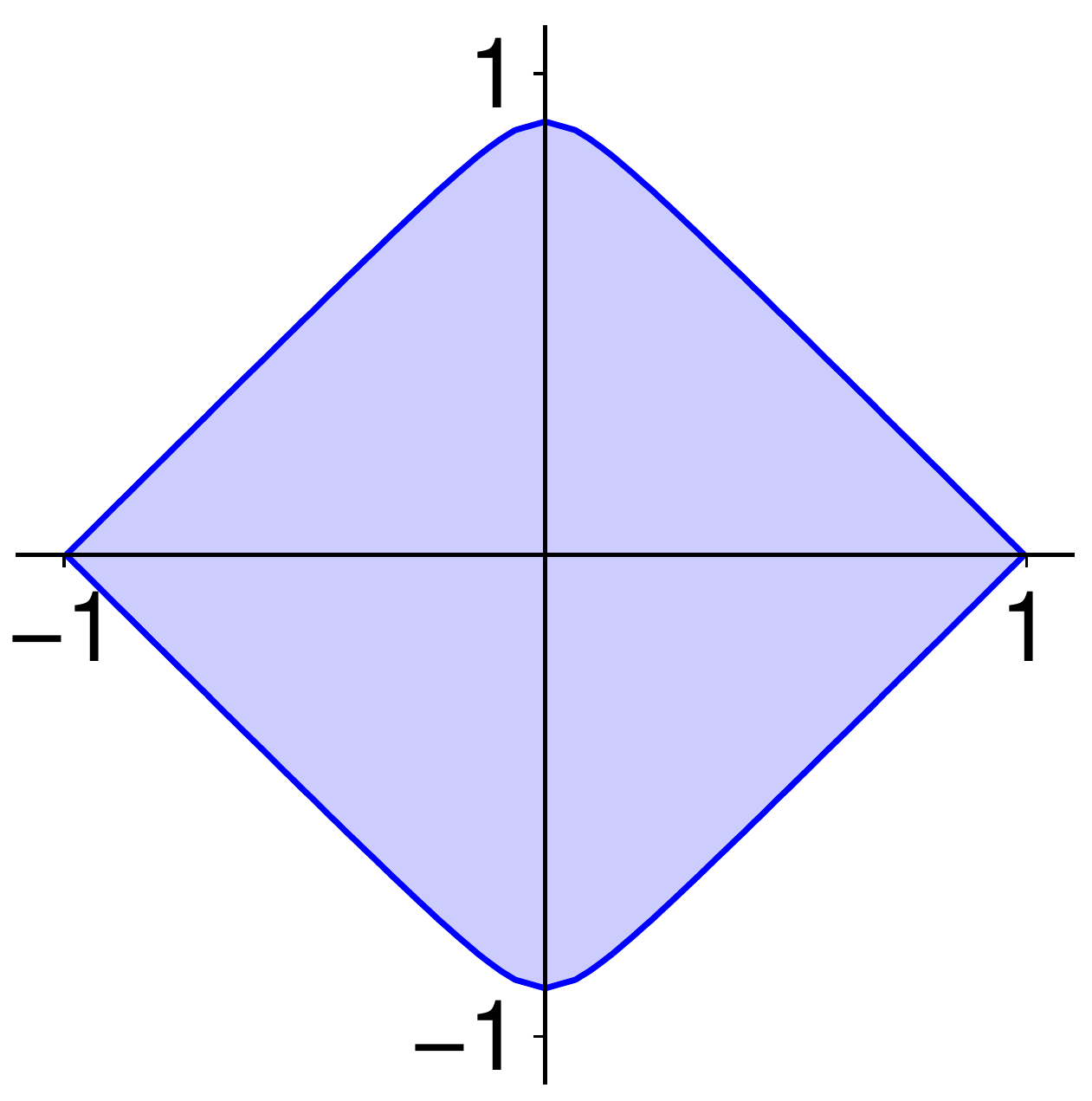}{$\beta_1^{(1)}$}{$\beta_2^{(1)}$} &
        \xylabel{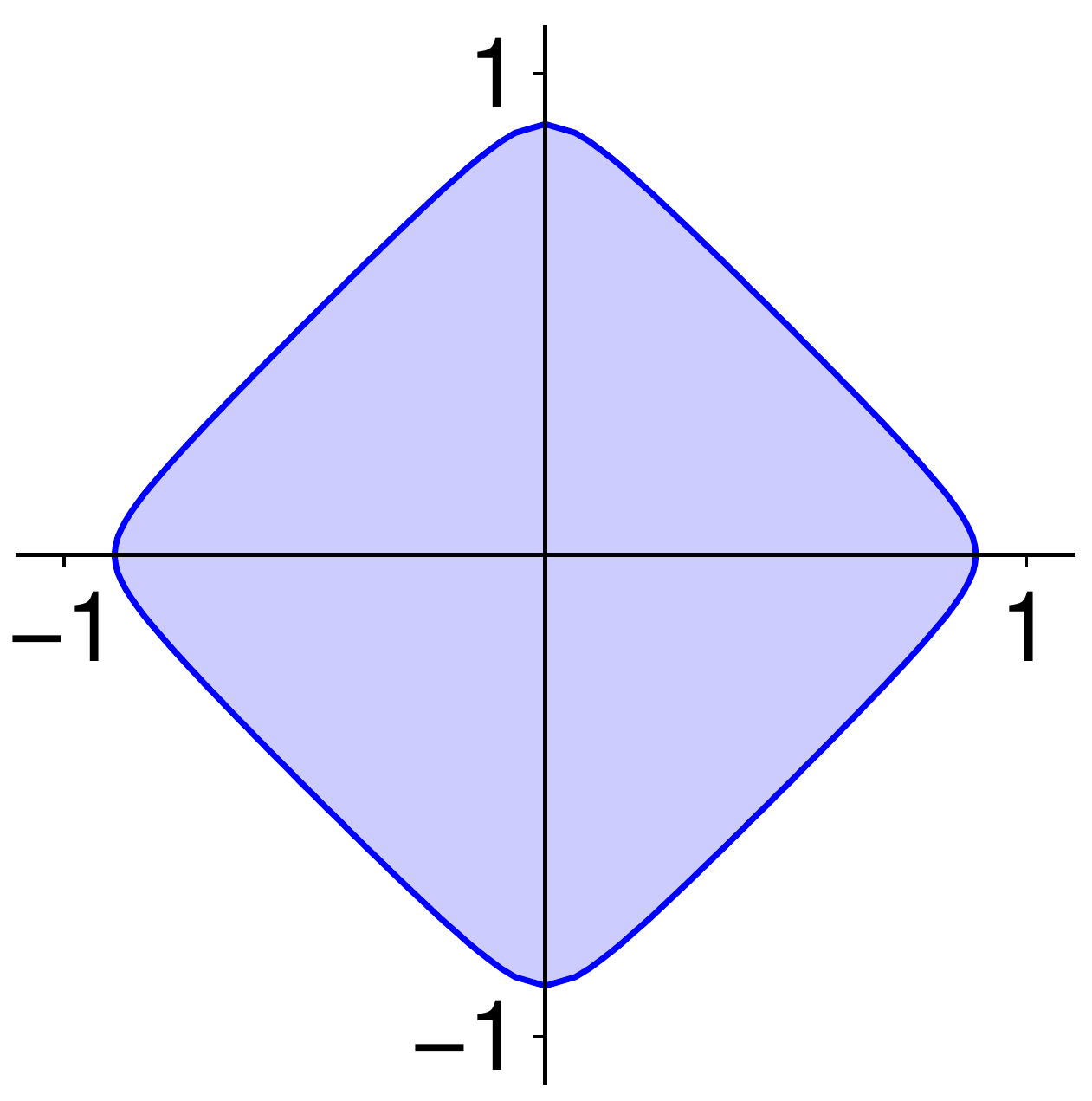}{$\beta_1^{(1)}$}{$\beta_2^{(1)}$} &
        \xylabel{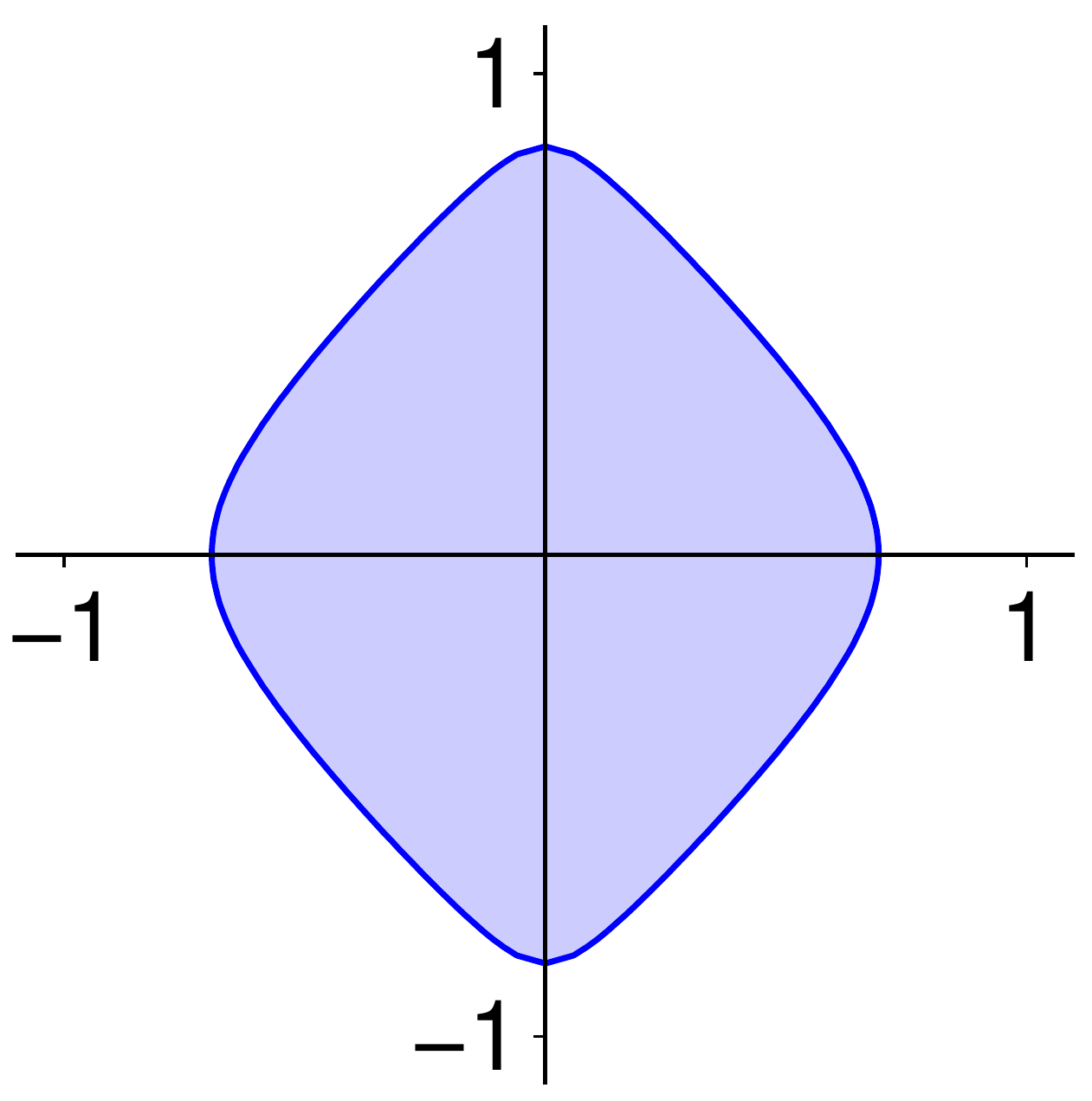}{$\beta_1^{(1)}$}{$\beta_2^{(1)}$} \rule[-5em]{0pt}{0pt} \\
        \hline & & & \\
        \rotatebox{90.0}{\makebox[0.3cm]{$\beta_1^{(2)}=0.3$}}&
        \xylabel{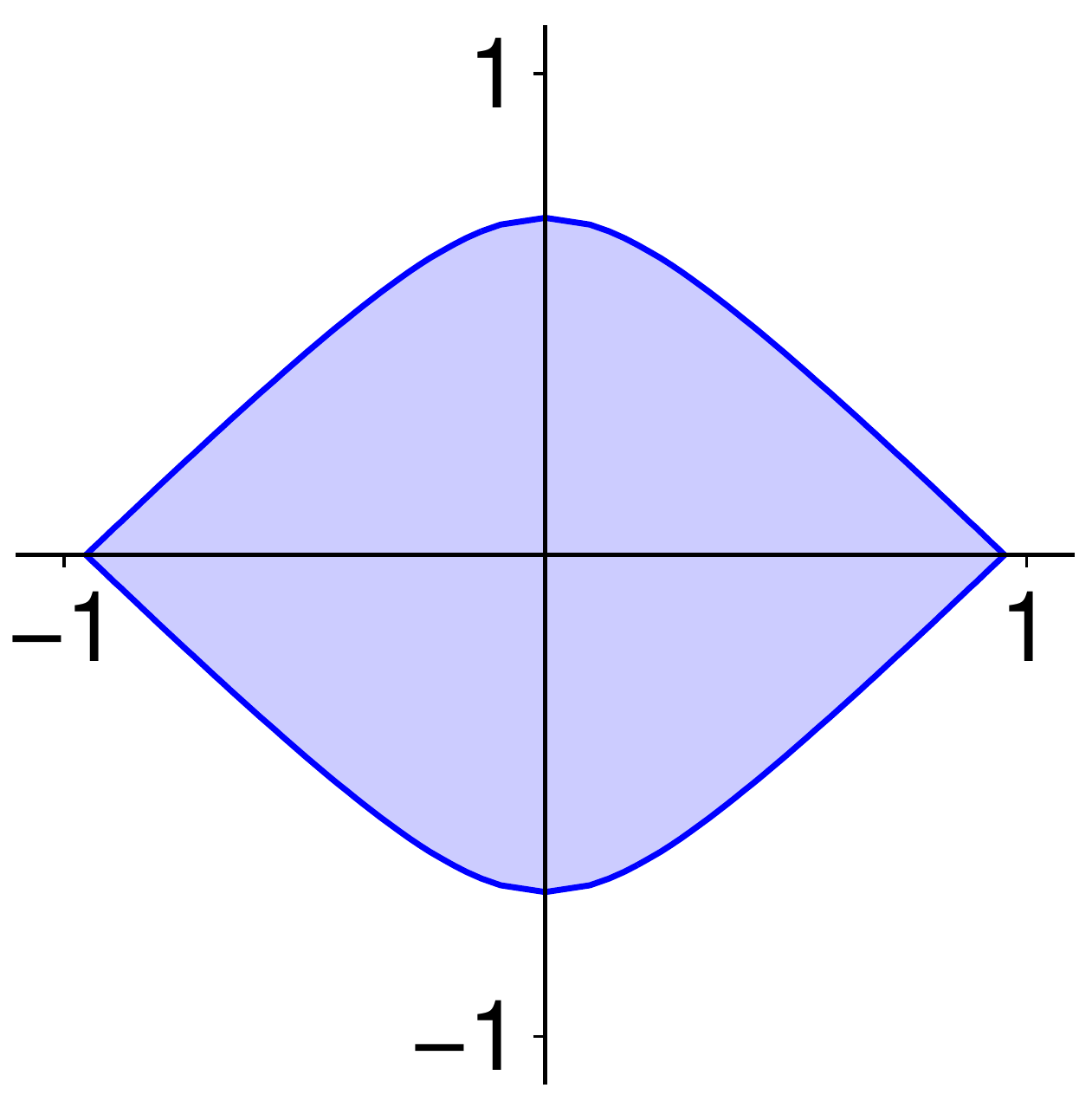}{$\beta_1^{(1)}$}{$\beta_2^{(1)}$} &
        \xylabel{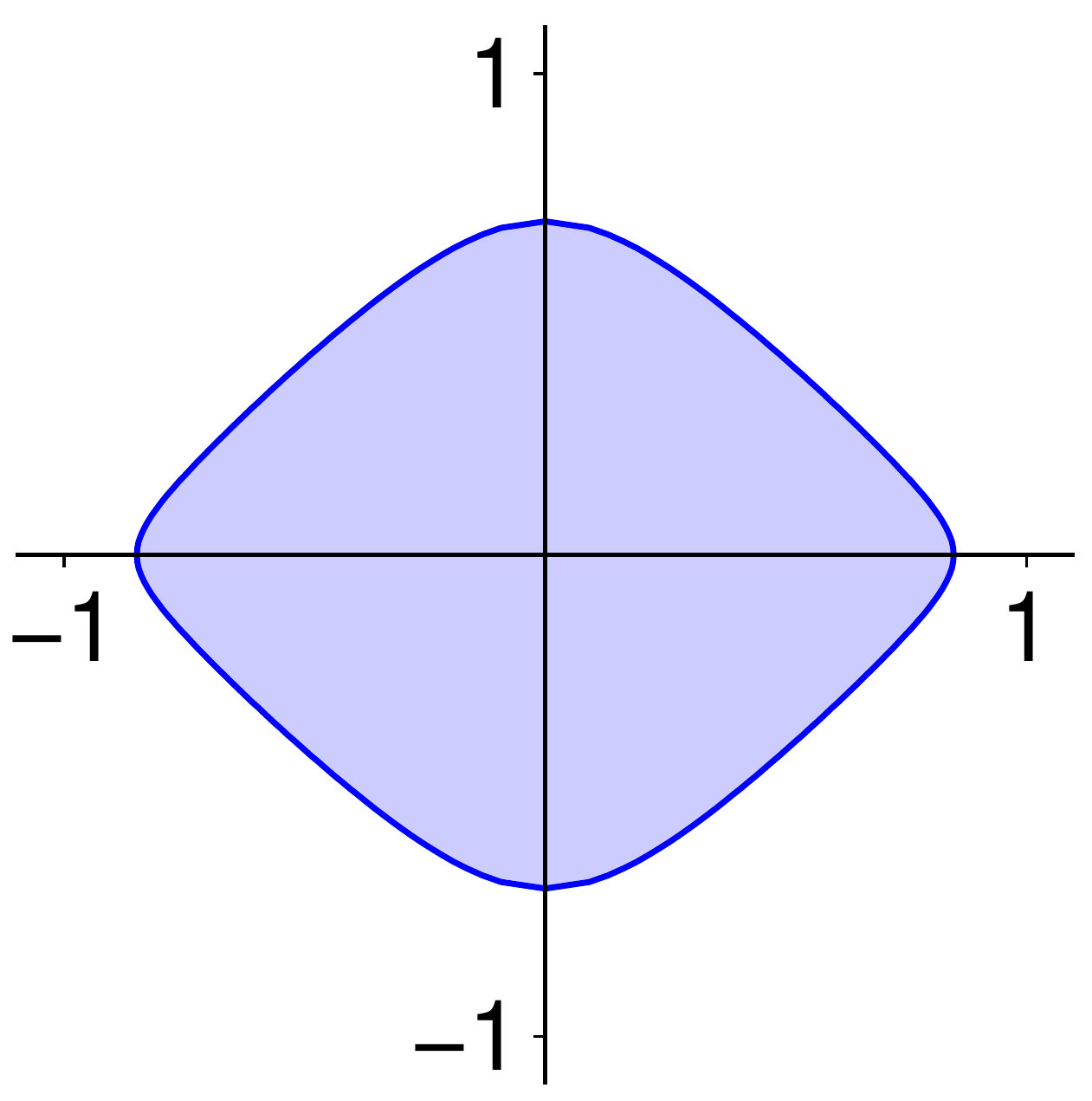}{$\beta_1^{(1)}$}{$\beta_2^{(1)}$} &
        \xylabel{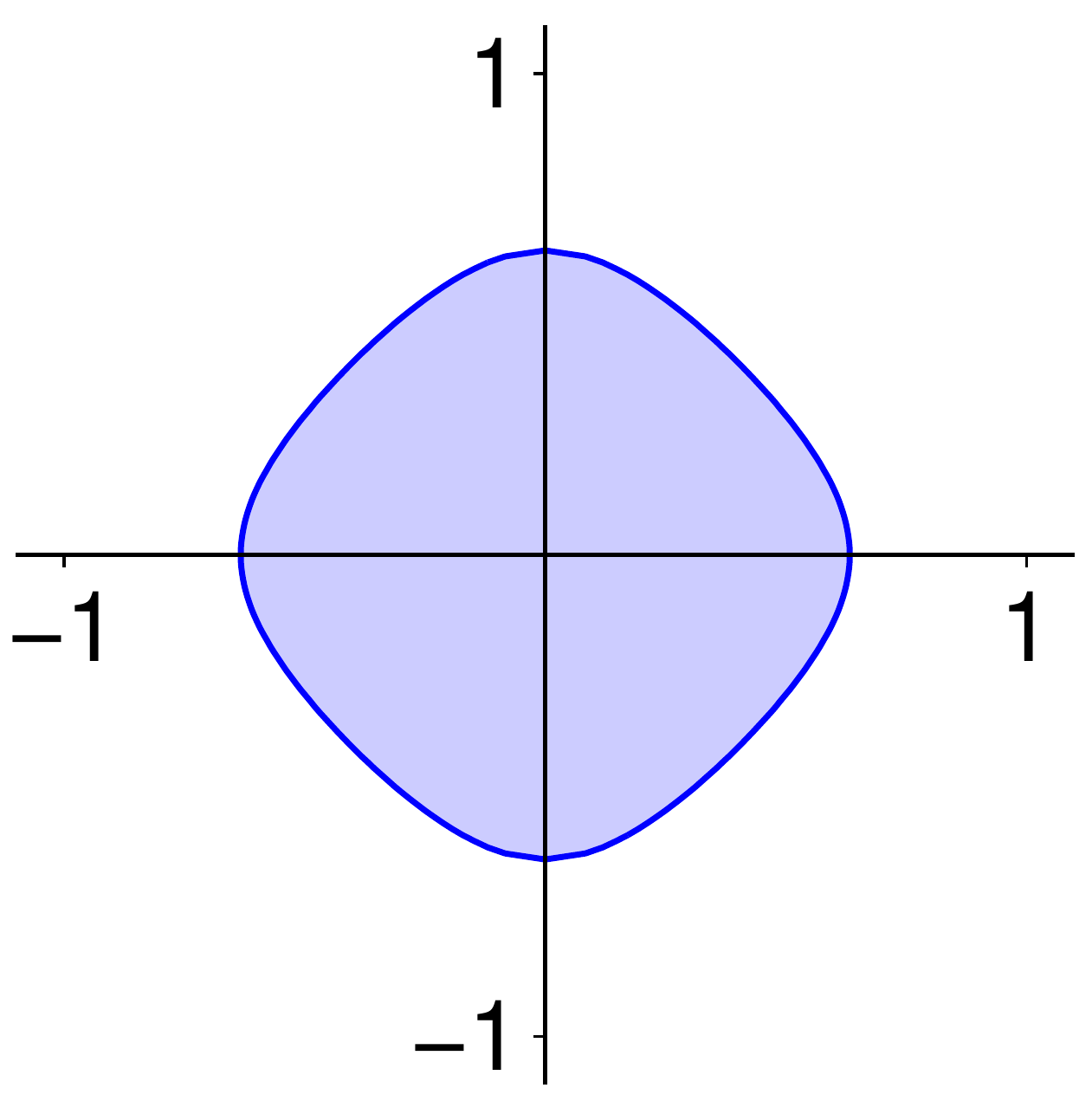}{$\beta_1^{(1)}$}{$\beta_2^{(1)}$}\rule[-5em]{0pt}{0pt} \\ 
        \hline 
      \end{tabular}
    \end{small}
  \caption{Admissible set for the Group-LASSO penalty
           for a problem with two tasks and two features.
           Top row: cuts of the unit ball through
           $(\beta_1^{(1)},\beta_1^{(2)},\beta_2^{(1)})$ for various values of
           $\beta_2^{(2)}$, where $(\beta_1^{(1)},\beta_1^{(2)})$ span the
           horizontal plane, and $\beta_2^{(1)}$ is on the vertical axis; bottom
           rows: cuts through $(\beta_1^{(1)},\beta_2^{(1)})$ for various values
           of $(\beta_1^{(2)}$ and $\beta_2^{(2)})$.}
  \label{fig:group-lasso}
\end{figure*}

\begin{figure*}
  \begin{center}
    \begin{small}
      \begin{tabular}{|*{4}{c@{\hspace*{0.25cm}}|}}
        \hline
        & $\beta_2^{(2)}=0$ & $\beta_2^{(2)}=0.1$ & $\beta_2^{(2)}=0.3$ \\ \hline
        & & & \\ 
        &
        \includegraphics[width=0.25\textwidth]{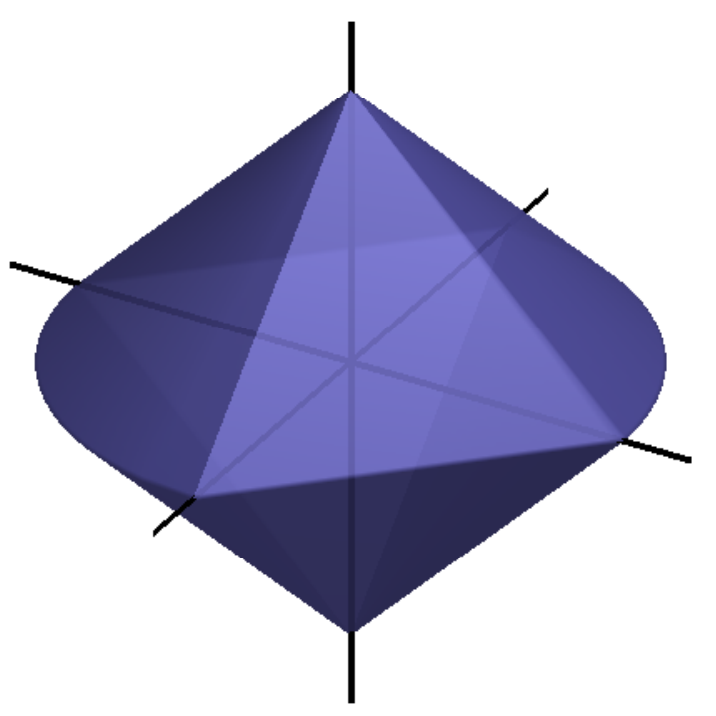} &
        \includegraphics[width=0.25\textwidth]{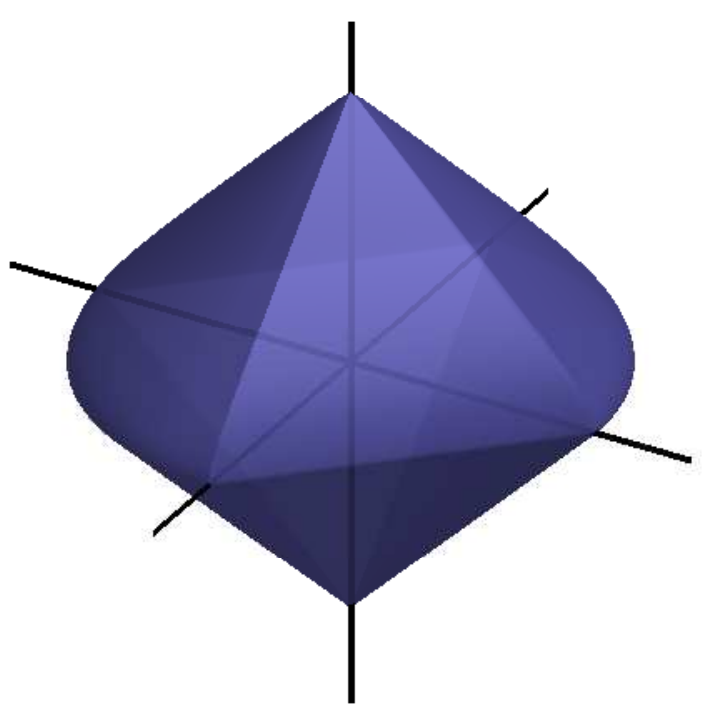} &
        \includegraphics[width=0.25\textwidth]{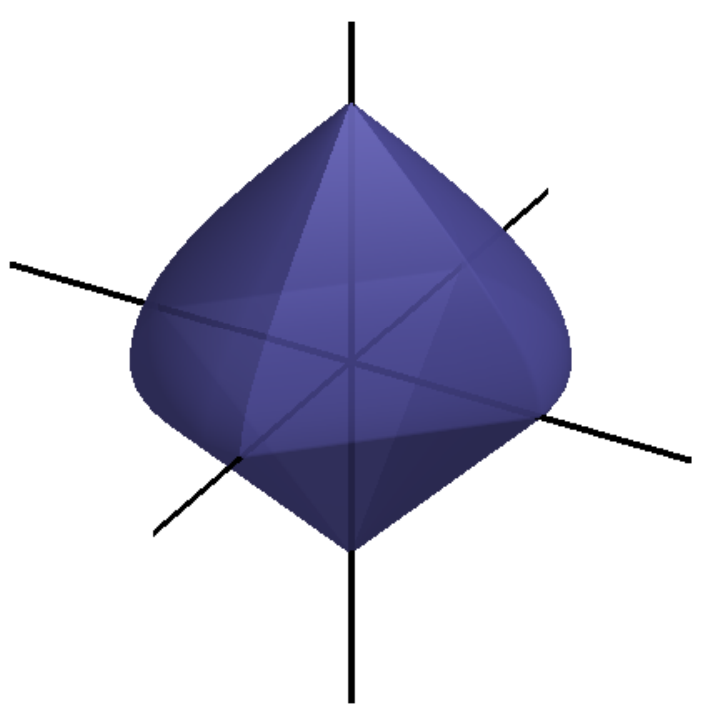} \\
        \hline & & & \\
        \rotatebox{90.0}{\makebox[0cm]{$\beta_1^{(2)}=0$}}&
        \xylabel{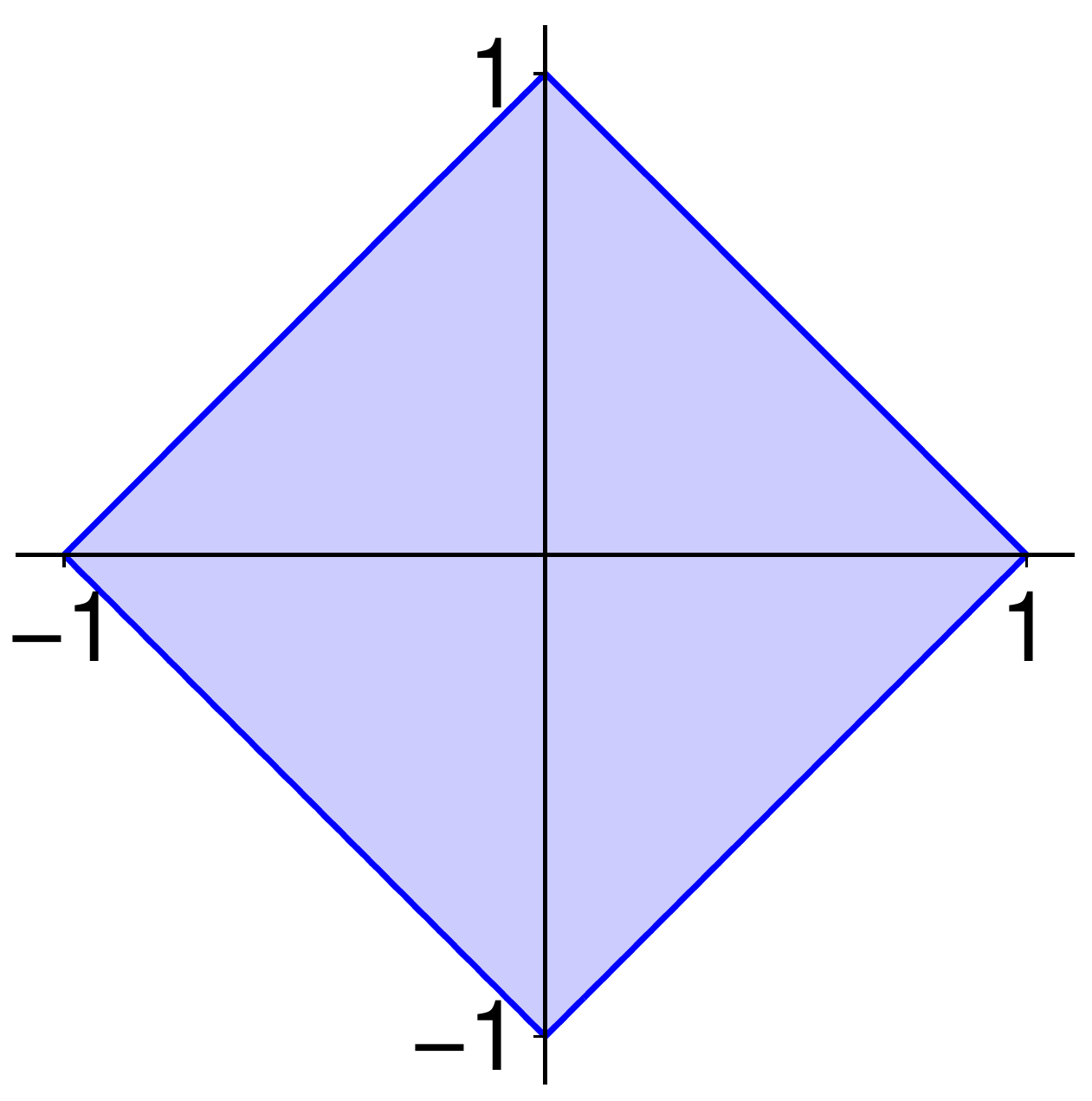}{$\beta_1^{(1)}$}{$\beta_2^{(1)}$} &
        \xylabel{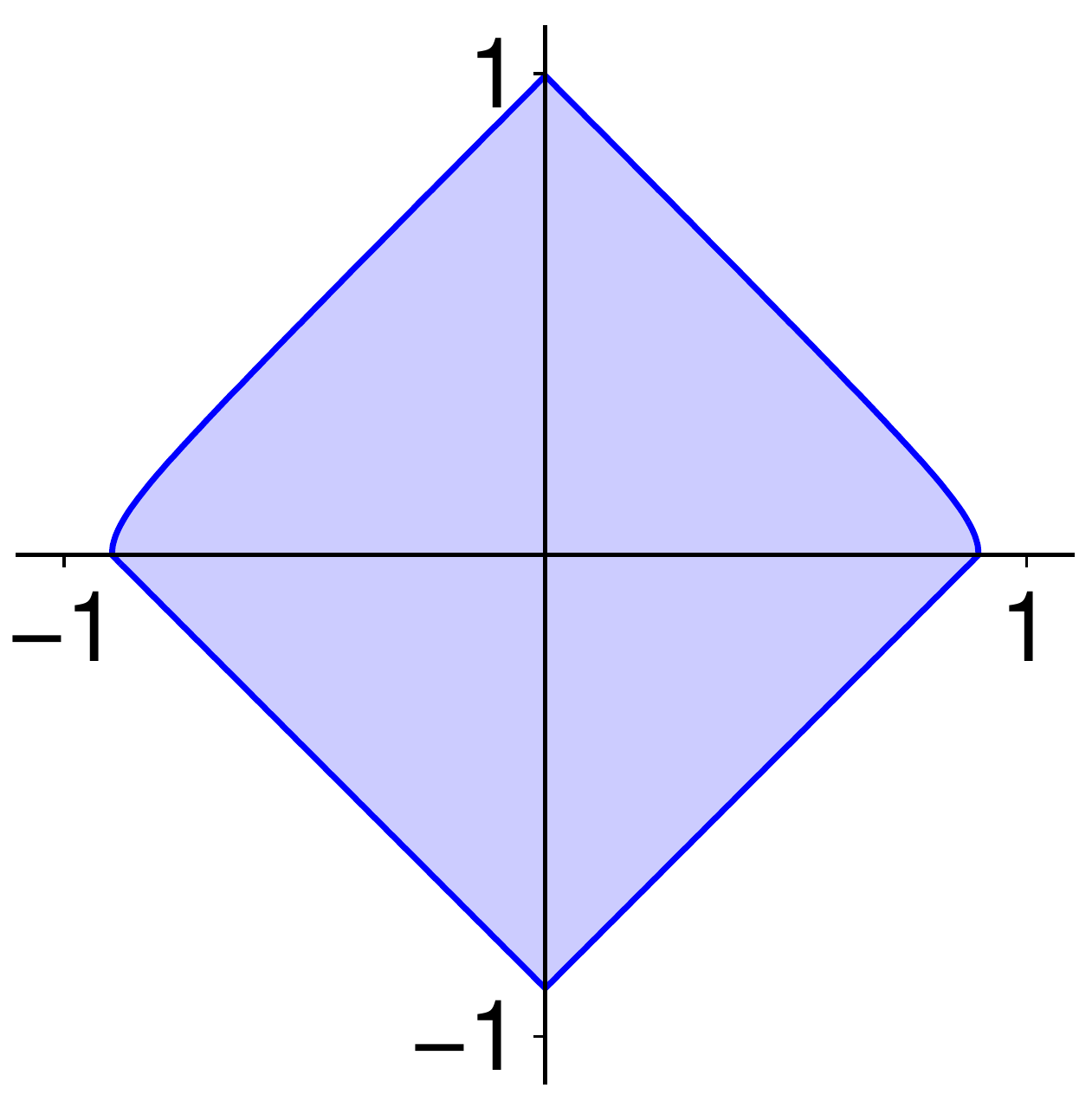}{$\beta_1^{(1)}$}{$\beta_2^{(1)}$} &
        \xylabel{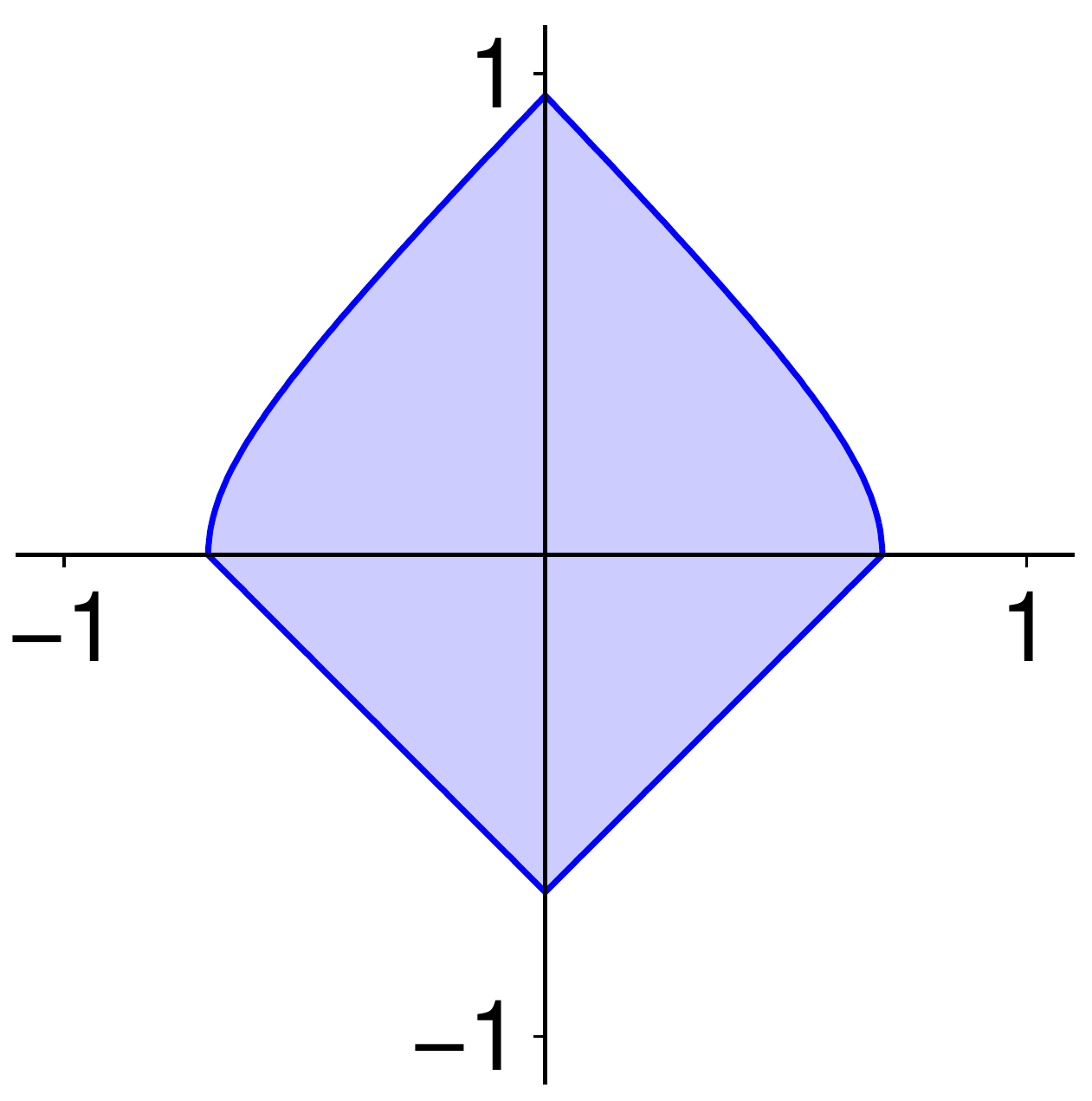}{$\beta_1^{(1)}$}{$\beta_2^{(1)}$}\rule[-5em]{0pt}{0pt} \\
        \hline & & & \\
        \rotatebox{90.0}{\makebox[0.3cm]{$\beta_1^{(2)}=0.1$}}&
        \xylabel{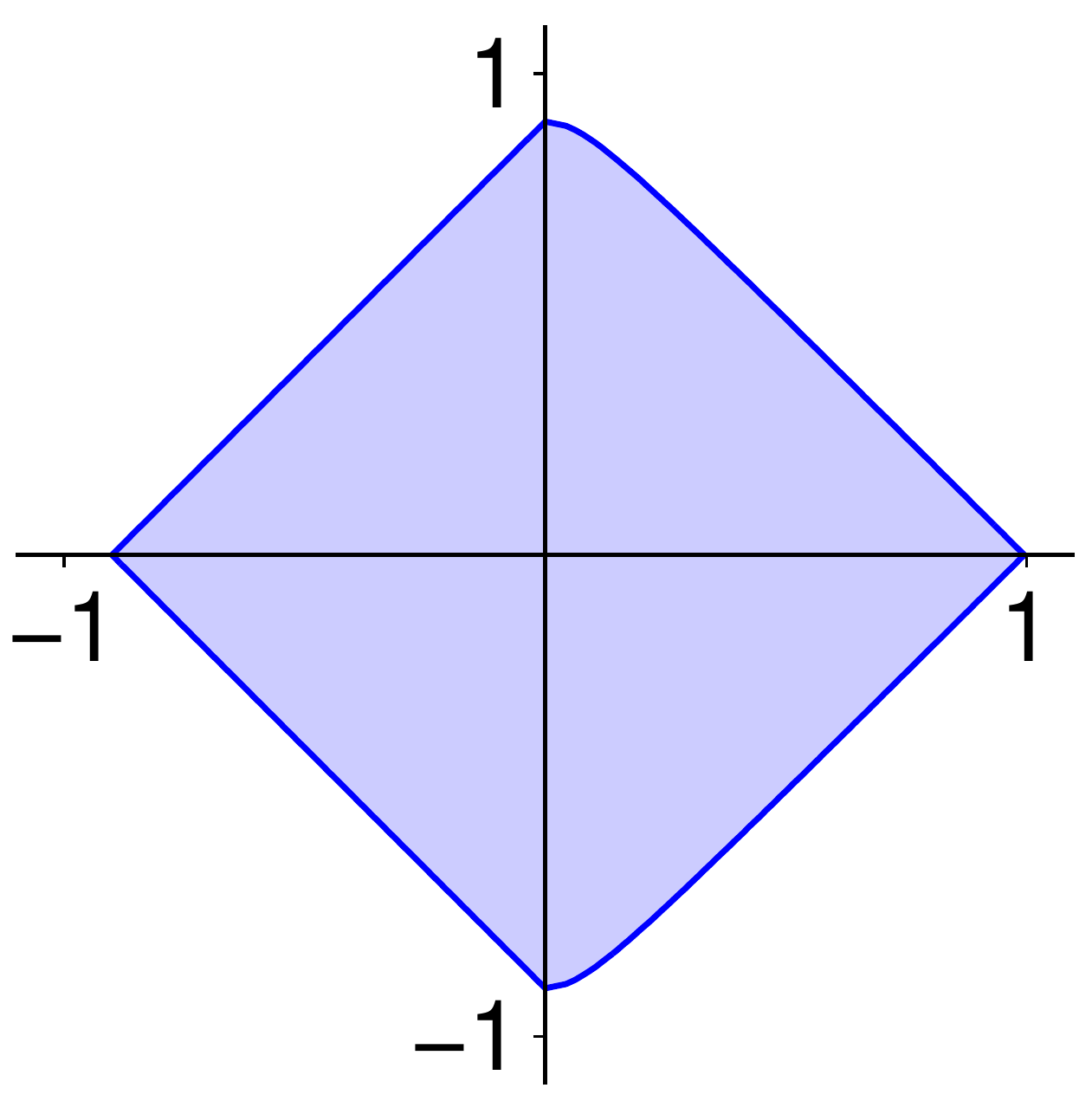}{$\beta_1^{(1)}$}{$\beta_2^{(1)}$} &
        \xylabel{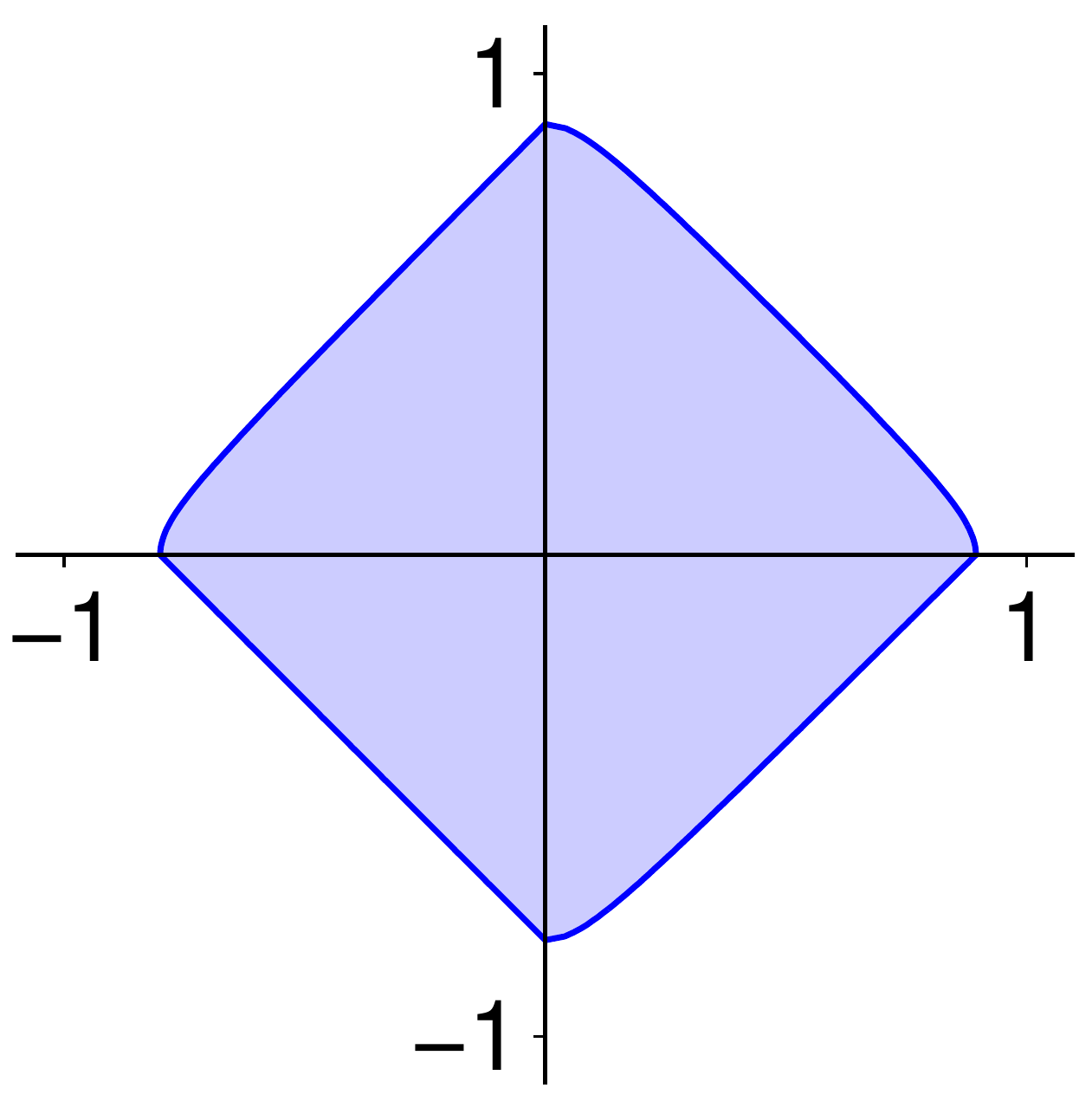}{$\beta_1^{(1)}$}{$\beta_2^{(1)}$} &
        \xylabel{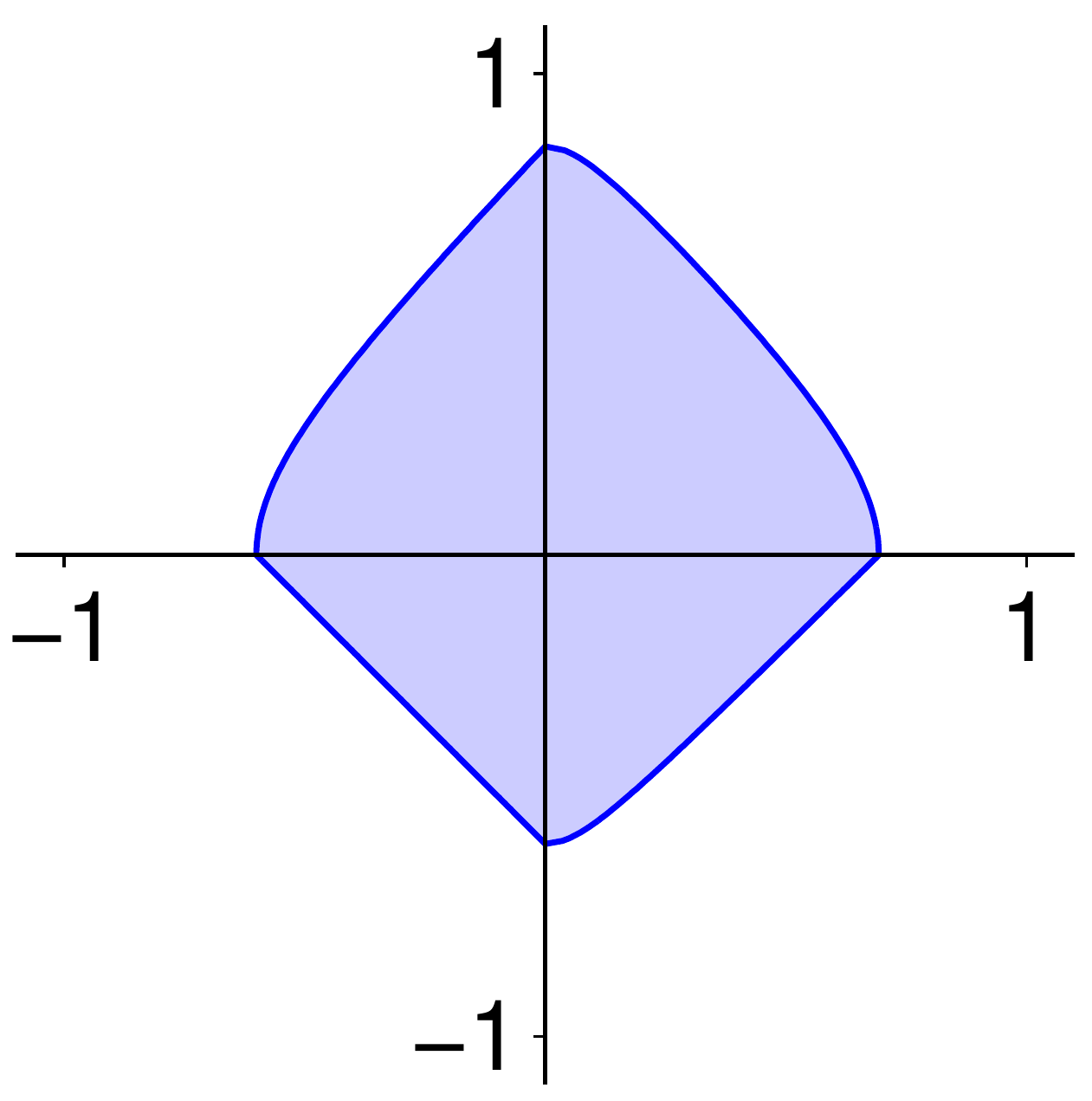}{$\beta_1^{(1)}$}{$\beta_2^{(1)}$} \rule[-5em]{0pt}{0pt} \\
        \hline & & & \\
        \rotatebox{90.0}{\makebox[0.3cm]{$\beta_1^{(2)}=0.3$}}&
        \xylabel{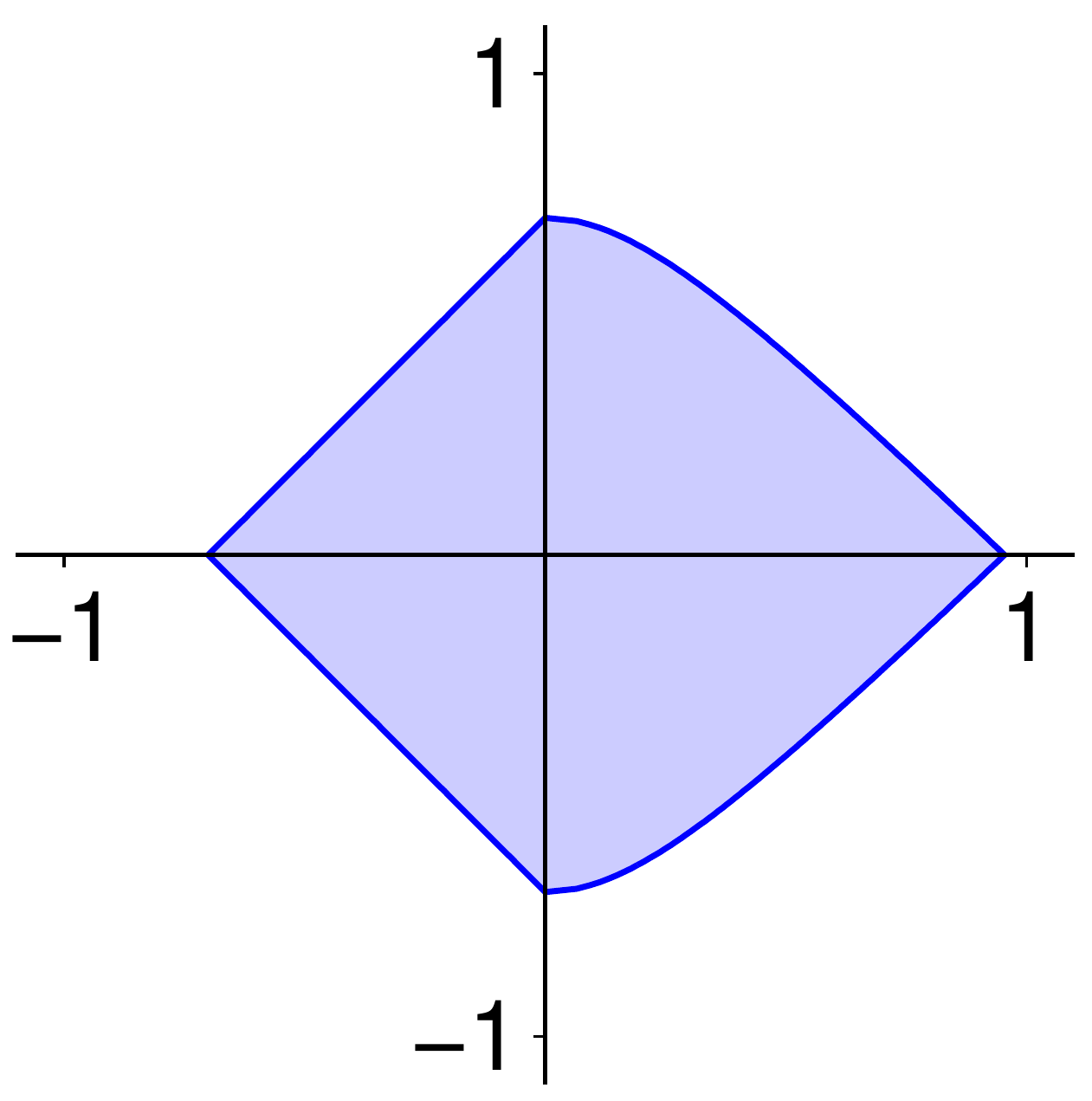}{$\beta_1^{(1)}$}{$\beta_2^{(1)}$} &
        \xylabel{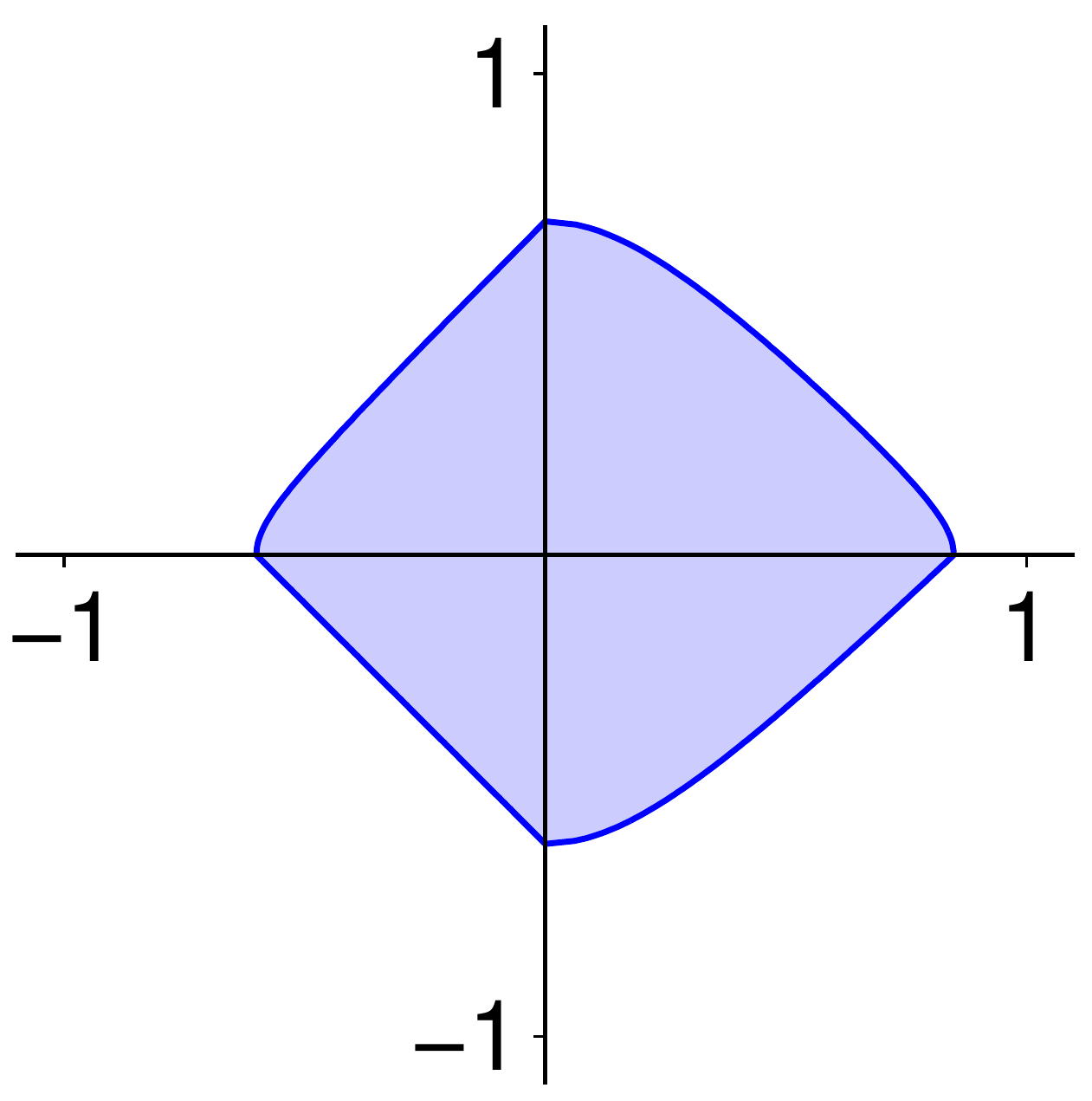}{$\beta_1^{(1)}$}{$\beta_2^{(1)}$} &
        \xylabel{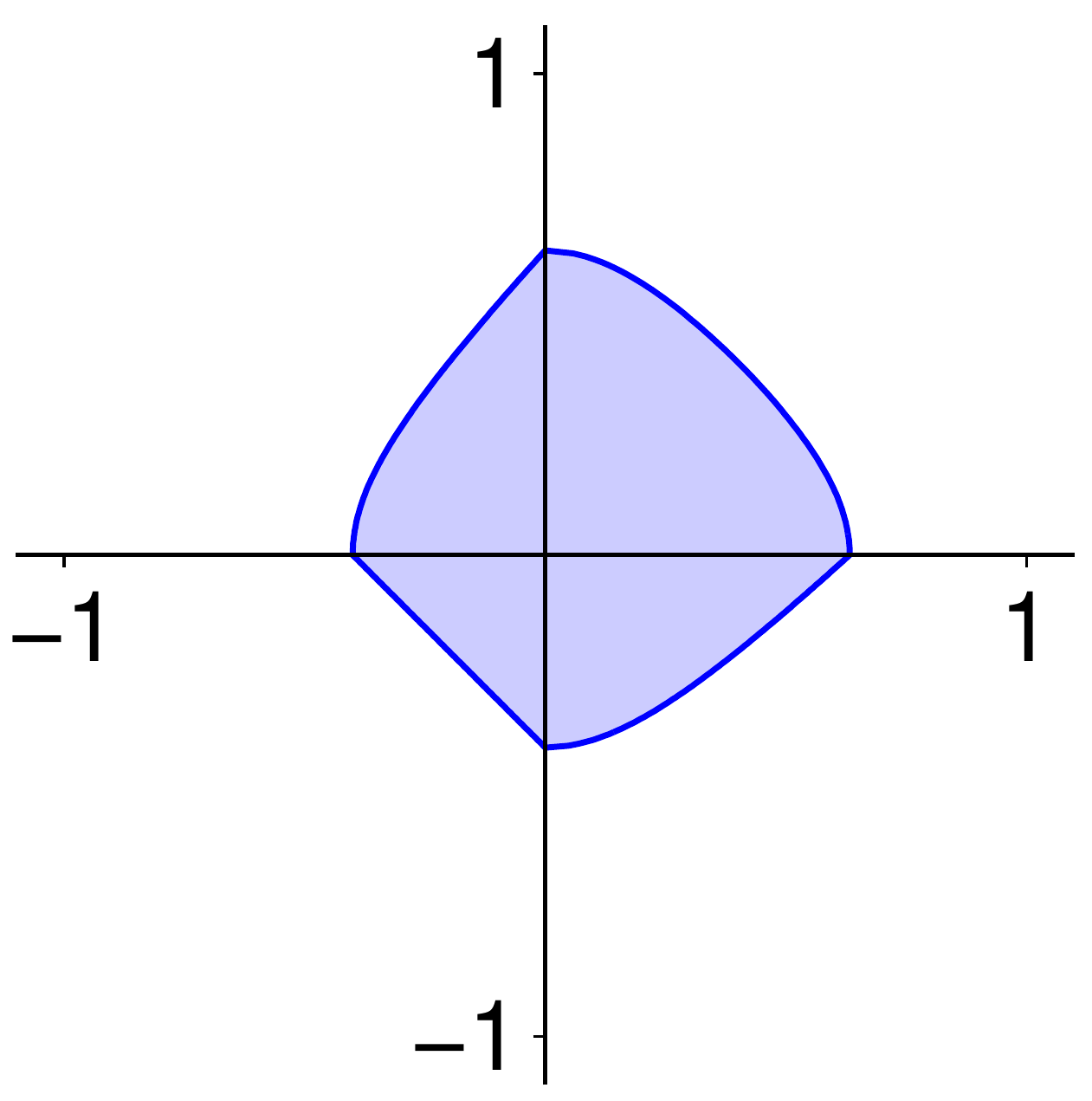}{$\beta_1^{(1)}$}{$\beta_2^{(1)}$}\rule[-5em]{0pt}{0pt} \\ 
        \hline 
      \end{tabular}
    \end{small}
  \end{center}
  \caption{Admissible set for the Cooperative-LASSO penalty for
           a problem with two tasks and two features.
           Top row: cuts of the unit ball through $(\beta_1^{(1)},\beta_1^{(2)},\beta_2^{(1)})$
           for various values of $\beta_2^{(2)}$, where $(\beta_1^{(1)},\beta_1^{(2)})$ 
           span the horizontal plane, and $\beta_2^{(1)}$ is on the vertical axis;
           bottom rows: cuts through $(\beta_1^{(1)},\beta_2^{(1)})$ for various
           values of $(\beta_1^{(2)}$ and $\beta_2^{(2)})$.}
  \label{fig:coop-lasso}
\end{figure*}

Figures \ref{fig:group-lasso} and \ref{fig:coop-lasso} illustrate 
the role of each penalty on a problem with $T=2$ tasks and $p=2$ variables.
They represent 
several views of the unit balls
\begin{multline*}
  \sum_{i=1}^2 \bigg( \sum_{t=1}^2 {\beta_i^{(t)}}^2 \bigg)^{1/2} \leq 1
  \enspace,\, \text{and} \enspace \\
  \sum_{i=1}^2 \bigg( \sum_{t=1}^2 \left(\beta_i^{(t)}\right)_{\!+}^2 \bigg)^{1/2}
      + \bigg( \sum_{t=1}^2 \left(-\beta_i^{(t)}\right)_{\!+}^2 \bigg)^{1/2}
      \leq 1\enspace,
\end{multline*}
that is, the admissible set for a
penalty for a problem with two tasks and two features.

These plots also provide some insight on the sparsity pattern that originate
from the penalty, since sparsity is related to the singularities at the boundary
of the admissible set \citep{Nikolova00}.  
In particular, the first column illustrates that, when $\beta_2^{(2)}$ is null,
$\beta_2^{(1)}$ may also be exactly zero, while the second column shows that
this event is improbable when $\beta_2^{(2)}$ differs from zero.  The second row
illustrates the same type of relationship between $\beta_1^{(2)}$ and
$\beta_1^{(1)}$ that are expected due to the symmetries of the unit ball.

Figure \ref{fig:coop-lasso}  corresponds to a Cooperative-LASSO penalty.
These plots should be compared with their Group-LASSO counterpart in Figure
\ref{fig:group-lasso}. 
We see that there are additional discontinuities in the unit ball resulting in
new vertices on the 3-D plots.
As before, we have that, when $\beta_2^{(2)}$ is null, $\beta_2^{(1)}$ may also
be exactly zero, but in addition, we may also have $\beta_1^{(1)}$ or
$\beta_1^{(2)}$ exactly null.  Accordingly, in the second and third row, we see
that we may have $\beta_2^{(1)}$ null when $\beta_2^{(2)}$ is non-zero.
These new edges will result in some new zeroes when the Group-LASSO would have
allowed a solution with opposite signs between tasks. 

The second main striking difference with Group-LASSO is the loss of the axial
symmetry of the Cooperative-LASSO when some variables are non-zero.
These plots illustrate that the decoupling of the positive and negative parts of
the regression coefficients in the penalty favors solutions where these
coefficients are of same sign across tasks.
The penalties are identical in the positive and negative orthant, but the
Cooperative-LASSO penalization is more stringent elsewhere, when there are some sign
mismatches between tasks.
The most extreme situation occurs when there is no sign agreement across tasks
for all variables.  In the
setup represented here, with only two tasks, the effective penalty
then reduces to the LASSO.

\section{Algorithms}

In this section, we describe the strategy proposed for solving the three
optimization problems introduced above, based upon the proposal of
\citet{2000_JCGS_Osborne} for solving the LASSO. This part also draws its
inspiration from \citet{2000_JNA_Osborne,2006_SS_Kim,2008_ICML_Roth}.

\subsection{Problem Decomposition}

The  multiple  independent  tasks  Problem \eqref{eq:T_indep}  can  be
solved    by    considering    either    $T$   single    tasks    like
\eqref{eq:pen_pseudologlikelihood} of size  $(p-1) \times p$ (each one
possibly decomposed  in $p$  LASSO sub-problems of  size $p-1$),  or a
single large  problem of size $T\times  (p-1) \times p$,  which can be
decomposed  into $p$  LASSO  sub-problems of  size  $(p-1) \times  T$,
through Proposition \ref{prop:blockwise_resolution}. 
This line of attack is not computationally efficient here, but it will become
advantageous when considering the penalties presented in Section
\ref{sec:cooplasso}.  It is introduced at this point to
provide a unified conceptual view of all algorithms.%

Consider the $(p \, T) \times (p \, T)$ block-diagonal matrix
$\mathbf{C}$  composed by  the empirical  covariance matrices  of each
tasks
\begin{equation*}
  \mathbf{C} = \begin{pmatrix}
    \mathbf{S}^{(1)} & & \text{\Large $0$} \\
    & \ddots & \\ 
    \text{\Large 0} & & \mathbf{S}^{(T)} \\
  \end{pmatrix},
\end{equation*}
and define
\begin{equation}
  \label{eq:block_mat_fullindep}
  \mathbf{C}_{\backslash i  \backslash i} = \begin{pmatrix}
    \mathbf{S}_{\backslash i  \backslash i}^{(1)} & & \text{\Large $0$} \\
    & \ddots & \\ 
    \text{\Large 0} & & \mathbf{S}_{\backslash i  \backslash i}^{(T)} \\
  \end{pmatrix}, \ 
  \mathbf{C}_{i  \backslash i} = \begin{pmatrix}
    \mathbf{S}_{i  \backslash i}^{(1)}\\
    \vdots \\
    \mathbf{S}_{i  \backslash i}^{(T)}\\
  \end{pmatrix}.
\end{equation}
The $(p-1) \, T \times (p-1) \, T $ matrix $\mathbf{C}_{\backslash i \backslash
i}$ is the matrix $\mathbf{C}$ where we removed each line and each column
pertaining to variable $i$.  
We define $\tilde{\mathbf{C}},
\tilde{\mathbf{C}}_{\backslash      i      \backslash     i}$      and
$\tilde{\mathbf{C}}_{i     \backslash      i}$     similarly,     with
$\mathbf{S}^{(t)}$  being replaced  by  $\tilde{\mathbf{S}}^{(t)}$ for
each $t=1, \dots, T$ in the above definitions.

Let ${\boldsymbol \beta}^{(t)}\in\mathbb{R}^{(p-1)}$ denote the vector
estimating   $\mathbf{K}_{i   \backslash   i}^{(t)}$,   defined   from
$\mathbf{K}^{(t)}$     as    in     \eqref{eq:K_block},     and    let
$\boldsymbol\beta\in\mathbb{R}^{T\times(p-1)}$  be the  vector  of the
concatenated estimates \mbox{$\boldsymbol\beta^\intercal =
  (   {{\boldsymbol  \beta}^{(1)}}^\intercal,   \cdots,  {{\boldsymbol
      \beta}^{(T)}}^\intercal)$}.        The      optimization      of
\eqref{eq:T_indep} is achieved by solving $p$ sub-problems in form:
\begin{equation}
  \label{eq:block_T_indep}
  \min_{\boldsymbol\beta\in\mathbb{R}^{T\times(p-1)}}    \frac{1}{2}
  \left\|            \mathbf{C}_{\backslash i  \backslash i}^{1/2}\boldsymbol\beta           +
    \mathbf{C}_{\backslash i  \backslash i}^{-1/2}   \mathbf{C}_{i  \backslash i}\right\|_2^2   +   \lambda
  \sum_{t=1}^T \frac{1}{n_t} \left\| \boldsymbol\beta \right\|_1
  \enspace.
\end{equation}
Note that we do not need to perform the costly matrix operations that are
expressed in the the first term of the objective function of 
Problem~\eqref{eq:block_T_indep}.
In practice, we compute
\begin{equation*}
  f({\boldsymbol\beta};\mathbf{C}) = \frac{1}{2} {\boldsymbol\beta}^\intercal
    \mathbf{C}_{\backslash i \backslash i} {\boldsymbol\beta} +
    {\boldsymbol\beta}^\intercal \mathbf{C}_{i  \backslash i}
  \enspace,
\end{equation*}
which only differs from the squared $\ell_2$ norm in \eqref{eq:block_T_indep} by
a constant that is irrelevant for the optimization process.

Accordingly, Problems \eqref{eq:criterion1}, \eqref{eq:criterion2} and
\eqref{eq:criterion3}   can  be   decomposed  into   $p$  minimization
sub-problems whose objective functions may be decomposed as
\begin{equation}
  \label{eq:lagrangian}
  L_k(\boldsymbol{\beta}) = f(\boldsymbol\beta) + \lambda
  g_k(\boldsymbol\beta)
  \enspace,
\end{equation} 
where, with a slight abuse of notation, $f(\boldsymbol\beta)$ is either
$f({\boldsymbol{\beta}};\tilde{\mathbf{C}})$
for Problem \eqref{eq:criterion1} or $f({\boldsymbol\beta};\mathbf{C})$ for
Problems \eqref{eq:criterion2} and \eqref{eq:criterion3}, and where
$g_k(\boldsymbol\beta)$ stands for the penalty functions respectively defined
below:

\begin{itemize}
\item for the graphical Intertwined LASSO
  \begin{equation*}
    g_1({\boldsymbol\beta}) =  
    \sum_{t=1}^T \frac{1}{n_t} \left\| \boldsymbol\beta^{(t)} \right\|_1
    \enspace;
  \end{equation*}
  \item for the graphical Group-LASSO
  \begin{equation*}
    g_2({\boldsymbol\beta}) = 
    \sum_{i=1}^{p-1} \left\|\boldsymbol\beta_i^{[1:T]} \right\|_2
    \enspace,
  \end{equation*}
  where ${\boldsymbol\beta_i^{[1:T]}} =
  \left(\beta_i^{(1)},\ldots,\beta_i^{(T)}\right)^\intercal \in\mathbb{R}^{T}$ is the
  vector of the $i$th component across tasks;
  \item for the graphical Cooperative-LASSO
  \begin{equation*}
    \hspace*{-1em}%
    g_3({\boldsymbol\beta}) = \sum_{i=1}^{p-1} \left(
    \left\| \left(\boldsymbol\beta_i^{[1:T]}\right)_+ \right\|_2 + \left\|
    \left(-\boldsymbol\beta_i^{[1:T]}\right)_+ \right\|_2 \right)
    \,.
  \end{equation*}
\end{itemize}

Since  $f$  is convex  with  respect  to  $\boldsymbol\beta$, and  all
penalties are norms, all these objective functions are convex and thus
easily amenable to optimization.  They are non-differentiable at zero,
due to the penalty terms,  which all favor zero coefficients.  Bearing
in mind the  typical problems in biological data,  where graphs have a
few  tens   or  hundreds  nodes,   and  where  connectivity   is  very
weak\footnote{%
  Typically, the expected number of vertices in graphs to scale as the
  number of  nodes, that is,  we expect order of  $\sqrt{pT}$ non-zero
  coefficients in each sub-problem of size $T\times(p-1)$.%
},  we   need  convex  optimization  tools  that   are  efficient  for
medium-size problems with extremely sparse solutions.  We thus chose a
greedy  strategy   that  aims  at  solving  a   series  of  small-size
sub-problems, and will offer a simple monitoring of convergence.

\subsection{Solving the Sub-Problems}

The minimizers $\boldsymbol\beta$ of the objective functions
\eqref{eq:lagrangian} are assumed to have
many zero coefficients.
The approach developed for the LASSO by \citet{2000_JCGS_Osborne} takes
advantage of this sparsity by solving a series of small linear systems, whose
size is incrementally increased/decreased, similarly to a column generation
algorithm.
The master problem is the original problem, but solved only with respect to the
subset of variables currently identified as non-zero $\boldsymbol\beta$
coefficients.
The subproblem of identifying new non-zero variables simply consists in
detecting the violations of the first-order optimality conditions with respect
to all variables.
When there are no more such violations, the current solution is optimal.

The objective functions $L_k(\boldsymbol{\beta})$ are convex and smooth except at
some locations with zero coefficients.
Thus, the minimizer is such that the null vector $\mathbf{0}\in\mathbb{R}^{p-1}$
is an element of the subdifferential
$\partial_{\boldsymbol{\beta}}L_k(\boldsymbol{\beta})$.
In our problems, the  subdifferential is given by 
\begin{equation}
  \label{eq:subdiff}
  \partial_{\boldsymbol{\beta}}L_k(\boldsymbol\beta) = 
  \nabla_{\boldsymbol\beta}            f(\boldsymbol\beta)           +
  \lambda \partial_{\boldsymbol\beta} g_k({\boldsymbol\beta})
  \enspace,
\end{equation}
where       $\nabla_{\boldsymbol\beta}      f(\boldsymbol\beta)      =
\mathbf{C}_{\backslash    i   \backslash    i}    \boldsymbol\beta   +
\mathbf{C}_{i    \backslash    i}$    and    
where the form of $\partial_{\boldsymbol{\beta}} g_k(\boldsymbol{\beta})$
differs for the three problems and will be detailed below.

The  algorithm  is started  from  a  sparse  initial guess,  that  is,
$\boldsymbol\beta=\mathbf{0}$ or, if  available, the solution obtained
on  a more constrained  problem with  a larger  penalization parameter
$\lambda$.  Then, one converges to the global solution iteratively, by
managing  the  index $\mathcal{A}$  of  the  non-zero coefficients  of
$\boldsymbol\beta$ and solving  the master problem over $\mathcal{A}$,
where the  problem is continuously differentiable.   The management of
$\mathcal{A}$  requires   two  steps:  the  first   one  removes  from
$\mathcal{A}$ the coefficients that  have been zeroed when solving the
previous master  problem, ensuring  its differentiability at  the next
iteration,  and  the  second   one  examines  the  candidate  non-zero
coefficients  that  could   enter  $\mathcal{A}$.   In  this  process,
summarized in Algorithm  \ref{algo:active_constraint}, the size of the
bigger master problems  is typically of the order  of magnitude of the
number of non-zero entries in the solution. Solving the master problem
with       respect       to       the      non-zero       coefficients
$\boldsymbol\beta_{\mathcal{A}}$   can   be   formalized  as   solving
$\min_{\mathbf{h}}     L_k(\boldsymbol\beta_{\mathcal{A}}+\mathbf{h})$,
where  $\mathbf{h}  \in   \mathbb{R}^{|\mathcal{A}|}$  is  optimal  if
$\mathbf{0} \in
\partial_{\mathbf{h}}L_k(\boldsymbol\beta_{\mathcal{A}}+\mathbf{h})$.

\begin{algorithm}
  \dontprintsemicolon
  
  \CommentSty{// 0. INITIALIZATION}\; 
  $\boldsymbol \beta \leftarrow \mathbf{0}$\;
  $\mathcal{A} \leftarrow \emptyset $\;
  
  \BlankLine
  
  \While{$\mathbf{0} \notin \partial_{\boldsymbol\beta}  L(\boldsymbol\beta)$}{

    \BlankLine
    
    \CommentSty{// 1. MASTER PROBLEM: OPTIMIZATION WITH RESPECT TO $\boldsymbol\beta_\mathcal{A}$}\; 
    Find a (approximate) solution $\mathbf{h}$ to the smooth problem
    \begin{equation*}
      \nabla_{\mathbf{h}} f(\boldsymbol\beta_{\mathcal{A}} + \mathbf{h}) +
      \lambda   \partial_{\mathbf{h}} g_k (\boldsymbol\beta_{\mathcal{A}}    +
      \mathbf{h}) = 0 \enspace .
    \end{equation*}
    \CommentSty{// where $\partial_{\mathbf{h}} g_k=\left\{\nabla_{\mathbf{h}}g_k\right\}$}\;

    \BlankLine
    $\boldsymbol\beta_{\mathcal{A}} \leftarrow \boldsymbol\beta_{\mathcal{A}} +\mathbf{h}$\;
    \BlankLine
    
    \CommentSty{// 2. IDENTIFY NEWLY ZEROED VARIABLES}\; 
    \While{$\exists i \in \mathcal{A}: \beta_i=0$ and ${\displaystyle
    \min_{\theta \in \partial_{\beta_i}g_k}} \left| \frac{\partial
    f(\boldsymbol\beta)}{\partial \beta_i} + \lambda \theta \right|=0$}{
      $\mathcal{A} \leftarrow \mathcal{A}\backslash\{i\}$\;
    \BlankLine    
    }
    \BlankLine

    \CommentSty{// 3. IDENTIFY NEW NON-ZERO VARIABLES}\;
    \CommentSty{// Select $i\in\mathcal{A}^c$ such that an infinitesimal
                   change of $\beta_i$ provides the highest reduction of $L_k$
                   }\;
    ${\displaystyle i \leftarrow \argmax_{j\in\mathcal{A}^c} v_j}$, 
    where ${\displaystyle v_j= \min_{\theta \in \partial_{\beta_j}g_k}}
    \left| \frac{\partial f(\boldsymbol\beta)}{\partial \beta_j} + \lambda \theta \right|$\;
    \eIf{$v_i\neq 0$}{
          $\mathcal{A} \leftarrow \mathcal{A} \cup \set{i} $\;
    }{
      Stop and return $\boldsymbol\beta$, which is optimal\;
    }
    \BlankLine        
  }  
  \BlankLine
\caption{General optimization algorithm}
\label{algo:active_constraint}
\end{algorithm}

\subsection{Implementation Details}

We provide below the implementation  details that are specific to each
optimization   problem.  Specificity   of  each   problem   relies  on
$\partial_{\boldsymbol\beta}     g_k({\boldsymbol\beta})$,     denoted
${\boldsymbol\theta}$ herein.

\paragraph*{Intertwined LASSO --} This LASSO problem is solved as proposed by
\citet{2000_JCGS_Osborne}, except that we consider here the Lagrangian
formulation with $\lambda$ fixed.

The   components  of   $\boldsymbol\theta$   in  the   subdifferential
\eqref{eq:subdiff} read
\begin{equation*}
  \text{if} \enspace \beta_{i}=0 \enspace \text{then} \enspace \theta_i \in [-1,1]
  \enspace \enspace,\ \text{else} \enspace \theta_i = \mathrm{sign}({\beta}_{i})
  \enspace.
\end{equation*}
Solving the master problem on $\mathcal{A}$ requires an estimate of
$\boldsymbol\theta_{\mathcal{A}}$ at $\boldsymbol\beta_{\mathcal{A}}+\mathbf{h}$\,.
It is computed based on a local approximation, where the components of
$\mathrm{sign}(\boldsymbol\beta_{\mathcal{A}} + \mathbf{h})$ are replaced by
$\mathrm{sign}(\boldsymbol\beta_{\mathcal{A}})$.~\footnote{%
  When $\mathcal{A}$ is updated and that $\beta_i=0$, the corresponding $\theta_i$ is
  set to
  $-\mathrm{sign}(\partial f(\boldsymbol\beta)/\partial \beta_i)$\,.
}
This leads to the following descent direction $\mathbf{h}$:
\begin{equation*}
  \mathbf{h} = -\boldsymbol\beta_{\mathcal{A}} - 
     \tilde{\mathbf{C}}_{\backslash i \backslash i}^{-1}(\mathcal{A},\mathcal{A}) 
    ( \tilde{\mathbf{C}}_{i \backslash i}(\mathcal{A})
  + \lambda \boldsymbol\theta_{\mathcal{A}})
  \enspace,
\end{equation*}
where, in order to avoid double subscripts, we use the notation
$\mathbf{M}(\mathcal{A},\mathcal{A})$ for the square submatrix of
$\mathbf{M}$ formed by the rows and columns indexed by $\mathcal{A}$,
and $\mathbf{v}(\mathcal{A})$ for the subvector formed by the columns of
$\mathbf{v}$ indexed by $\mathcal{A}$.

Then, before updating $\boldsymbol\beta_{\mathcal{A}}$, one checks whether the
local approximation used to compute $\mathbf{h}$ is consistent with the sign of
the new solution.  If not the case, one looks for the largest step size $\rho$
in direction $\mathbf{h}$ such that $\boldsymbol\beta_{\mathcal{A}}^+ =
\boldsymbol\beta_{\mathcal{A}} + \rho\mathbf{h}$ is sign-consistent with
$\boldsymbol\beta_{\mathcal{A}}$.  This amounts to zero a coefficient, say
$\beta_{i}$, and $i$ is removed from $\mathcal{A}$ if
$|\partial f(\boldsymbol\beta^+)/\partial \beta_i|<\lambda$\,,
otherwise, the corresponding $\theta_i$ is set to
$-\mathrm{sign}(\partial f(\boldsymbol\beta^+)/\partial \beta_i)$\,.
In any case, a new direction $\mathbf{h}$ is computed as above, and
$\boldsymbol\beta_{\mathcal{A}}$ is updated until the
optimality conditions are  reached within $\mathcal{A}$.

Finally, the global optimum is attained if the first-order optimality conditions
are met for all the components of $\boldsymbol\beta$, that is, if
$\widehat{\boldsymbol\beta}$ verifies
\begin{equation*}
  \mathbf{0} \in \widetilde{\mathbf{C}}_{\backslash i \backslash i}
  \widehat{\boldsymbol\beta} + \widetilde{\mathbf{C}}_{i \backslash i} + \lambda
  \boldsymbol\theta \enspace,
\end{equation*}
where ${\boldsymbol\theta}$ is such
\begin{equation*}
  \boldsymbol\theta_{\mathcal{A}}=\mathrm{sign}(\widehat{\boldsymbol\beta}_{\mathcal{A}})
  \quad \text{and} \quad 
  \left\|\boldsymbol\theta_{\mathcal{A}^c} \right\|_\infty \leq 1 \enspace.
\end{equation*}

\paragraph*{Graphical Group-LASSO --} In this problem, the subdifferential
\eqref{eq:subdiff} is conditioned on the norm of $\boldsymbol\beta_i^{[1:T]}$,
the vector of the $i$th component across tasks.
Let $\boldsymbol\theta_i^{[1:T]} = \left(\theta_i^{(1)},\ldots,\theta_i^{(T)}\right)^\intercal
\in\mathbb{R}^{T}$ be defined similarly to $\boldsymbol\beta_i^{[1:T]}$, we have
that,
\begin{multline*}
  \text{if} \enspace \left\|\boldsymbol\beta_i^{[1:T]}\right\|_2=0 \enspace
  \text{then}
  \enspace \left\| \boldsymbol \theta_i^{[1:T]}  \right\|_2 \leq 1
  \enspace \enspace,\ \\ \text{else} \enspace 
  \boldsymbol\theta_i^{[1:T]} = \left\| \boldsymbol\beta_i^{[1:T]} \right\|_2^{-1}
                         \boldsymbol\beta_i^{[1:T]}
  \enspace,
\end{multline*}
where, here and in what follows, $0/0$ is defined by continuation as $0/0=0$.
As the subgradient w.r.t. $\boldsymbol\beta_i^{[1:T]}$ reduces to a gradient
whenever one component of $\boldsymbol\beta_i^{[1:T]}$ is non-zero, the
management of the null variables is done here by subsets of $T$ variables,
according to 
$\left\| \nabla_{\boldsymbol\beta_i^{[1:T]}} f(\boldsymbol\beta) \right\|_2$, 
instead of the one by one basis of the LASSO. Hence, we only need to index 
the groups $i \in \{1,\ldots,p-1\}$ in $\mathcal{A}$.

Here also, solving the master problem on $\mathcal{A}$ requires an estimate of
$\boldsymbol\theta_{\mathcal{A}}^{[1:T]}$ at
$\boldsymbol\beta_{\mathcal{A}}^{[1:T]}+\mathbf{h}$\,.  
Provided that $\left\|\boldsymbol\beta_i^{[1:T]}\right\|_2 \neq 0$ for all
$i\in\mathcal{A}$, $\boldsymbol\theta_{\mathcal{A}}^{[1:T]}$ is differentiable
w.r.t.
$\boldsymbol\beta_{\mathcal{A}}^{[1:T]}$.
It will thus be approximated by a first-order Taylor expansion, resulting in a
Newton-Raphson or quasi-Newton step. Here, we used quasi-Newton with BFGS
updates. 
Note that, whenever $\boldsymbol\beta_i^{[1:T]}=\mathbf{0}$, that is, when a new
group of variables has just been activated or is about to be deactivated, the
corresponding $\boldsymbol\theta_i^{[1:T]}$ is set so that
\begin{equation}\label{eq:grouplasso:activation}
\left\| 
  \nabla_{\boldsymbol\beta_i^{[1:T]}} f(\boldsymbol\beta) + 
  \lambda\boldsymbol\theta_i^{[1:T]} 
\right\|_2
\end{equation}
is minimum (that is, with $\boldsymbol\theta_i^{[1:T]}$ proportional to
$\nabla_{\boldsymbol\beta_i^{[1:T]}} f$). 
The updates of $\mathcal{A}$ are also based on the minimal value of
\eqref{eq:grouplasso:activation}.

\paragraph*{Graphical Cooperative-LASSO --} 
As the Group-LASSO, the Cooperative-LASSO considers a group structure, but its
implementation differs considerably from the former in the management of
$\mathcal{A}$.
Though several variables are usually activated or deactivated at the same time,
they typically correspond to subsets of $\boldsymbol\beta_i^{[1:T]}$, and these
subsets are context-dependent; they are not defined beforehand.
As a result, the index of non-zero $\boldsymbol\beta_i^{[1:T]}$ is better
handled by considering two sets: the index of $\boldsymbol\beta_i^{[1:T]}$ with
positive and negative components:
\begin{equation*}
  \begin{array}{rl@{\hfill}}
    & \mathcal{A}_+ = \left\{
    i\in\left\{1,\ldots,p-1\right\}: 
    \left\| \left(\boldsymbol\beta_i^{[1:T]} \right)_+ \right\|_2 > 0
  \right\} \enspace, \\[2.5ex]
  \text{and} & \mathcal{A}_- = \left\{
    i\in\left\{1,\ldots,p-1\right\}: 
    \left\| \left(-\boldsymbol\beta_i^{[1:T]} \right)_+\right\|_2 > 0
  \right\} \enspace.
  \end{array}
\end{equation*}
Let $\mathcal{T}$ denote the index of non-zero entries of
$\boldsymbol\beta_i^{[1:T]}$, with complement $\mathcal{T}^c$; the
subdifferential at the current solution is such that:
\begin{itemize}
\item if $i \in \mathcal{A}_+^c \cap \mathcal{A}_-^c$, then 
  \begin{equation*}
    \max\left(\left\| \left( \boldsymbol\theta_i^{[1:T]} \right)_+ \right\|_2,
      \left\| \left(-\boldsymbol\theta_i^{[1:T]} \right)_+ \right\|_2 \right)
    \leq
    1 \,;
  \end{equation*}
\item if $i \in \mathcal{A}_+^c \cap \mathcal{A}_-$ then
  \begin{gather*}
    \boldsymbol\theta_i^{\mathcal{T}}             =            \left\|
      \left(-\boldsymbol\beta_i^{\mathcal{T}}\right)_+ \right\|_2^{-1}
    \boldsymbol\beta_i^{\mathcal{T}} \,, \\
    \boldsymbol\theta_i^{\mathcal{T}^c} :
    \left\|   \left(   \boldsymbol\theta_i^{\mathcal{T}^c}   \right)_+
    \right\|_2     \leq      1     \     \text{and}      \     \left\|
      \left(-\boldsymbol\theta_i^{\mathcal{T}^c} \right)_+
    \right\|_2 = 0 \,;
  \end{gather*}
  \item $i \in \mathcal{A}_+ \cap \mathcal{A}_-^c$, then
    \begin{gather*}
      \boldsymbol\theta_i^{\mathcal{T}}            =           \left\|
        \left(                \boldsymbol\beta_i^{\mathcal{T}}\right)_+
      \right\|_2^{-1}
      \boldsymbol\beta_i^{\mathcal{T}} \,, \\
      \boldsymbol\theta_i^{\mathcal{T}^c}           :          \left\|
        \left(-\boldsymbol\theta_i^{\mathcal{T}^c}            \right)_+
      \right\|_2     \leq     1     \     \text{and}     \     \left\|
        \left( \boldsymbol\theta_i^{\mathcal{T}^c} \right)_+
      \right\|_2 = 0 \,;
    \end{gather*}
  \item if $i \in \mathcal{A}_+ \cap \mathcal{A}_-$, then
    \begin{equation*}
    \theta_i^{(t)} = 
    \left \| \left(\mathrm{sign}\big({\beta}_{i}^{(t)}\big)\boldsymbol\beta_i^{[1:T]}
      \right)_+\right\|_2^{-1} \beta_i^{(t)}
    \enspace,\ t=1,\ldots,T \,.
  \end{equation*}
\end{itemize}
Once $\mathcal{A}^+$ and $\mathcal{A}^-$ are determined, the master problem is
solved as for the Group-LASSO, with BFGS updates, with box constraints to
ensure sign feasible $\boldsymbol\beta_i^{[1:T]}$ for $i$
such that $i\in\mathcal{A}_+^c\cap\mathcal{A}_-$ or
$i\in\mathcal{A}_+\cap\mathcal{A}_-^c$.
When a new variable has just been activated or is about to be
deactivated, the corresponding  $\boldsymbol\theta_i^{[1:T]}$ is set so that
\begin{multline}\label{eq:cooplasso:activation}
\left\| \left(
  \nabla_{\boldsymbol\beta_i^{[1:T]}} f(\boldsymbol\beta) + 
  \lambda\boldsymbol\theta_i^{[1:T]} 
\right)_+ \right\|_2 \\ + 
\left\| \left(-
  \nabla_{\boldsymbol\beta_i^{[1:T]}} f(\boldsymbol\beta) - 
  \lambda\boldsymbol\theta_i^{[1:T]} 
\right)_+ \right\|_2
\end{multline}
is minimum. 
The updates of $\mathcal{A}$ are also based on the minimal value of
\eqref{eq:cooplasso:activation}.


\section{Experiments}

In most real-life applications, the major part of the inferred graphs are
unknown, with little available information on the
presence/absence  of  edges.   We  essentially  face  an  unsupervised
learning problem,  where there is  no objective criterion  allowing to
compare   different    solutions.    As   a    result,   setting   the
hyper-parameters is particularly troublesome, alike, say, choosing the
number of components  in a mixture model, and it  is a common practice
to visualize several networks corresponding to a series of penalties.

Regarding  the first  issue, we  chose to  present here  synthetic and
well-known  real  data  that   allow  for  an  objective  quantitative
evaluation.   Regarding  the second  issue,  the  problem of  choosing
penalty parameters can be guided by theoretical results that provide a
bound     on     the      rate     of     false     edge     discovery
\citep{2006_AS_Meinshausen,2008_JMLR_Banerjee,2009_article_Ambroise},
or by  more traditional information criteria  targeting the estimation
of    $\mathbf{K}$   \citep{2007_Biometrika_Yuan,2008_preprint_Rocha}.
However,  these proposals tend  to behave  poorly, and  it is  a usual
practice  to  compare  the   performance  of  learning  algorithms  by
providing  a series  of  results, such  as  precision-recall plots  or
ROC-curves, letting the choice of  penalty parameters as a mostly open
question for future research.   Although the shortcomings of this type
of comparison are  well-known \citep{Drummond06,Bengio05}, we will use
precision {\em vs.} recall plots,  which can be considered as valuable
exploratory tools.

Precision is the ratio  of the number of true selected
  edges to the total number  of selected edges in the inferred graphs.
  Recall is  the ratio  of the  number of true  selected edges  in the
  inferred graphs to the total number of edges in the true graphs.  In
  a statistical framework,  the recall is equivalent to  the power and
  the  precision  is  equivalent  to  one minus  the  false  discovery
  proportion.

\subsection{Synthetic Data}\label{sec:synthetic}

\subsubsection{Simulation Protocol}

\begin{figure}
  \begin{center}
  \includegraphics[width=.2\textwidth]{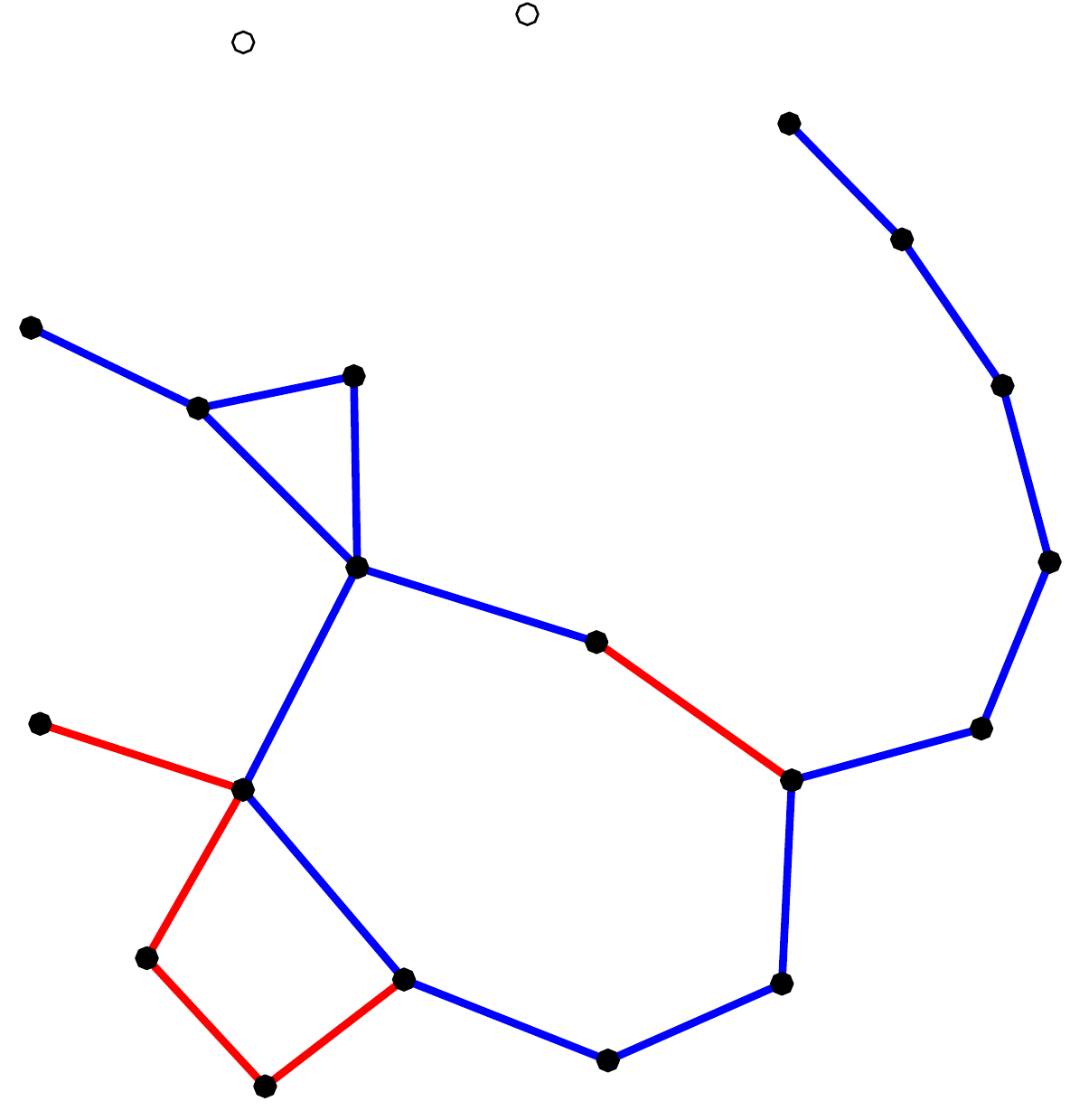} \\ 
  \includegraphics[width=.2\textwidth]{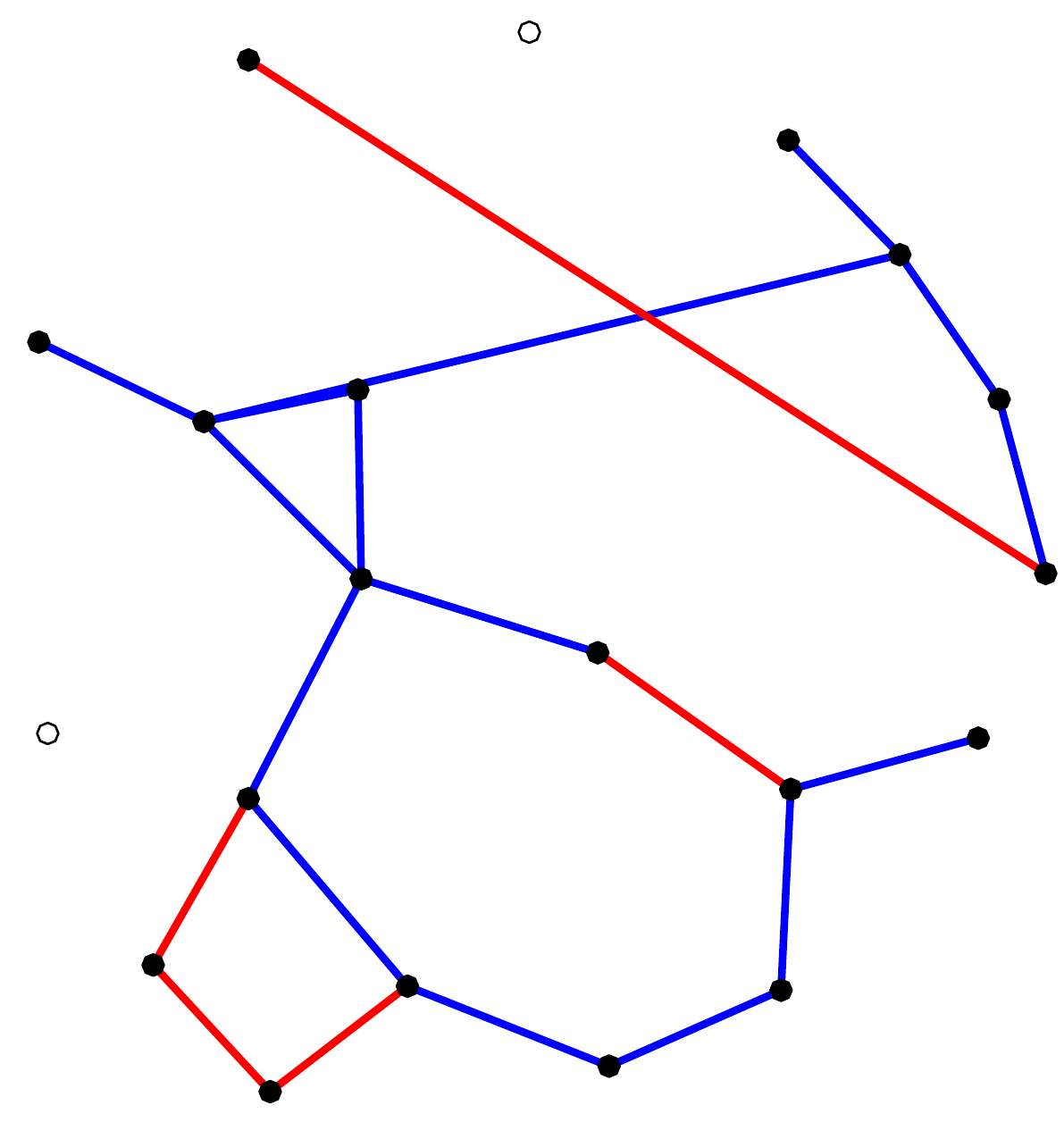}  \hspace*{3em}
  \includegraphics[width=.2\textwidth]{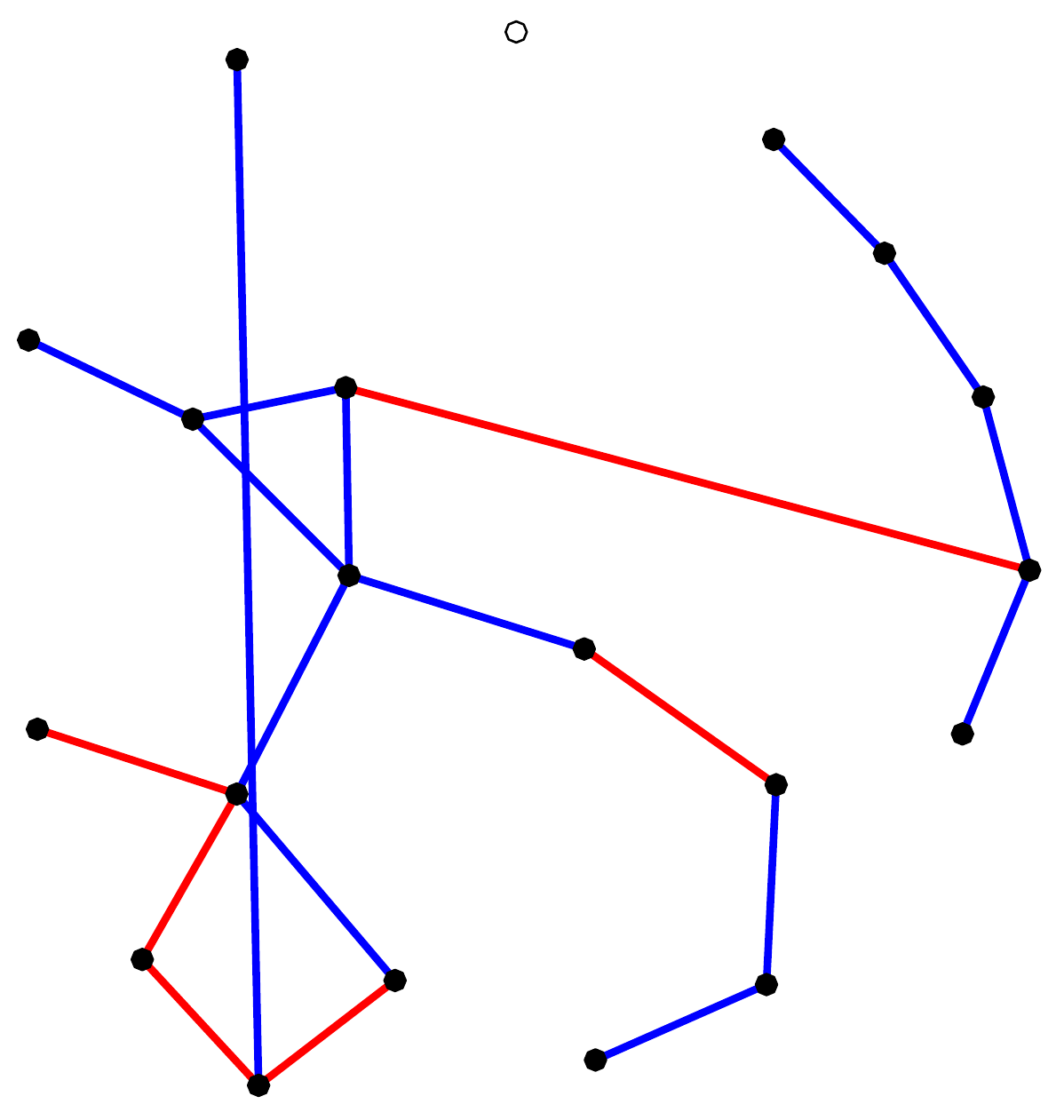}  
  \end{center}
  \caption{Set of simulated graphs: ancestor (top) and two
    children (bottom) engendered by a $\delta=2$
    perturbation.
   \label{fig:networks}
    }
\end{figure}

To generate $T$ samples stemming from a similar graph, we first
draw an ``ancestor'' graph with $p$ nodes and $k$ edges according to the Erd\H
os-R\'enyi model.
Here, we consider a simple setting with $T=4$ and a network with $p=20$ nodes
and $k=20$ edges, as illustrated in Figure \ref{fig:networks}.
Then, $T$ children graphs are produced by random addition and deletion of
$\delta$ edges in the ancestor graph.
The $T$ concentration matrices are built from the normalized
graph Laplacians, whose off-diagonal elements are slightly deflated to produce
strictly diagonally dominant matrices.
To allow for positively and negatively correlated variables, we generate a
strictly triangular matrix of random signs drawn from a Rademacher
distribution.  This matrix is symmetrized, complemented with ones on the
diagonal, and its component-wise multiplication
with the deflated Laplacians produces the ground-truth for the
concentration matrices $\mathbf{K}^{(1)},\dots,\mathbf{K}^{(T)}$.
Each $\mathbf{K}^{(t)}$ is finally used to generate $n$ Gaussian vectors with
zero mean and covariance ${\mathbf{K}^{(t)}}^{-1}$.

\subsubsection{Experimental Setup}

The precision-recall plots are computed by considering the cumulative number of
true and false edge detections among the $T=4$ children networks. 
Let ${\cal E}^{(t)}$ be the set of edges for children network $t$, precision and 
recall are respectively formaly defined as: 
\[
  \frac{\displaystyle \sum_{t=1}^T \sum_{(i,j)\in {\cal E}^{(t)}} \mathbf{1}(\widehat{K}_{ij}^{(t)})}
       {\displaystyle \sum_{t=1}^T \sum_{i > j}\mathbf{1}(\widehat{K}_{ij}^{(t)})} 
  \enspace
  \mbox{and}
  \enspace
  \frac{\displaystyle \sum_{t=1}^T \sum_{(i,j)\in {\cal E}^{(t)}} \mathbf{1}(\widehat{K}_{ij}^{(t)})}
       {\displaystyle \sum_{t=1}^T \left| {\cal E}^{(t)} \right|}
  \enspace,
\]
where  $\widehat{K}_{ij}^{(t)}$ is  the estimated  partial correlation
between variables $i$ and $j$ for network $t$, $\mathbf{1}(u)$ returns
$1$ if $u\neq  0$ and $0$ otherwise, and $\left|  {\cal E} \right|$ is
the cardinal of ${\cal E}$.

To  ensure   representativeness,  the  precision-recall   figures  are
averaged over  100 random draws  of the ancestor graph,  the averaging
being  performed  for  fixed  values  of  the  penalization  parameter
$\lambda$.    That   is,    each    point   in    the
  (precision,recall) plane  is the average of the  100 points obtained
  for each random  draw of the ancestor graph,  for a given estimation
  method and  for a  given value of  the penalization  parameter.  We
compare    our   proposals,    namely   the    Graphical   Intertwined
($\alpha=1/2$),  Cooperative and  Group  LASSO to  two baselines:  the
original      neighborhood       selection       of
\citet{2006_AS_Meinshausen}, either  applied separately to  each graph
(annotated ``independent''),  or computed on the data  set merging the
data originating from all graphs (annotated ``pooled'').

\subsubsection{Results}

\begin{figure}
  \begin{center}
    \begin{tabular}{@{}c@{\hspace{1.5ex}}c@{\hspace{1ex}}c@{}}
      \rotatebox{90.0}{\makebox[0cm]{$\delta=1$}} &
      \rotatebox{90.0}{\makebox[0cm]{Precision}} &
      \raisebox{-0.2\textwidth}{\includegraphics[width=0.4\textwidth]{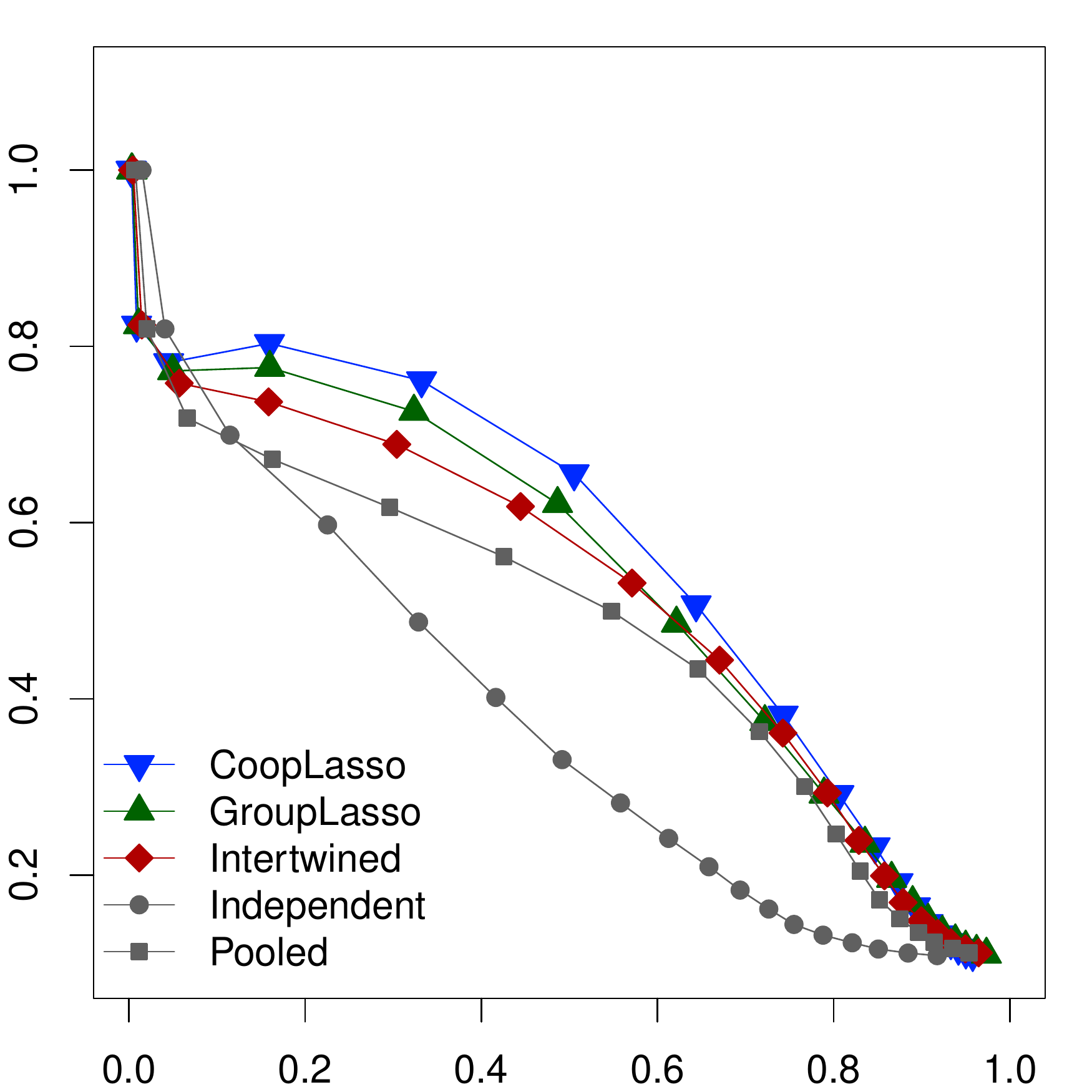}} \\
      & & Recall\\
      \rotatebox{90.0}{\makebox[0cm]{$\delta=3$}} &
      \rotatebox{90.0}{\makebox[0cm]{Precision}} &
      \raisebox{-0.2\textwidth}{\includegraphics[width=0.4\textwidth]{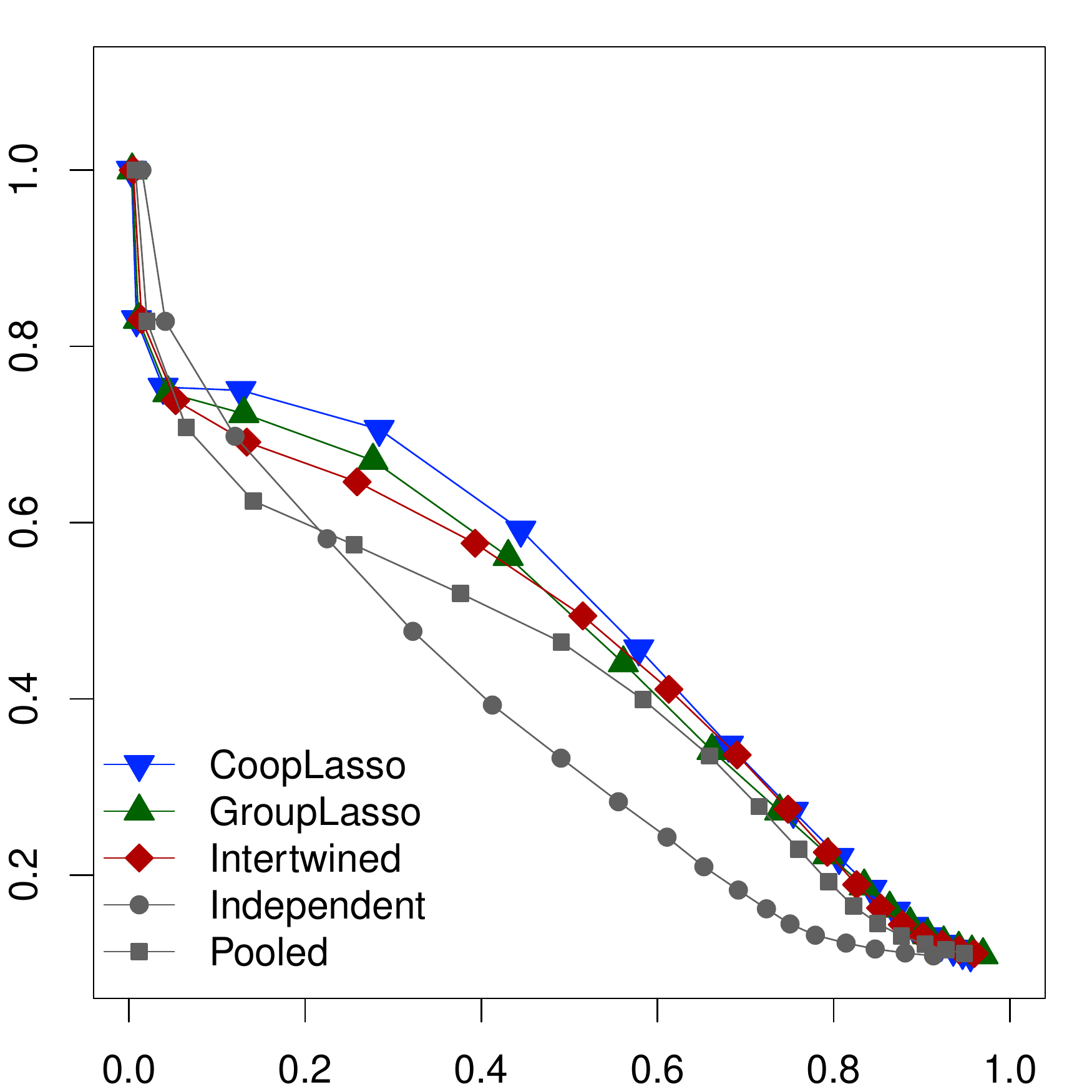}} \\
      & & Recall\\
      \rotatebox{90.0}{\makebox[0cm]{$\delta=5$}} &
      \rotatebox{90.0}{\makebox[0cm]{Precision}} &
      \raisebox{-0.2\textwidth}{\includegraphics[width=0.4\textwidth]{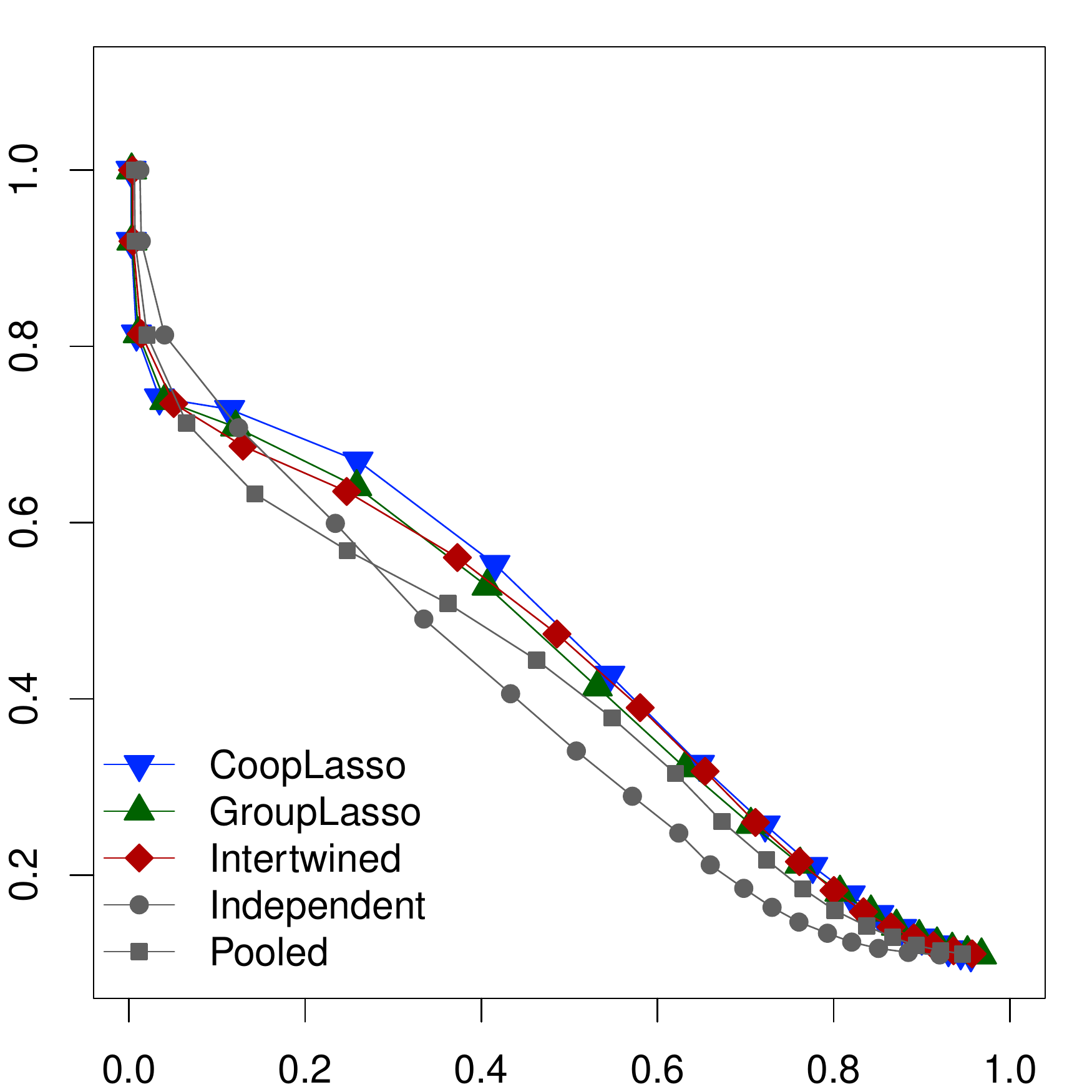}} \\
      & & Recall
    \end{tabular}
  \end{center}
  \caption{Precision-recall curves for the Intertwined, Cooperative, Group
    and the two baseline LASSO, for inferring four graphs (each with $p=20$
    nodes, $k=20$ edges and a perturbation $\delta$ from the ancestor graph)
    from four samples of size $n_t=25$.
    \label{fig:prcurvesa}}
\end{figure}

\begin{figure}
  \begin{center}
    \begin{tabular}{@{}c@{\hspace{1.5ex}}c@{\hspace{1ex}}c@{}}
      \rotatebox{90.0}{\makebox[0cm]{$\delta=1$}} &
      \rotatebox{90.0}{\makebox[0cm]{Precision}} &
      \raisebox{-0.2\textwidth}{\includegraphics[width=0.4\textwidth]{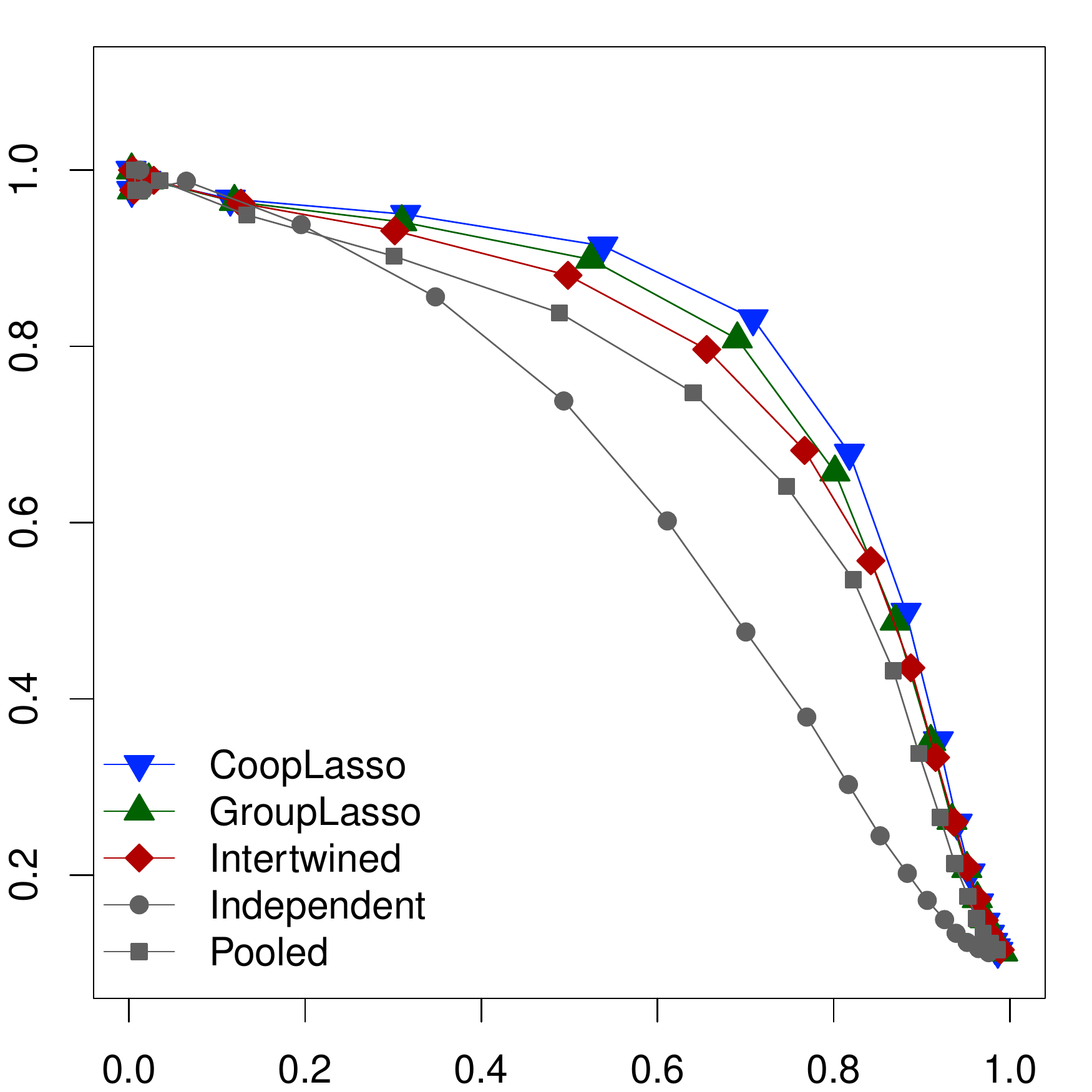}} \\
      & & Recall\\
      \rotatebox{90.0}{\makebox[0cm]{$\delta=3$}} &
      \rotatebox{90.0}{\makebox[0cm]{Precision}} &
      \raisebox{-0.2\textwidth}{\includegraphics[width=0.4\textwidth]{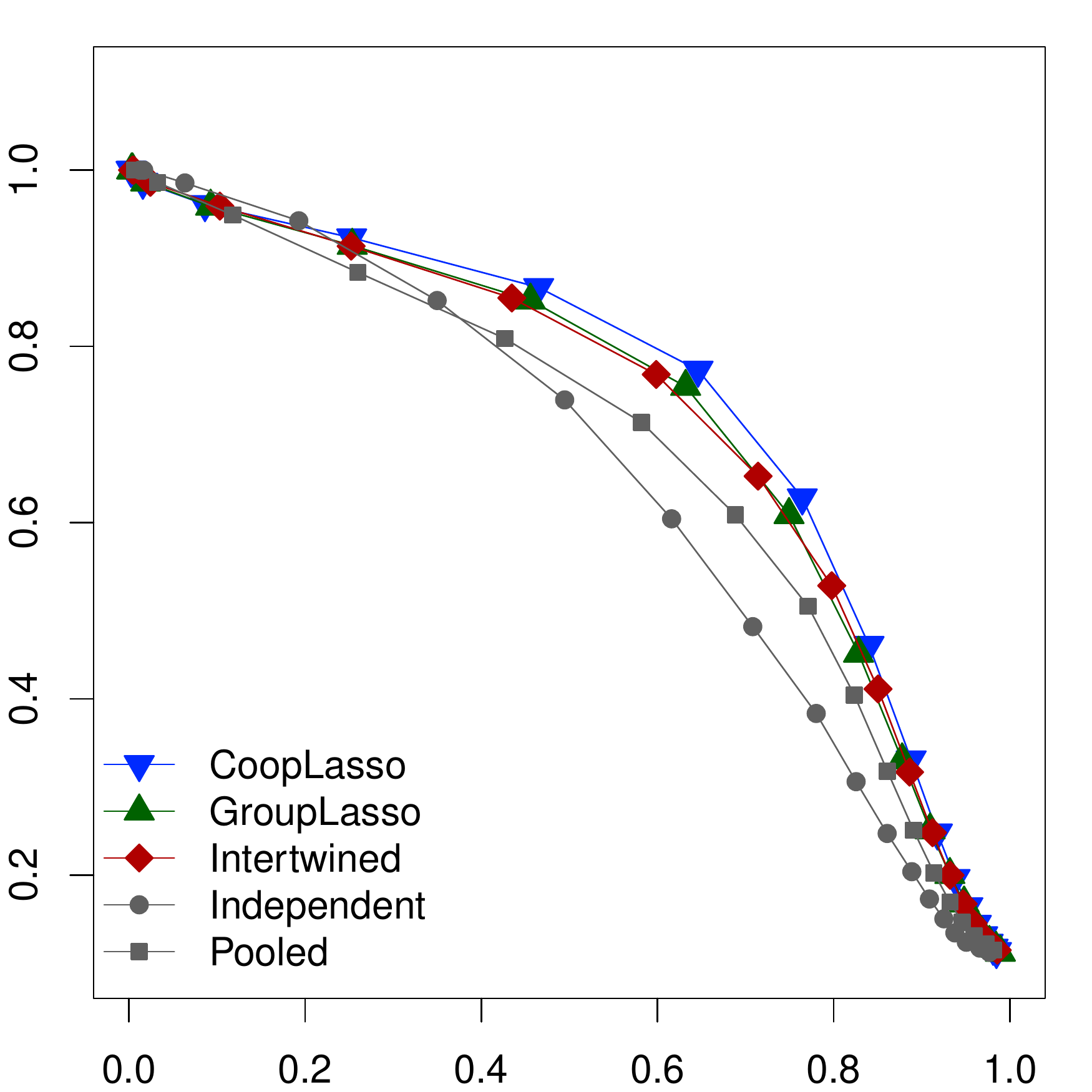}} \\
      & & Recall\\
      \rotatebox{90.0}{\makebox[0cm]{$\delta=5$}} &
      \rotatebox{90.0}{\makebox[0cm]{Precision}} &
      \raisebox{-0.2\textwidth}{\includegraphics[width=0.4\textwidth]{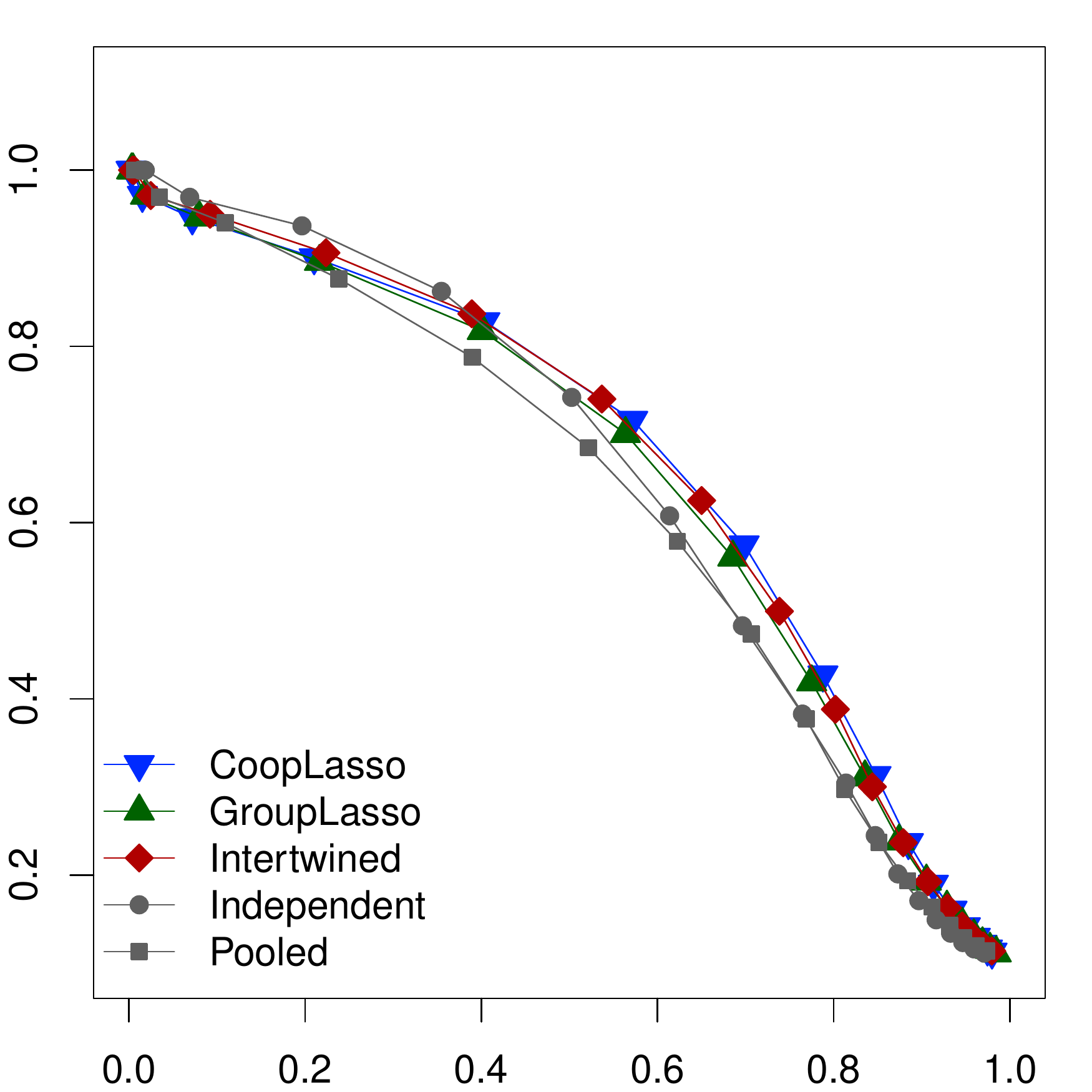}} \\
      & & Recall
    \end{tabular}
  \end{center}
    \caption{Precision-recall curves for the Intertwined, Cooperative, Group
    and the two baseline LASSO, for inferring four graphs (each with $p=20$
    nodes, $k=20$ edges and a perturbation $\delta$ from the ancestor graph)
    from four samples of size $n_t=50$.
    \label{fig:prcurvesb}}
\end{figure}

\begin{figure}
  \begin{center}
    \begin{tabular}{@{}c@{\hspace{1.5ex}}c@{\hspace{1ex}}c@{}}
      \rotatebox{90.0}{\makebox[0cm]{$\delta=1$}} &
      \rotatebox{90.0}{\makebox[0cm]{Precision}} &
      \raisebox{-0.2\textwidth}{\includegraphics[width=0.4\textwidth]{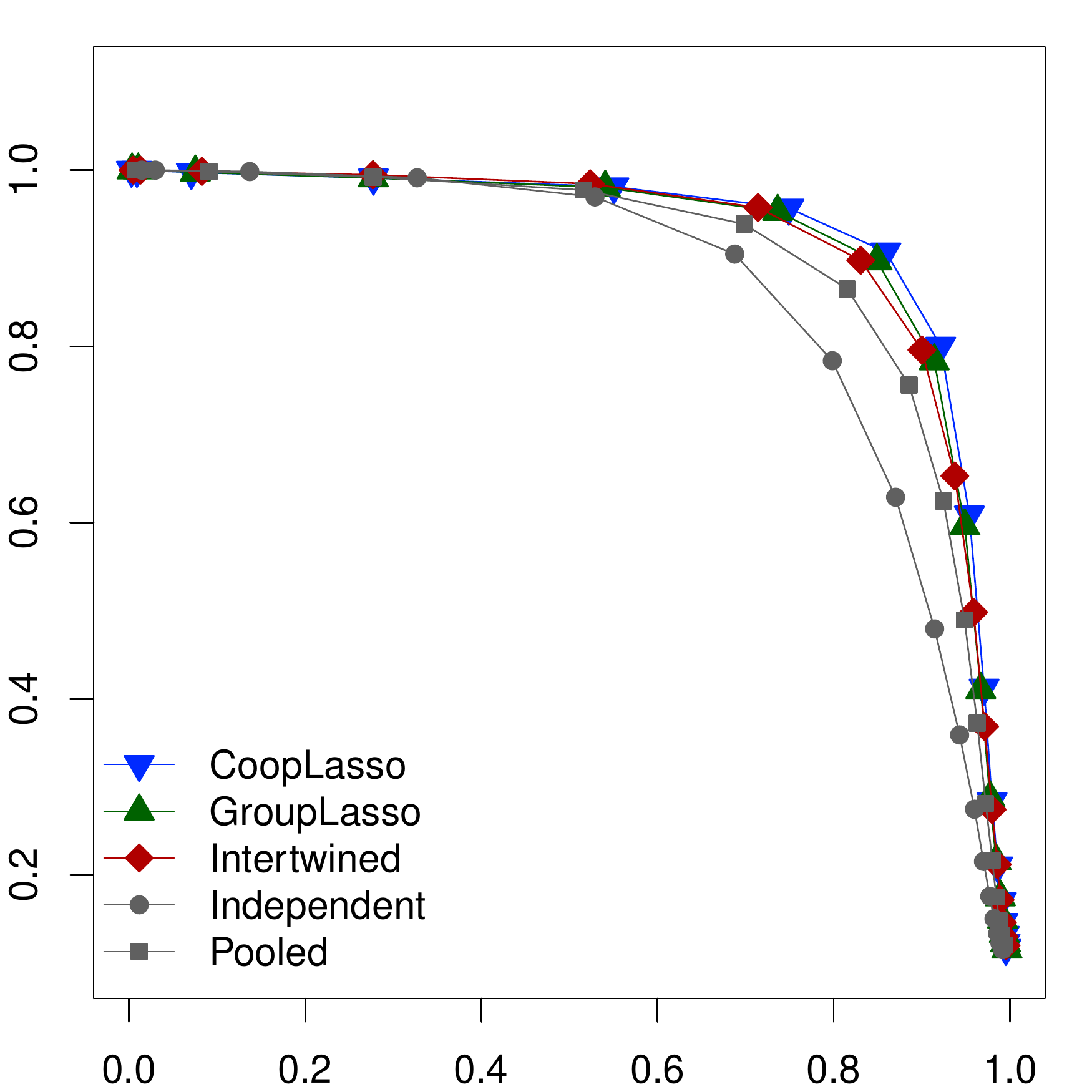}} \\
      & & Recall\\
      \rotatebox{90.0}{\makebox[0cm]{$\delta=3$}} &
      \rotatebox{90.0}{\makebox[0cm]{Precision}} &
      \raisebox{-0.2\textwidth}{\includegraphics[width=0.4\textwidth]{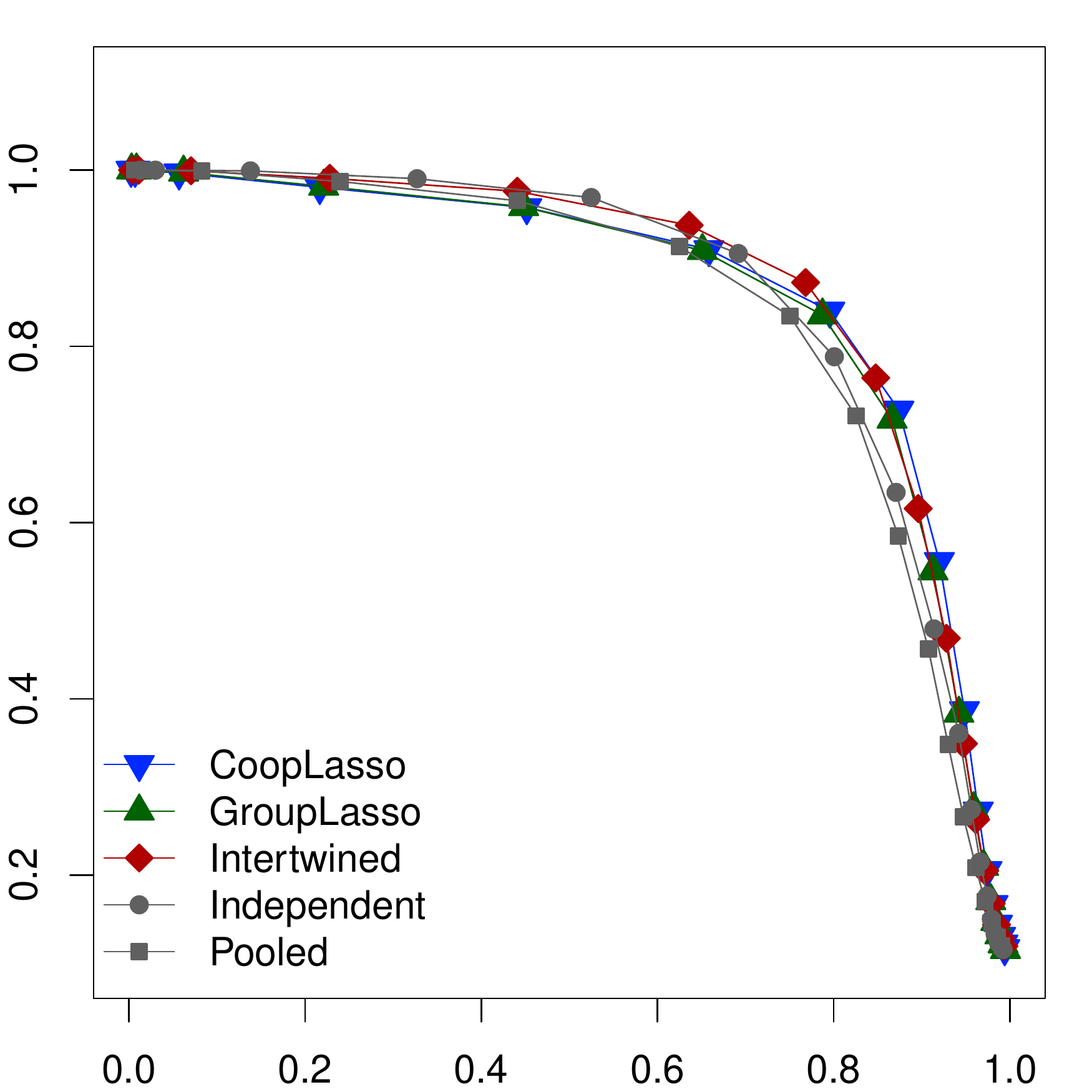}} \\
      & & Recall\\
      \rotatebox{90.0}{\makebox[0cm]{$\delta=5$}} &
      \rotatebox{90.0}{\makebox[0cm]{Precision}} &
      \raisebox{-0.2\textwidth}{\includegraphics[width=0.4\textwidth]{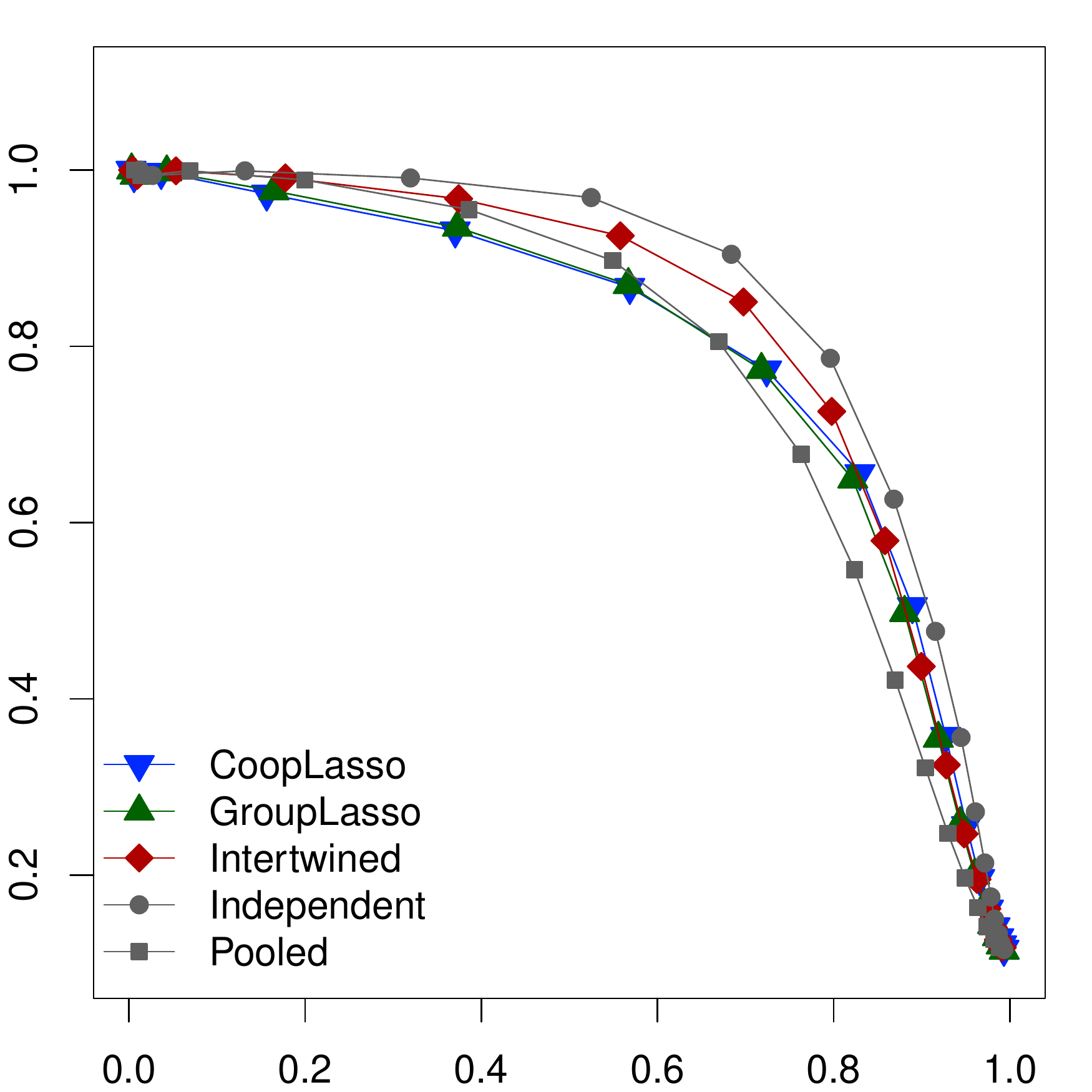}} \\
      & & Recall
    \end{tabular}
  \end{center}
  \caption{Precision-recall curves for the Intertwined, Cooperative, Group
    and the two baseline LASSO, for inferring four graphs (each with $p=20$
    nodes, $k=20$ edges and a perturbation $\delta$ from the ancestor graph)
    from four samples of size $n_t=100$.
    \label{fig:prcurvesc}}
\end{figure}

Figures       \ref{fig:prcurvesa},       \ref{fig:prcurvesb}       and
\ref{fig:prcurvesc}   display    precision-recall   plots   for   nine
prototypical situations.
From  Figure \ref{fig:prcurvesa}  to  \ref{fig:prcurvesc}, the  sample
size increases,  and from top  to bottom, the  differentiation between
networks increases.
First, note  that the  independent strategy is  not influenced  by the
level of perturbation, yet only by the sub-sample size, as expected. 

The   top   graph  in   Figure   \ref{fig:prcurvesa}  represents   the
small-sample low-perturbation situation, where  merging data sets is a
good  strategy,  leveraging  the  independent  analysis.   The  latter
performs   poorly,  and  our   multi-task  approaches
  dominate the  pooled strategy, the  Cooperative-LASSO being superior
  to  the Group-LASSO,  which  has the  advantage  on the  Intertwined
  LASSO.  The medium/large-sized-sample low-perturbation (top graph in
  Figures      \ref{fig:prcurvesb}      and      \ref{fig:prcurvesc}),
  small/medium-sized-sample   medium-perturbation  (middle   graph  in
  Figures    \ref{fig:prcurvesa}    and    \ref{fig:prcurvesb})    and
  small-sized-sample   large-perturbation  (bottom  graph   in  Figure
  \ref{fig:prcurvesa})   are    qualitatively   similar,   with   less
  differences  between  all  competitors. For  the
  large-sample    medium-perturbation   (middle   graph    in   Figure
  \ref{fig:prcurvesc})    and   medium-sized-sample   low-perturbation
  (bottom  graph  in Figure  \ref{fig:prcurvesb})  cases, all  methods
  perform similarly.   There is a slight advantage  for the multi-task
  strategies  for  high  recalls,  and  a  slight  advantage  for  the
  independent   analysis  for   low   recalls  (that   is,  for   high
  penalization, where  there is less  effective degrees of  freedom to
  determine).  The   bottom  graph  in   Figure  \ref{fig:prcurvesc}
represents the large-sample high-perturbation situation, where merging
data is  a bad strategy,  since the networks differ  significantly and
there  is enough  data to  estimate each  network  independently.  The
independent strategy  works best, closely followed  by the Intertwined
LASSO.    The  Cooperative  and   Group-LASSO  behave
  equally well for high recalls (low penalization parameters), but for
  highly  penalized solutions  (low recalls),  they  eventually become
  slightly worse than the pooled estimation.

These  experiments   show  that  our   proposals are
  valuable,  especially in the  most common  situation where  data are
  scarce.  Among  the baselines, the usual pooled  sample strategy is
good   in  the   small-sample  low-perturbation,   and   the  opposite
independent strategy  is better in  the large-sample high-perturbation
case.  The  intertwined LASSO is very  robust, in the
  sense  that  it  always  performs  favorably compared  to  the  best
  baseline method over the whole spectrum of situations.  Furthermore,
  except for the  large-sample high-perturbation case, the Group-LASSO
  performs even better, and the Cooperative-LASSO improves further the
  supremacy of the multiple graph inference approach.
    
\subsection{Protein Signaling Network}\label{sec:real}

Only a few real data sets come with a reliable and exhaustive ground-truth  
allowing quantitative assessments.
We make use of a multivariate flow cytometry data set pertaining to a
well-studied human T-cell signaling pathway \citep{2005_Science_Sachs}. 
The latter involves 11
signaling molecules (phospholipids and phosphorylated proteins) and 20
interactions described in the literature. 
The signaling network is
perturbed by activating or inhibiting the production of a given
molecule. Fourteen assays have been conducted, aiming to reveal different
part of the network. 
Here, we used only four assays (inhibition of PKC, activation of PKC, inhibition
of AKT, activation of PKA).

\begin{figure*}[htpb!]
  \centering
  \begin{tabular}{@{}c@{\hspace{1cm}}c@{}}
    \includegraphics[width=.325\textwidth]{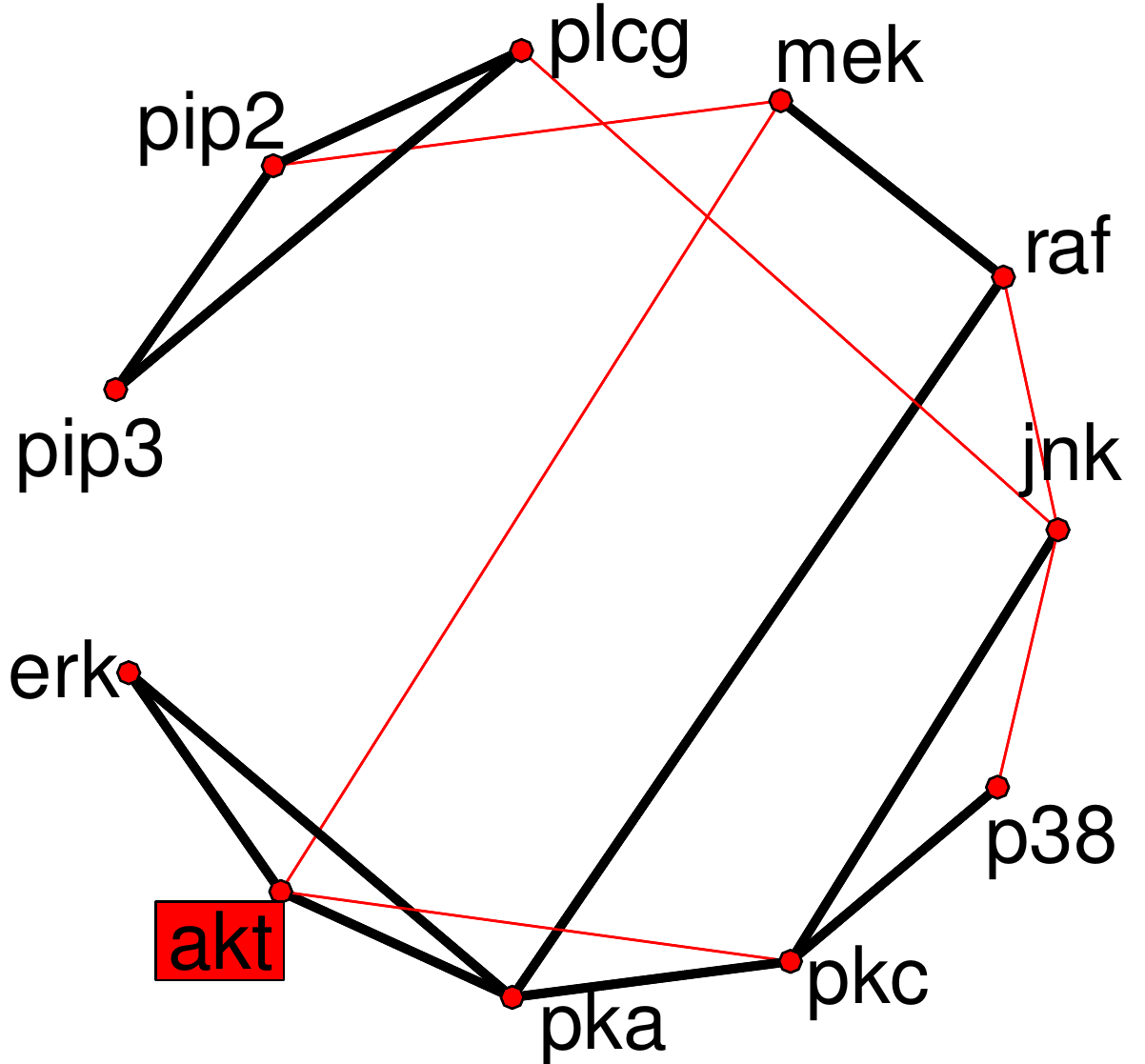} &
    \includegraphics[width=.325\textwidth]{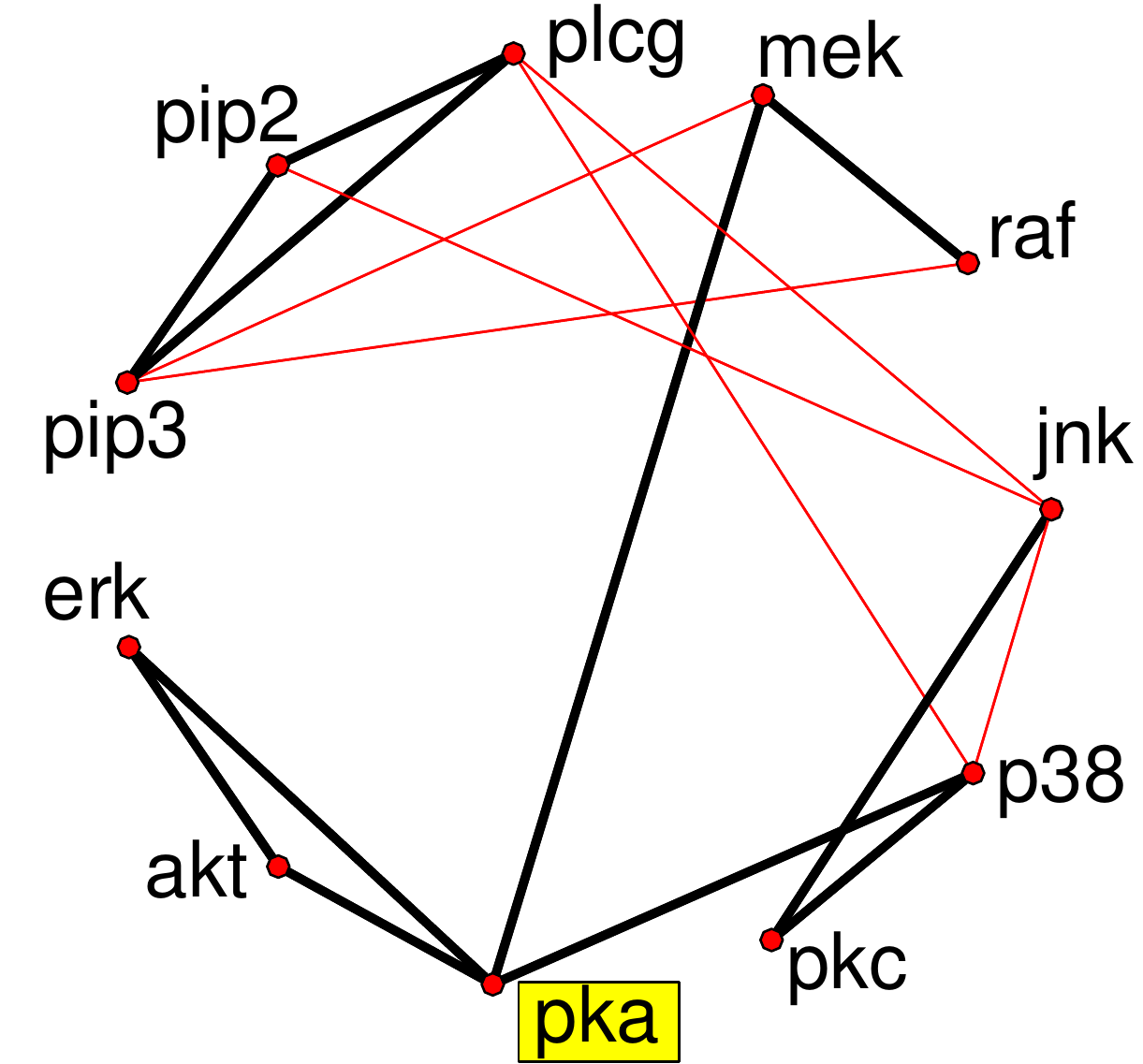} \\
    \includegraphics[width=.325\textwidth]{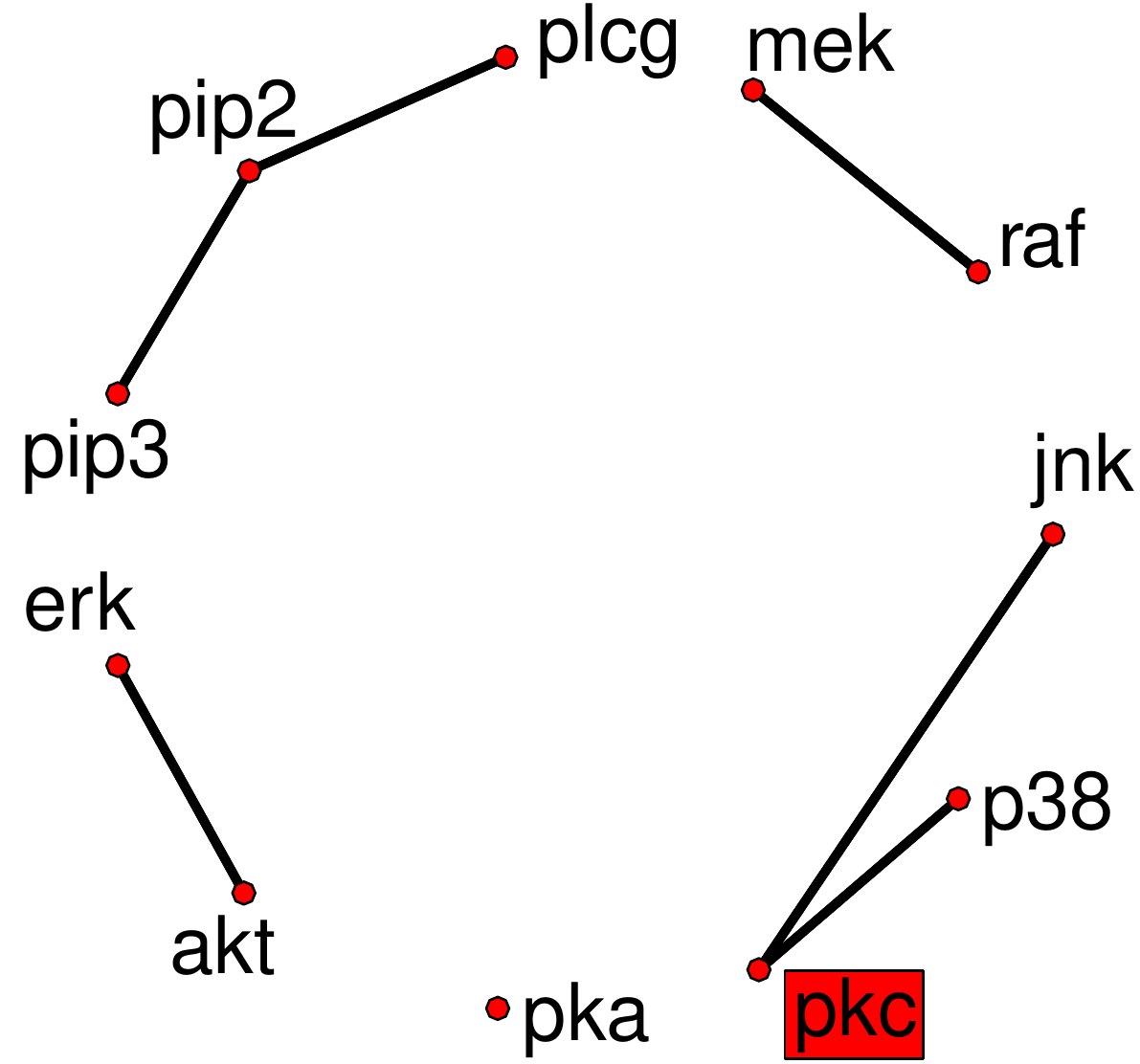} &
    \includegraphics[width=.325\textwidth]{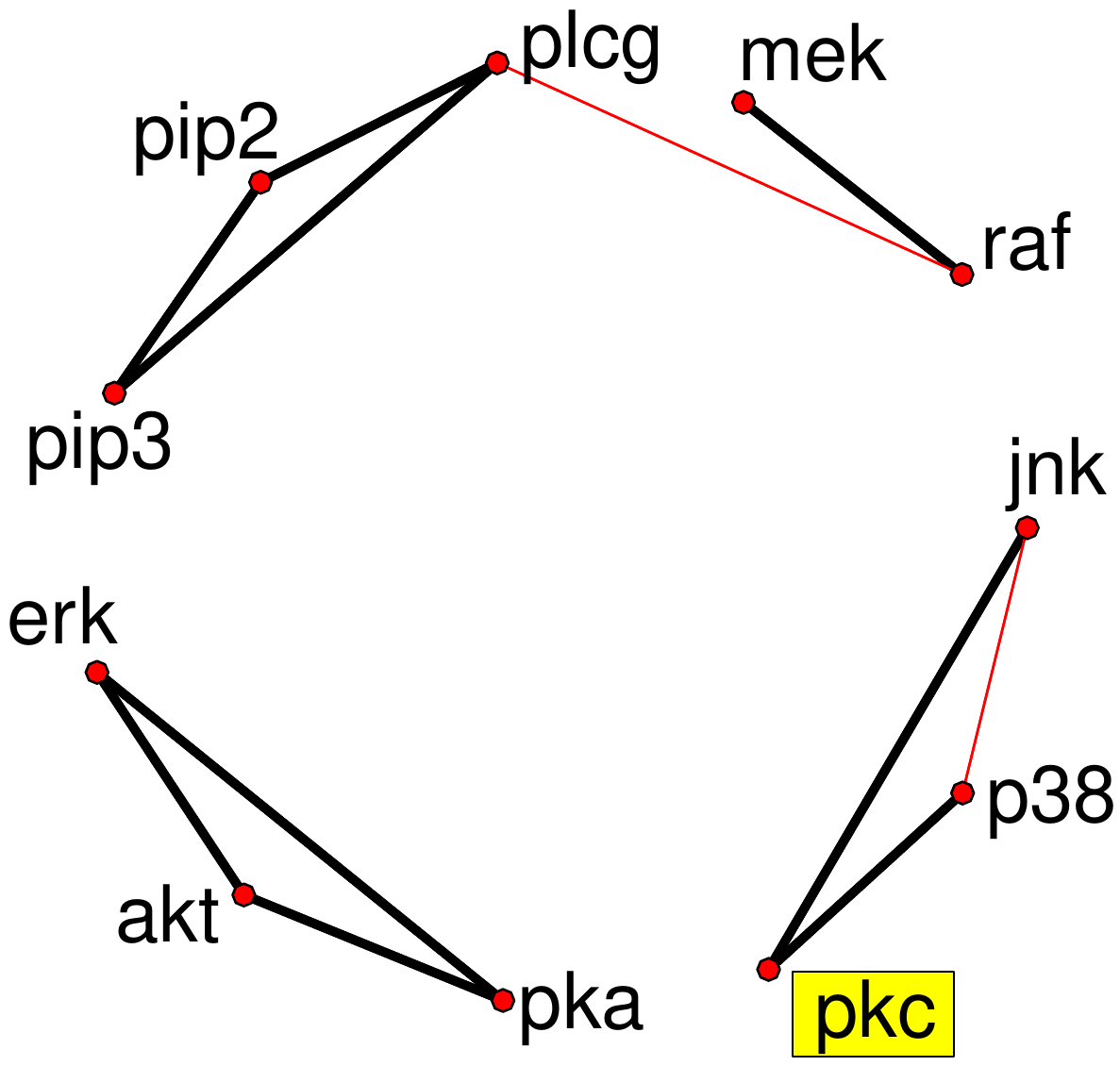}
  \end{tabular}
  \caption{Four graphs inferred from single assay. From left to right,
    top to bottom, we have respectively graphs inferred from an assay:
    inhibiting  akt,   activating  pka,  inhibiting   pkc,  activating
    pkc. Thick black lines represent  true positive and thin red lines
    are false positive. 
    \label{fig:sachs_indep}}
\end{figure*}

Graphs inferred  using only one assay  at a time show  that each assay
really   focus  on  different   part  of   the  network   (see  Figure
\ref{fig:sachs_indep}).

\begin{figure*}[htbp!]
  \centering
  \includegraphics[width=.325\textwidth]{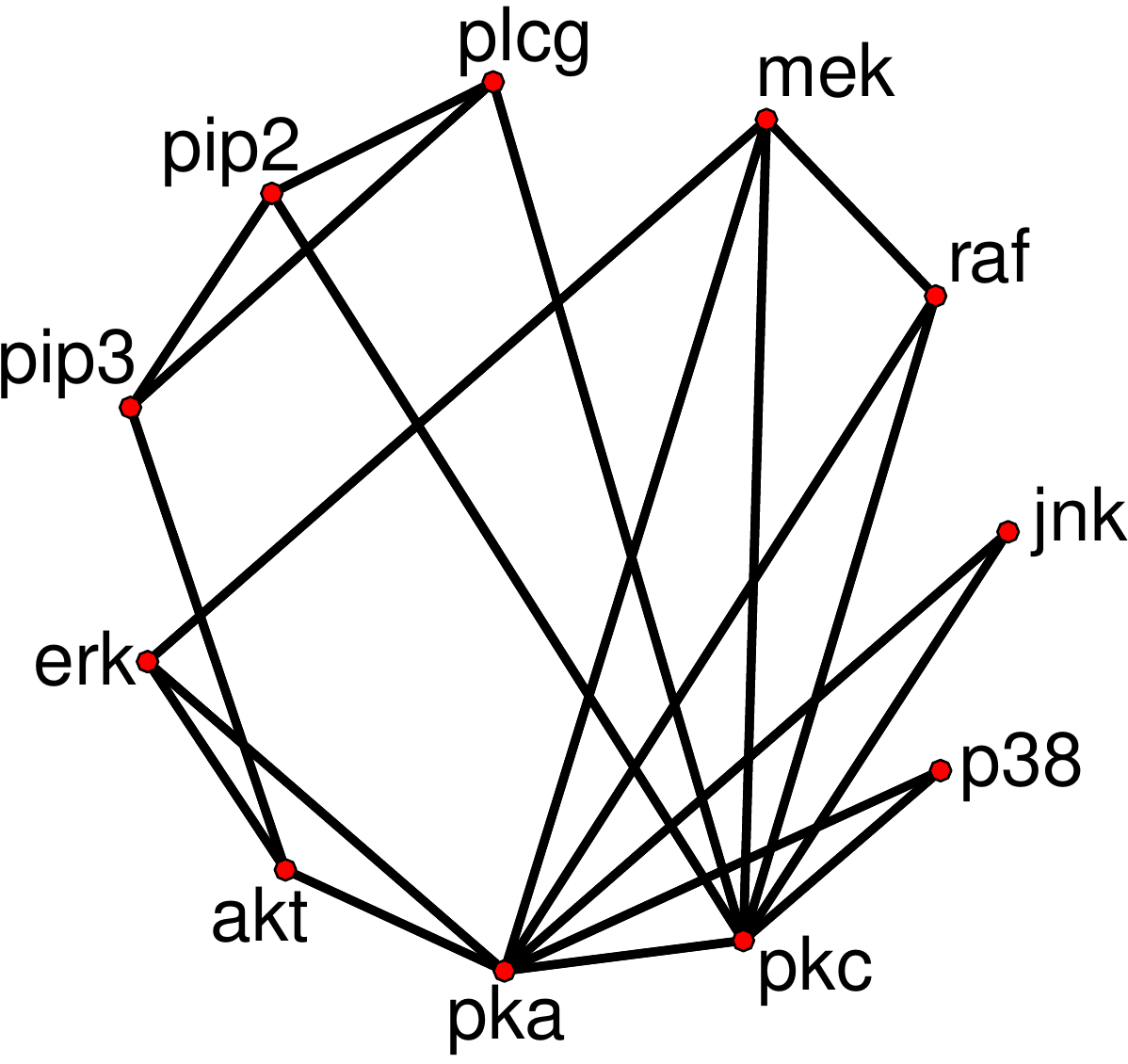} \hspace*{3em}
  \includegraphics[width=.325\textwidth]{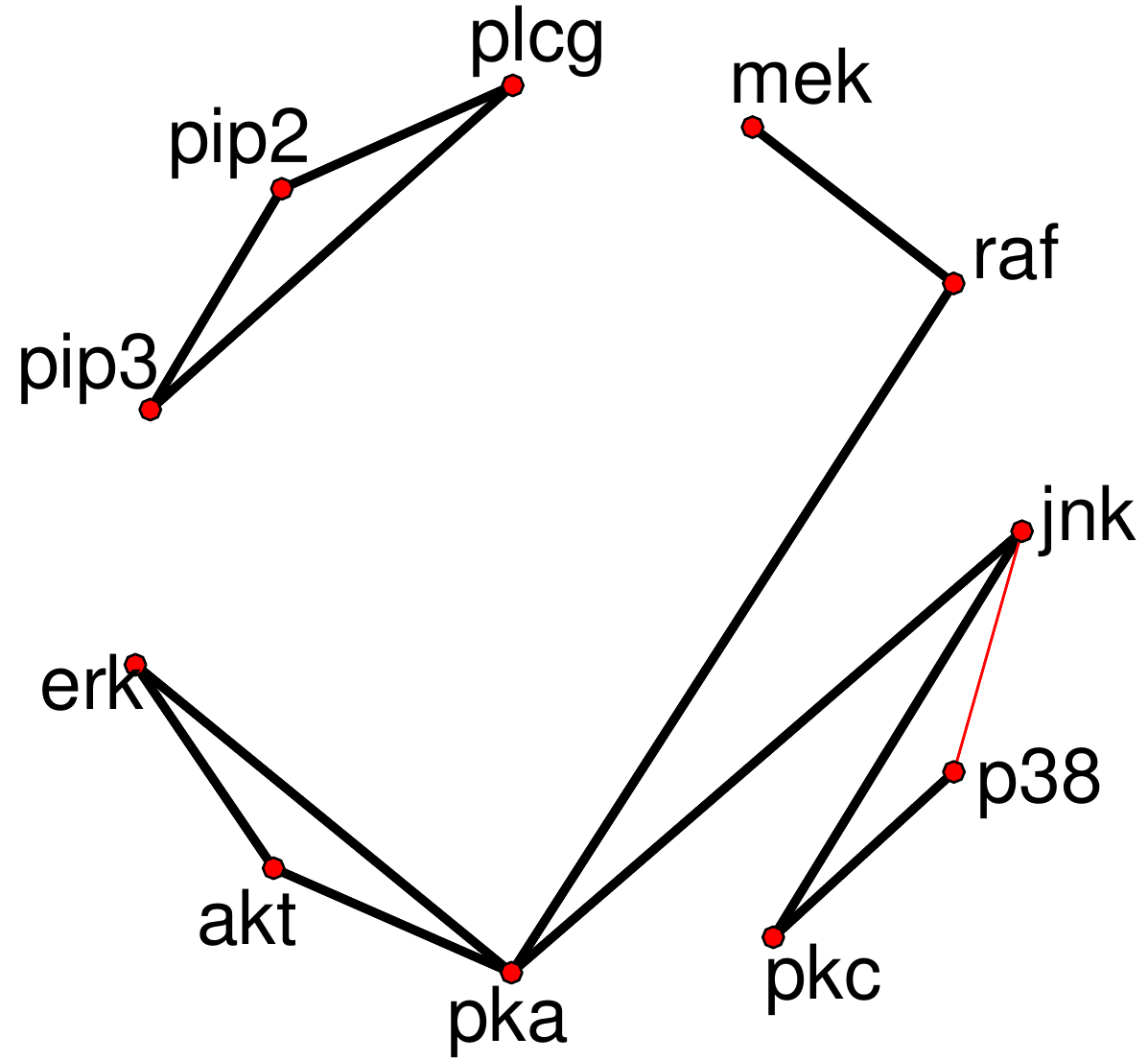} 
  \caption{Ground-truth pathway (left) and graph sum of the four graphs estimated
    by Intertwined LASSO using all data (right).
    Thick black lines represent true positive and thin red lines are false
    positive.
    \label{fig:sachs}}
\end{figure*}

When considering  a strategy based on inference  from multiple assays,
the first  false positive inferred by the  Intertwined Graphical LASSO
occurs when  11 true interactions out  of 20 are  detected (see Figure
\ref{fig:sachs}). This edge, between p38 and Jnk, is
  in  fact due  to an  indirect connection  via unmeasured  MAP kinase
  kinases  \citep{2005_Science_Sachs}, which is  a typical  problem of
  confounding   arising   in   this  context.    Considering   partial
  correlations within  the subset of available variables,  the edge is
  correctly detected, but  it is a false positive  with respect to the
  biological  ground   truth.   Furthermore,  in   larger  biological
networks, the absence of edge in the ground truth pathway often merely
means that there is yet no evidence that the co-regulation exists.  As
a result,  most real  data evaluation of  graph inference  methods are
based on qualitative subjective assessments by experts.

This caveat being, the various inference algorithms behave here as in the
synthetic experiments:
all inference methods perform about equally well for large samples (each assay
consists here of about 1000 repeated measurements).  

\begin{figure}
  \begin{center}
    \begin{tabular}{@{}c@{\hspace{1.5ex}}c@{\hspace{1ex}}c@{}}
      \rotatebox{90.0}{\makebox[0cm]{$n_t=7$}} &
      \rotatebox{90.0}{\makebox[0cm]{Precision}} &
      \raisebox{-0.2\textwidth}{\includegraphics[width=0.4\textwidth]{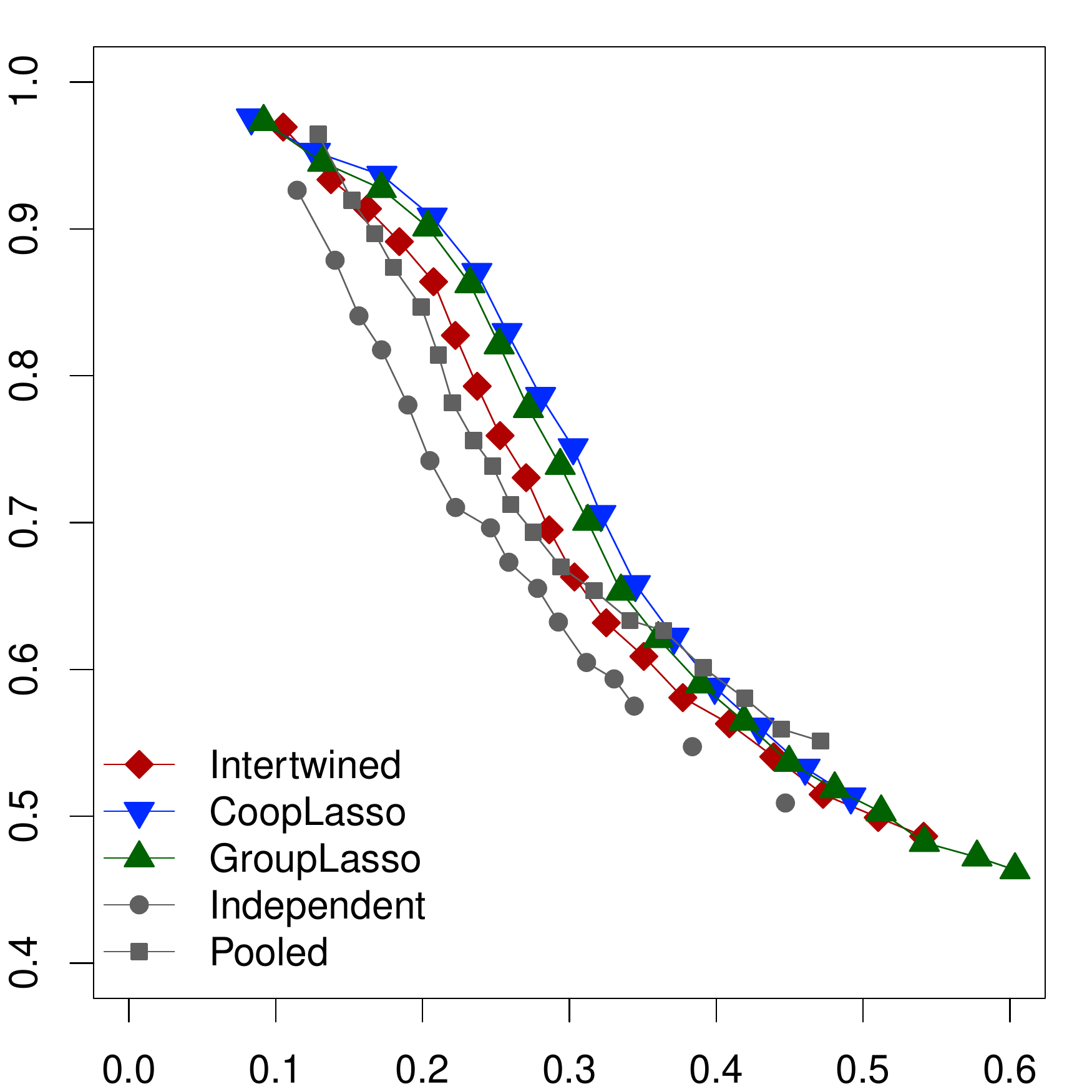}} \\
      & & Recall\\
      \rotatebox{90.0}{\makebox[0cm]{$n_t=10$}} &
      \rotatebox{90.0}{\makebox[0cm]{Precision}} &
      \raisebox{-0.2\textwidth}{\includegraphics[width=0.4\textwidth]{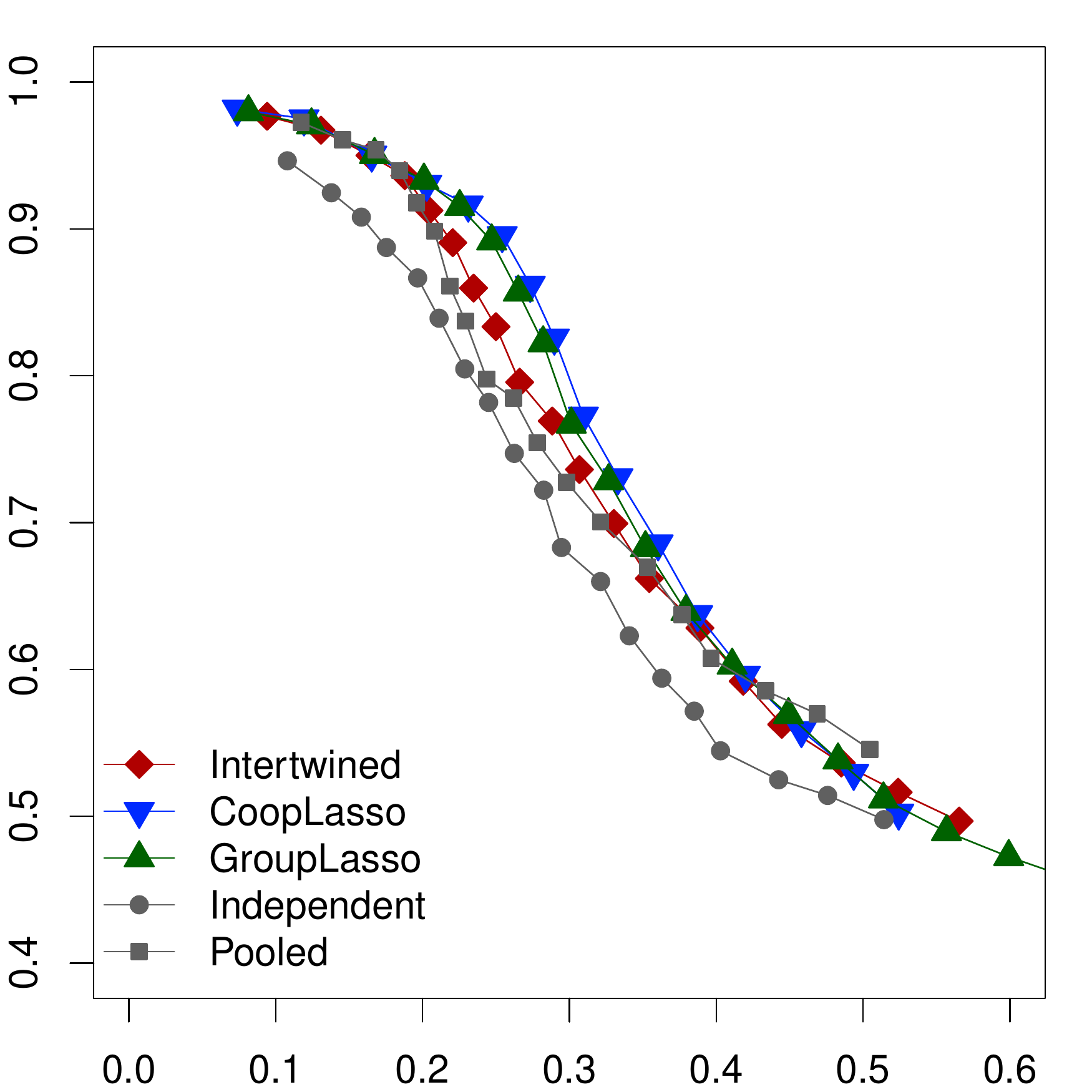}} \\
      & & Recall\\
      \rotatebox{90.0}{\makebox[0cm]{$n_t=20$}} &
      \rotatebox{90.0}{\makebox[0cm]{Precision}} &
      \raisebox{-0.2\textwidth}{\includegraphics[width=0.4\textwidth]{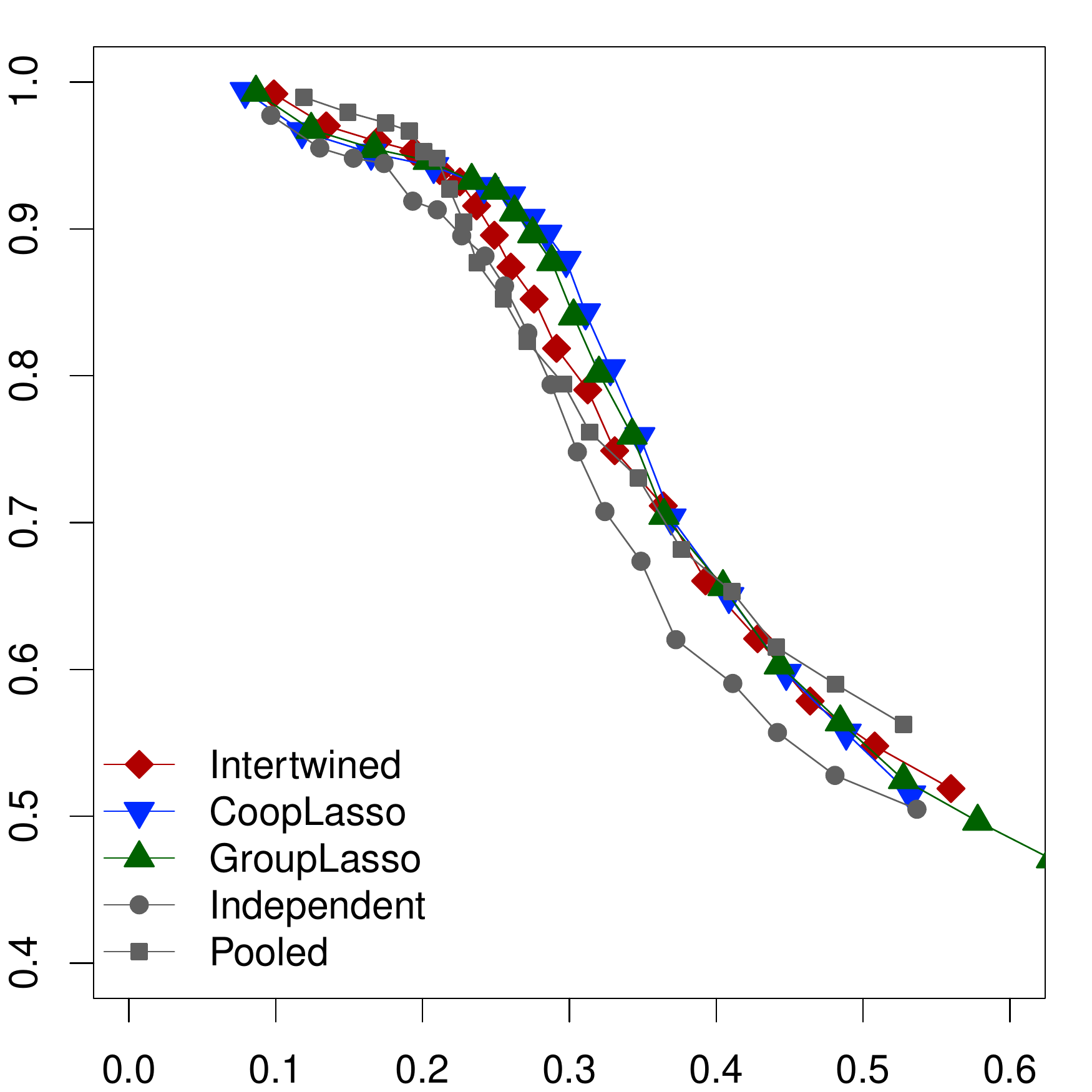}} \\
      & & Recall
    \end{tabular}
  \end{center}
    \caption{Precision-recall curves for the Intertwined, Cooperative, Group
    and the two baseline LASSO, for inferring the graphs on four assays of
    Sachs' data from four samples of size $n_t=7, 10$ and $20$.\label{fig:sachs_pr}}
\end{figure}

Figure  \ref{fig:sachs_pr}  displays the  results  obtained for  small
sample sizes.  Here also, the precision-recall plots are averaged over
100  independent  random draws  of  samples  of  size $n_t$,  that  is
$n=4n_t$      observations      over      the     four      considered
assays. It  is worth  noticing that the  large sample
  size limit is  almost obtained for $n_t=20$.  As  for the synthetic
experiments,  the  averaging is  performed  for  fixed  values of  the
penalization  parameter $\lambda$.  In  this situation,  our proposals
dominate  the  best baseline  strategy,  which  is pooled  estimation.
Again, the Intertwined LASSO  is very robust, but the
  Group-LASSO, and  to a greater extent  the Cooperative-LASSO perform
  better in the small-sample-size regime.


\section{Conclusion}

This paper  presents the first  methods dedicated to the  inference of
multiple   graphs   in  the   Gaussian   Graphical  Model   framework.
In this setup, the two baseline approaches consist in
  either handling the inference problems separately or as a single one
  by  merging the available  data sets.   Our proposals,  motivated by
  bioinformatics   applications,   were   devised  to   describe   the
  dependencies  between  pairs  of  variables in  analogous  operating
  conditions, such as measurements recorded in different assays.  This
  situation occurs routinely with omics data.

Our   approaches  are   based  on   the  neighborhood
  selection  of  \citet{2006_AS_Meinshausen}.   The  first  one,  the
Intertwined Graphical LASSO, relaxes the uniqueness constraint that is
implicit when the tasks are  processed as a single one, merely biasing
the  results  towards  a  common  answer.  Our  second  approach,  the
Graphical Cooperative-LASSO,  is based on a  group-penalty that favors
similar graphs,  with homogeneous dependencies between  the same pairs
of  variables.  Homogeneity is  quantified here  by the  magnitude and
sign  of partial  correlations.  The  Cooperative-LASSO  contrasts the
Group-LASSO in  being able to infer differing  graph structures across
tasks.  Our experimental results  show that our proposals are valuable
and robust,  consistently performing at least  as well as  the best of
the two baseline solutions.

The  algorithms  developed  in  this paper  are  made
  available within the \texttt{R}-package \texttt{simone} from version
  1.0-0  and  later.   This  package  also  embeds  extension  of  the
  multi-task   framework   to   time-course   data,  that   is,   when
  transcriptomics   data  are  collected   by  considering   the  same
  individual   across  time.   This   implementation  builds   on  the
  $\ell_1$-penalized       VAR(1)       model       described       in
  \cite{2010_SAGMB_Charbonnier}.

As  future  work,  we  will  provide  a  theoretical
  analysis  of  the  Cooperative-LASSO  regarding  uniqueness  of  the
  solution   and   selection   consistency,  or   sparsistence,   that
  corresponds  here  to  the  asymptotic  convergence of  the  set  of
  detected edges towards the set of true edges.


\section*{Acknowledgments}

Yves Grandvalet was partially supported by the PASCAL2 Network of Excellence,
the European ICT FP7 under grant No 247022 - MASH, and the French National
Research Agency (ANR) under grant ClasSel ANR-08-EMER-002.
Christophe Ambroise and Julien Chiquet were partially supported by the ANR under
grants GD2GS ANR-05-MMSA-0013 and NeMo ANR-08-BLAN-0304-01.

\appendix
\section{Proofs}

\subsection{Derivation of the pseudo-log-likelihood}
\label{sec:derivation:pseudolikelihood}

We show here that the pseudo-log-likelihood
  \begin{equation}
    \label{eq:prop:pseudologlikelihood_def}
    \mathcal{L}(\mathbf{K}|\mathbf{X}) =  \sum_{i=1}^p \left(\sum_{k=1}^n
      \log \mathbb{P}(X_i^k|X^k_{\backslash i};\mathbf{K}_i)\right)
    \enspace,
  \end{equation}
  associated to a sample of size $n$ drawn independently from the multivariate
  Gaussian vector
  $X\sim\mathcal{N}(\mathbf{0}_p,{\boldsymbol\Sigma})$ reads
  \begin{equation*}
    \mathcal{L}(\mathbf{K}|\mathbf{X})    =   \frac{n}{2}   \log\det
    (\mathbf{D}) -\frac{n}{2} 
    \mathrm{Tr}\left(\mathbf{D}^{-\frac{1}{2}}\mathbf{K}     \mathbf{S}
      \mathbf{K}\mathbf{D}^{-\frac{1}{2}}\right)     -    \frac{np}{2}
    \log(2\pi)
  \enspace,
  \end{equation*}
  where $\mathbf{S}  = n^{-1} \mathbf{X}^\intercal  \mathbf{X}$ is the
  empirical variance-covariance matrix and $\mathbf{D}$ is the diagonal
  matrix such that $D_{ii} = K_{ii}$, for $i=1,\dots,p$.

\begin{proof}
  
  Since the joint distribution of $X^k$ is Gaussian, the distributions of
  $X_i^k$ conditioned on the remaining variables $X_{\backslash i}^k$ are also
  Gaussian.  Their parameters $(\mu_i^k,\sigma_i)$ are given by
  \begin{equation}
    \label{eq:mui_sigmai}
    \mu_i^k = {\boldsymbol \Sigma}_{i\backslash i}^\intercal {\boldsymbol \Sigma}^{-1}_{\backslash i \backslash i} X_{\backslash i}^k \enspace, \qquad \sigma_i
    = {\Sigma}_{ii} - {\boldsymbol \Sigma}_{i \backslash i}^\intercal {\boldsymbol \Sigma}^{-1}_{\backslash i \backslash i} {\boldsymbol \Sigma}_{i \backslash i}
    \enspace. 
  \end{equation}
  where ${\boldsymbol \Sigma}_{\backslash i\backslash i}$ is matrix ${\boldsymbol \Sigma}$ deprived
  of its $i$th column and its $i$th line, ${\boldsymbol \Sigma}_{i \backslash i}$ is the
  $i$th column of matrix ${\boldsymbol \Sigma}$ deprived of its $i$th element.

  As $\mathbf{K}=\mathbf{{\boldsymbol \Sigma}}^{-1}$, reordering the rows and
  columns of the matrices yields

  \begin{equation*}
    \begin{bmatrix}
      {\boldsymbol \Sigma}_{\backslash i\backslash i} & {\boldsymbol \Sigma}_{i \backslash i} \\
      {\boldsymbol \Sigma}_{i \backslash i}^\intercal & {\Sigma}_{ii} \\
    \end{bmatrix}
    \times
    \begin{bmatrix}
      \mathbf{K}_{\backslash i\backslash i} & \mathbf{K}_{i \backslash i} \\
      \mathbf{K}_{i \backslash i}^\intercal & K_{ii} \\
   \end{bmatrix}
    =
    \begin{bmatrix}
      I_{p-1} & 0\\
      0       & 1 \\
    \end{bmatrix},
  \end{equation*}
  where $\mathbf{K}_{\backslash i\backslash i}$ is matrix $\mathbf{K}$ deprived
  of its $i$th column and its $i$th line, $\mathbf{K}_{i \backslash i}$ is the
  $i$th column of matrix $\mathbf{K}$ deprived of its $i$th element, and $I_{p-1}$
  is the identity matrix of size $p-1$.
  Two of these blockwise equalities are rewritten as follows:
  \begin{gather*}
    \Sigma_{ii} = (1- {\boldsymbol \Sigma}_{i \backslash i}^\intercal 
                      \mathbf{K}_{i \backslash i})/K_{ii} \enspace,\\
    {\boldsymbol \Sigma}_{\backslash i \backslash i}^{-1} 
    {\boldsymbol \Sigma}_{i\backslash i} = 
     -\mathbf{K}_{i \backslash i}/K_{ii} \enspace.
  \end{gather*}
  Using the above identities in \eqref{eq:mui_sigmai}, we obtain
  \begin{gather*}
    \sigma_i = (1- {\boldsymbol \Sigma}_{i \backslash i}^\intercal 
      \mathbf{K}_{i \backslash i})/K_{ii} + {\boldsymbol \Sigma}_{i\backslash i}^\intercal
      \mathbf{K}_{i \backslash i} /K_{ii} =
    1/K_{ii} \enspace,\\
    {\boldsymbol \mu}_i   =   -\mathbf{K}_{i    \backslash
      i}^\intercal {\mathbf X}_{\backslash i}^\intercal /K_{ii}.
  \end{gather*}
  where ${\boldsymbol \mu}_i= (\mu_i^1,\dots,\mu_i^n)^\intercal$.

  Using these notations and the corresponding blockwise notations for
  $\mathbf{S}$ ($S_{ii} = n^{-1} {\mathbf X}_i^\intercal {\mathbf X}_i$,
  $\mathbf{S}_{i\backslash i} = n^{-1} {\mathbf X}_{\backslash i}^\intercal
  {\mathbf X}_i$ and
  $\mathbf{S}_{\backslash i\backslash i} = n^{-1} 
  {\mathbf X}_{\backslash i}^\intercal {\mathbf X}_{\backslash i}$), Equation
  \eqref{eq:prop:pseudologlikelihood_def} reads
  \begin{align}
    \mathcal{L}(\mathbf{K}|{\mathbf X}) & =  -\frac{n}{2} \sum_{i=1}^p \log \sigma_i
    -   \sum_{i=1}^p  \frac{1}{2\sigma_i}({\mathbf   X}_i-  {\boldsymbol
      \mu}_i)^\intercal ({\mathbf X}_i- {\boldsymbol \mu}_i) - \frac{np}{2}\log(2\pi)
      \nonumber \\
    & = \ \frac{n}{2} \sum_{i=1}^p \log K_{ii} - \frac{np}{2}\log(2\pi)
      \nonumber \\
    & \hspace{3em} -  \frac{n}{2} \sum_{i=1}^p K_{ii}\left(S_{ii}  + \frac{2}{K_{ii}}
      \mathbf{S}_{i  \backslash i}^\intercal  \mathbf{K}_{i \backslash
        i}  + \frac{1}{K_{ii}^2}\mathbf{K}_{i  \backslash i}^\intercal
      \mathbf{S}_{\backslash i \backslash i} \mathbf{K}_{i
        \backslash i} \right)
      \label{eq:separablelikelihood} \\
    & = \ \frac{n}{2}  \log \det  \mathbf{D}  - \frac{np}{2}\log(2\pi)  -
    \frac{n}{2}                                            \sum_{i=1}^p
    \frac{1}{K_{ii}}\left(     \mathbf{K}_{i}^\intercal     \mathbf{S}
      \mathbf{K}_{i} \right),
    \nonumber 
  \end{align}
  where  $\mathbf{K}_{i}$  is the  $i$th  column  of $\mathbf{K}$  and
  $\mathbf{D}$ is the diagonal matrix such that $D_{ii} = K_{ii}$.
  Finally, we use that $
  \sum_{i=1}^p\frac{1}{K_{ii}}\left(           \mathbf{K}_{i}^\intercal
    \mathbf{S}            \mathbf{K}_{i}           \right)           =
  \mathrm{Tr}(\mathbf{D}^{-\frac{1}{2}}\mathbf{K}            \mathbf{S}
  \mathbf{K}\mathbf{D}^{-\frac{1}{2}})$ to conclude the proof.
\end{proof}

\subsection{Blockwise Optimization of the pseudo-log-likelihood}
\label{sec:blockdecomposition}

\begin{proof}[Proof of Proposition~\ref{prop:blockwise_resolution}]
  From \eqref{eq:separablelikelihood}, we have
  \begin{equation}
    \label{eq:pseudolikelihood_result}
    \mathcal{L}(\mathbf{K}|\mathbf{S}) = 
    - \frac{n}{2}\sum_{i=1}^p \left(
      2 \mathbf{S}_{i\backslash i}^\intercal\mathbf{K}_{i\backslash i}
      +  \frac{1}{K_{ii}} \mathbf{K}_{i\backslash  i}^\intercal \mathbf{S}_{\backslash
        i\backslash i} \mathbf{K}_{i\backslash i}\right) + c,
  \end{equation}
  where $c$ does not depend on $K_{ij}$ with $j\neq i$.  Thus, if we discard the
  symmetry     constraint    on    $\mathbf{K}$,     maximizing    the
  pseudo-likelihood \eqref{eq:pseudolikelihood_result} with respect to
  the  non-diagonal  entries  of  $\mathbf{K}$ amounts to solve $p$
  independent  maximization  problems  with  respect  to 
  $\mathbf{K}_{i\backslash i}$\,, $i=1,\ldots,p$.  The summands of
  \eqref{eq:pseudolikelihood_result} can be rewritten as
  \begin{multline*}
    -\frac{n}{2K_{ii}}   \left(  2   K_{ii}  \mathbf{S}_{i  \backslash i}^\intercal
      \mathbf{K}_{i \backslash i} + \mathbf{K}_{i \backslash i}^\intercal
      \mathbf{S}_{\backslash i \backslash i}
      \mathbf{K}_{i  \backslash i}\right)
    \\
    = - \frac{nK_{ii}}{2} \left\|  K_{ii}^{-1} \mathbf{S}_{\backslash i \backslash i}^{1/2}
    \mathbf{K}_{i \backslash i} +\mathbf{S}_{\backslash i \backslash i}^{-1/2}
      \mathbf{S}_{i  \backslash i}\right\|_2^2 + c',
  \end{multline*}
  where $c'= {n}/{2} \, K_{ii} \mathbf{S}_{ i \backslash i} ^\intercal
  \mathbf{S}_{\backslash i \backslash i}\mathbf{S}_{ i \backslash i}$
  does not depend on $K_{ij}$ with $j\neq i$.
  Adding an $\ell_1$ penalty term on $\mathbf{K}_{i \backslash i}$ and
  defining $\boldsymbol\beta = K_{ii}^{-1} \mathbf{K}_{i \backslash i}$
  leads to the objective function of Problem
  \eqref{eq:pseudolikelihood_penalized_blockwise}, which concludes the
  proof.
\end{proof}

\subsection{Subdifferential for the Cooperative-LASSO}

By definition, for a convex function $g$, the subdifferential is
\[
\left.\partial g \right|_{\boldsymbol\beta_{0}}= 
  \left\{
    \boldsymbol\theta : \forall \boldsymbol\beta \,,\ g(\boldsymbol\beta) - g(\boldsymbol\beta_{0}) \geq
    \boldsymbol\theta^\top (\boldsymbol\beta- \boldsymbol\beta_{0})
  \right\}
\]
The function $g(\boldsymbol\beta) = 
  \left\| \left( \boldsymbol\beta\right)_+ \right\|_2 +
  \left\| \left(-\boldsymbol\beta\right)_+ \right\|_2$ 
has kinks whenever $\boldsymbol\beta$ has at least one zero component and that
it has either no positive or no negative component. 
There are thus three situations where the subdifferential does not reduce to the
gradient : 
\begin{enumerate}
  \item 
  $\left\| \left( \boldsymbol\beta_{0}\right)_+ \right\|_2=0$ and 
  $\left\| \left(-\boldsymbol\beta_{0}\right)_+ \right\|_2\neq0$,
  \item 
  $\left\| \left( \boldsymbol\beta_{0}\right)_+ \right\|_2\neq0$ and 
  $\left\| \left(-\boldsymbol\beta_{0}\right)_+ \right\|_2=0$,
  \item 
  $\left\| \left( \boldsymbol\beta_{0}\right)_+ \right\|_2=0$ and 
  $\left\| \left(-\boldsymbol\beta_{0}\right)_+ \right\|_2=0$, {\em i.e.}
  $\boldsymbol\beta_{0}=0$.
\end{enumerate}

For the first situation, denoting $\mathcal{A}$ the index of non-zero entries of
$\boldsymbol\beta_{0}$ and $\mathcal{A}^c$ its complement, the subdifferential
is defined as
\begin{multline}\label{eq:subdiff1:def}
  \Big\{
  \left\| \left(-\boldsymbol\beta_{0}\right)_+ \right\|_2^{-1} \boldsymbol\beta_{0}
  + \boldsymbol\theta
  : \boldsymbol\theta_{\mathcal{A}} = \boldmath{0} \\ 
  \text{and} \ 
 \forall \boldsymbol\beta_{\mathcal{A}^c} \,,\ \left\| \left( \boldsymbol\beta_{\mathcal{A}^c} \right)_+ \right\|_2 
\geq
    \boldsymbol\theta_{\mathcal{A}^c}^\top \boldsymbol\beta_{\mathcal{A}^c}
  \Big\}
  \enspace.
\end{multline}
The set of admissible $\boldsymbol\theta$ is explicitly given by
\begin{equation}\label{eq:subdiff1:sol}
  \left\{
  \boldsymbol\theta : 
  \boldsymbol\theta_{\mathcal{A}} = \boldmath{0} \,,\  
  \left\| \left( \boldsymbol\theta_{\mathcal{A}^c} \right)_+ \right\|_2 \leq 1
  \enspace \text{and} \enspace
  \left\| \left(-\boldsymbol\theta_{\mathcal{A}^c} \right)_+ \right\|_2 = 0
  \right\}
  \enspace.
\end{equation}
\begin{proof}
We first show that, for any $\boldsymbol\theta$ in the set defined
in~\eqref{eq:subdiff1:sol}, the inequality in definition~\eqref{eq:subdiff1:def} 
always holds. Dropping the subscript $\mathcal{A}^c$ for readability, we have:
\begin{eqnarray*}
  \boldsymbol\theta^\top \boldsymbol\beta & = &  
  \left( \boldsymbol\theta\right)_+^\top \boldsymbol\beta -  
  \left(-\boldsymbol\theta\right)_+^\top \boldsymbol\beta \\
  & = &  
  \left( \boldsymbol\theta\right)_+^\top \boldsymbol\beta \\
  & = &  
  \left( \boldsymbol\theta\right)_+^\top \left( \boldsymbol\beta\right)_+ -
  \left( \boldsymbol\theta\right)_+^\top \left(-\boldsymbol\beta\right)_+ \\
  & \leq &
  \left( \boldsymbol\theta\right)_+^\top \left( \boldsymbol\beta\right)_+ 
  \leq
  \left\|\left( \boldsymbol\theta\right)_+ \right\|_2 
  \left\|\left( \boldsymbol\beta\right)_+ \right\|_2  
  \leq
  \left\|\left( \boldsymbol\beta\right)_+ \right\|_2  
  \enspace.
\end{eqnarray*}

To finish the proof, it is sufficient to exhibit some
$\boldsymbol\beta$ such that the inequality in
definition~\eqref{eq:subdiff1:def} does not hold when
$\left\| \left( \boldsymbol\theta \right)_+ \right\|_2 > 1$ or when
$\left\| \left(-\boldsymbol\theta \right)_+ \right\|_2 > 0$.
For $\left\| \left( \boldsymbol\theta \right)_+ \right\|_2 > 1$, we choose 
$\boldsymbol\beta =  \left( \boldsymbol\theta \right)_+$,
yielding  
$\boldsymbol\theta^\top \boldsymbol\beta = 
 \left\| \left( \boldsymbol\theta \right)_+ \right\|_2^2$, and
\mbox{$\left\| \left( \boldsymbol\beta \right)_+ \right\|_2 = 
 \left\| \left( \boldsymbol\theta \right)_+ \right\|_2 <
 \left\| \left( \boldsymbol\theta \right)_+ \right\|_2^2$,}
hence 
$\left\| \left( \boldsymbol\beta \right)_+ \right\|_2 <
\boldsymbol\theta^\top \boldsymbol\beta$;
for $\left\| \left(-\boldsymbol\theta \right)_+ \right\|_2 >0$, we choose 
$\boldsymbol\beta =  -\left(-\boldsymbol\theta \right)_+$,
yielding  
$\boldsymbol\theta^\top \boldsymbol\beta = 
 \left\| \left(-\boldsymbol\theta \right)_+ \right\|_2^2 > 0$, and
$\left\| \left( \boldsymbol\beta \right)_+ \right\|_2 = 0$,
hence 
$\left\| \left( \boldsymbol\beta \right)_+ \right\|_2 <
\boldsymbol\theta^\top \boldsymbol\beta$.
\end{proof}

The second situation is treated as the 
first one, yielding 
\begin{multline*}
  \left.\partial g \right|_{\boldsymbol\beta_{0}}= 
  \Big\{
  \left\| \left( \boldsymbol\beta_{0}\right)_+ \right\|_2^{-1} \boldsymbol\beta_{0}
  + \boldsymbol\theta
  : \boldsymbol\theta_{\mathcal{A}} = \boldmath{0} \,,\  
  \left\| \left(-\boldsymbol\theta_{\mathcal{A}^c} \right)_+ \right\|_2 \leq 1
  \\  \text{and} \ 
  \left\| \left( \boldsymbol\theta_{\mathcal{A}^c} \right)_+ \right\|_2 = 0
  \Big\}
  \ .
\end{multline*}

For the last situation, the subdifferential, defined as
\begin{equation}\label{eq:subdiff2:def}
\left.\partial g \right|_{\boldsymbol\beta_{0}}= 
  \left\{
    \boldsymbol\theta : \forall \boldsymbol\beta \,,\ 
    \left\| \left( \boldsymbol\beta \right)_+ \right\|_2 +
    \left\| \left(-\boldsymbol\beta \right)_+ \right\|_2
\geq
    \boldsymbol\theta^\top \boldsymbol\beta
  \right\}
  \enspace,
\end{equation}
reads
\begin{equation}\label{eq:subdiff2:sol}
\left.\partial g \right|_{\boldsymbol\beta_{0}}= 
  \left\{
    \boldsymbol\theta : 
    \max\left(\left\| \left( \boldsymbol\theta \right)_+ \right\|_2,
              \left\| \left(-\boldsymbol\theta \right)_+ \right\|_2 \right)
  \leq
    1
  \right\}
  \enspace,
\end{equation}
\begin{proof}
We first show that, for all the elements of $\partial g$ as explicitly defined
in~\eqref{eq:subdiff2:sol}, the inequality in definition~\eqref{eq:subdiff2:def} 
always holds:
\begin{eqnarray*}
  \boldsymbol\theta^\top \boldsymbol\beta & = &  
  \left( \boldsymbol\theta\right)_+^\top \boldsymbol\beta -  
  \left(-\boldsymbol\theta\right)_+^\top \boldsymbol\beta \\
  & \leq &
  \left( \boldsymbol\theta\right)_+^\top \left( \boldsymbol\beta\right)_+ +  
  \left(-\boldsymbol\theta\right)_+^\top \left(-\boldsymbol\beta\right)_+ \\
  & \leq &
 \left\|\left( \boldsymbol\theta\right)_+ \right\|_2 
 \left\|\left( \boldsymbol\beta\right)_+ \right\|_2 +
 \left\|\left(-\boldsymbol\theta\right)_+ \right\|_2 
 \left\|\left(-\boldsymbol\beta\right)_+ \right\|_2 \\
 & \leq &
 \max\left(\left\| \left( \boldsymbol\theta \right)_+ \right\|_2,
           \left\| \left(-\boldsymbol\theta \right)_+ \right\|_2 \right)
 \left(
 \left\|\left( \boldsymbol\beta\right)_+ \right\|_2 +
 \left\|\left(-\boldsymbol\beta\right)_+ \right\|_2
 \right)
  \enspace.
\end{eqnarray*}
To finish the proof, it is sufficient to exhibit some
$\boldsymbol\beta$ such that the inequality in
definition~\eqref{eq:subdiff2:def} does not hold for
$\max\left(\left\| \left( \boldsymbol\theta \right)_+ \right\|_2, 
\left\| \left(-\boldsymbol\theta \right)_+ \right\|_2 \right) >1
$.
Without loss of generality, we assume
$\left\| \left( \boldsymbol\theta \right)_+ \right\|_2 > 1$,
and choose 
$\boldsymbol\beta =  \left( \boldsymbol\theta \right)_+$,
yielding  
$\boldsymbol\theta^\top \boldsymbol\beta = 
 \left\| \left( \boldsymbol\theta \right)_+ \right\|_2^2$, and
$\left\| \left( \boldsymbol\beta \right)_+ \right\|_2 +
 \left\| \left(-\boldsymbol\beta \right)_+ \right\|_2 = 
 \left\| \left( \boldsymbol\beta \right)_+ \right\|_2 =
  \left\| \left( \boldsymbol\theta \right)_+ \right\|_2 <
  \left\| \left( \boldsymbol\theta \right)_+ \right\|_2^2$,
hence 
$\left\| \left( \boldsymbol\beta \right)_+ \right\|_2  +
 \left\| \left(-\boldsymbol\beta \right)_+ \right\|_2<
\boldsymbol\theta^\top \boldsymbol\beta$.
\end{proof}


\bibliography{biblio_common}

\begin{thebibliography}{26}
\providecommand{\natexlab}[1]{#1}
\providecommand{\url}[1]{\texttt{#1}}
\expandafter\ifx\csname urlstyle\endcsname\relax
  \providecommand{\doi}[1]{doi: #1}\else
  \providecommand{\doi}{doi: \begingroup \urlstyle{rm}\Url}\fi

\bibitem[Ambroise et~al.(2009)Ambroise, Chiquet, and
  Matias]{2009_article_Ambroise}
C.~Ambroise, J.~Chiquet, and C.~Matias.
\newblock Inferring sparse {G}aussian graphical models with latent structure.
\newblock \emph{Electron. J. Stat.}, 3:\penalty0 205--238, 2009.

\bibitem[Argyriou et~al.(2008)Argyriou, Evgeniou, and Pontil]{Argyriou08}
A.~Argyriou, T.~Evgeniou, and M.~Pontil.
\newblock Convex multi-task feature learning.
\newblock \emph{Mach. Learn.}, 73\penalty0 (3):\penalty0 243--272, 2008.

\bibitem[Banerjee et~al.(2008)Banerjee, {El Ghaoui}, and
  d'Aspremont]{2008_JMLR_Banerjee}
O.~Banerjee, L.~{El Ghaoui}, and A.~d'Aspremont.
\newblock Model selection through sparse maximum likelihood estimation for
  multivariate {G}aussian or binary data.
\newblock \emph{J. Mach. Learn. Res.}, 9:\penalty0 485--516, 2008.

\bibitem[Baxter(2000)]{2000_JAIR_Baxter}
J.~Baxter.
\newblock A model of inductive bias learning.
\newblock \emph{J. Artif. Int. Res.}, 12\penalty0 (1):\penalty0 149--198, 2000.

\bibitem[Bengio et~al.(2005)Bengio, Mari{\'e}thoz, and Keller]{Bengio05}
S.~Bengio, J.~Mari{\'e}thoz, and M.~Keller.
\newblock The expected performance curve.
\newblock In \emph{In {ICML} Workshop on ROC Analysis in Machine Learning},
  2005.

\bibitem[Caruana(1997)]{Caruana97}
R.~Caruana.
\newblock Multitask learning.
\newblock \emph{Mach. Learn.}, 28\penalty0 (1):\penalty0 41--75, 1997.

\bibitem[Charbonnier et~al.(2010)Charbonnier, Chiquet, and
  Ambroise]{2010_SAGMB_Charbonnier}
C.~Charbonnier, J.~Chiquet, and C.~Ambroise.
\newblock Weighted-lasso for structured network inference from time course
  data.
\newblock \emph{Statistical Applications in Genetics and Molecular Biology},
  9\penalty0 (1), 2010.

\bibitem[Drummond and Holte(2006)]{Drummond06}
C.~Drummond and R.~C. Holte.
\newblock Cost curves: An improved method for visualizing classifier
  performance.
\newblock \emph{Mach. Learn.}, 65\penalty0 (1):\penalty0 95--130, 2006.

\bibitem[Efron(2009)]{2009_report_Efron}
B.~Efron.
\newblock The future of indirect evidence.
\newblock Technical Report 250, Division of Biostatistics, Stanford University,
  2009.

\bibitem[Friedman et~al.(2008)Friedman, Hastie, and
  Tibshirani]{2007_BS_Friedman}
J.~Friedman, T.~Hastie, and R.~Tibshirani.
\newblock Sparse inverse covariance estimation with the graphical lasso.
\newblock \emph{Biostatistics}, 9\penalty0 (3):\penalty0 432--441, 2008.

\bibitem[Friedman(1989)]{1989_JASA_Friedman}
J.~H. Friedman.
\newblock Regularized discriminant analysis.
\newblock \emph{J. Amer. Statist. Assoc.}, 84\penalty0 (405):\penalty0
  165--175, 1989.

\bibitem[Kim et~al.(2006)Kim, Kim, and Kim]{2006_SS_Kim}
Y.~Kim, J.~Kim, and Y.~Kim.
\newblock Blockwise sparse regression.
\newblock \emph{Statistica Sinica}, 16:\penalty0 375--390, 2006.

\bibitem[Kolar et~al.(2009)Kolar, Le~Song, and Xing]{2009_AAS_Kolar}
M.~K. Kolar, A.~A. Le~Song, and E.~P. Xing.
\newblock Estimating time-varying networks.
\newblock \emph{Ann. Appl. Stat.}, 2009.

\bibitem[Meinshausen and B{\"u}hlmann(2006)]{2006_AS_Meinshausen}
N.~Meinshausen and P.~B{\"u}hlmann.
\newblock High-dimensional graphs and variable selection with the lasso.
\newblock \emph{Ann. Statist.}, 34\penalty0 (3):\penalty0 1436--1462, 2006.

\bibitem[Nikolova(2000)]{Nikolova00}
M.~Nikolova.
\newblock Local strong homogeneity of a regularized estimator.
\newblock \emph{SIAM J. Appl. Math.}, 61\penalty0 (2):\penalty0 633--658, 2000.

\bibitem[Osborne et~al.(2000{\natexlab{a}})Osborne, Presnell, and
  Turlach]{2000_JCGS_Osborne}
M.~R. Osborne, B.~Presnell, and B.~A. Turlach.
\newblock On the {LASSO} and its dual.
\newblock \emph{J. Comput. Graph. Statist.}, 9\penalty0 (2):\penalty0 319--337,
  2000{\natexlab{a}}.

\bibitem[Osborne et~al.(2000{\natexlab{b}})Osborne, Presnell, and
  Turlach]{2000_JNA_Osborne}
M.~R. Osborne, B.~Presnell, and B.~A. Turlach.
\newblock A new approach to variable selection in least squares problems.
\newblock \emph{IMA J. Numer. Anal.}, 20\penalty0 (3):\penalty0 389--403,
  2000{\natexlab{b}}.

\bibitem[Ravikumar et~al.(2010)Ravikumar, Wainwright, and
  Lafferty]{2010_AS_Ravikumar}
P.~Ravikumar, M.~J. Wainwright, and J.~Lafferty.
\newblock High-dimensional {I}sing model selection using $\ell_1$-regularized
  logistic regression.
\newblock \emph{Ann. Statist.}, 38:\penalty0 1287--1319, 2010.

\bibitem[Rocha et~al.(2008)Rocha, Zhao, and Yu]{2008_preprint_Rocha}
G.~V. Rocha, P.~Zhao, and B.~Yu.
\newblock A path following algorithm for sparse pseudo-likelihood inverse
  covariance estimation (splice), 2008.

\bibitem[Roth and Fischer(2008)]{2008_ICML_Roth}
V.~Roth and B.~Fischer.
\newblock The group-lasso for generalized linear models: uniqueness of
  solutions and efficent algorithms.
\newblock In \emph{International Conference on Machine Learning}, 2008.

\bibitem[Sachs et~al.(2005)Sachs, Perez, Pe'er, Lauffenburger, and
  Nolan]{2005_Science_Sachs}
K.~Sachs, O.~Perez, D.~Pe'er, D.A. Lauffenburger, and G.P. Nolan.
\newblock Causal protein-signaling networks derived from multiparameter
  single-cell data.
\newblock \emph{Science}, 308:\penalty0 523--529, 2005.

\bibitem[Sch{\"a}fer and Strimmer(2005)]{2005_SAGMB_Schafer}
J.~Sch{\"a}fer and K.~Strimmer.
\newblock A shrinkage approach to large-scale covariance matrix estimation and
  implications for functional genomics.
\newblock \emph{Stat. Appl. Genet. Mol. Biol.}, 4\penalty0 (1), 2005.

\bibitem[Toh and Horimoto(2002)]{2002_Toh}
H.~Toh and K.~Horimoto.
\newblock Inference of a genetic network by a combined approach of cluster
  analysis and graphical gaussian modeling.
\newblock \emph{Bioinformatics}, 18:\penalty0 287--297, 2002.

\bibitem[Villers et~al.(2008)Villers, Schaeffer, Bertin, and
  Huet]{2008_SAGM_Villers}
F.~Villers, B.~Schaeffer, C.~Bertin, and S.~Huet.
\newblock Assessing the validity domains of graphical {G}aussian models in
  order to infer relationships among components of complex biological systems.
\newblock \emph{Stat. Appl. Genet. Mol. Biol.}, 7\penalty0 (2), 2008.

\bibitem[Yuan and Lin(2006)]{2006_Yuan}
M.~Yuan and Y.~Lin.
\newblock Model selection and estimation in regression with grouped variables.
\newblock \emph{J. R. Stat. Soc. Ser. B Stat. Methodol.}, 68\penalty0
  (1):\penalty0 49--67, 2006.

\bibitem[Yuan and Lin(2007)]{2007_Biometrika_Yuan}
M.~Yuan and Y.~Lin.
\newblock Model selection and estimation in the {G}aussian graphical model.
\newblock \emph{Biometrika}, 94\penalty0 (1):\penalty0 19--35, 2007.

\end{thebibliography}

\end{document}